\documentclass[english]{article}
\usepackage{lmodern}

\usepackage[T1]{fontenc}
\usepackage[latin9]{inputenc}
\usepackage{geometry}
\usepackage{color}
\usepackage{babel}
\usepackage{verbatim}
\usepackage{rotating}
\usepackage{float}
\usepackage{units}
\usepackage{multirow}
\usepackage{amsmath}
\usepackage{amsthm}
\usepackage{amssymb}
\usepackage{graphicx}
\usepackage[unicode=true,pdfusetitle,
 bookmarks=true,bookmarksnumbered=false,bookmarksopen=false,
 breaklinks=false,pdfborder={0 0 1},backref=page,colorlinks=true]
 {hyperref}
\usepackage{breakurl}

\makeatletter

\floatstyle{ruled}
\newfloat{algorithm}{tbp}{loa}
\providecommand{\algorithmname}{Algorithm}
\floatname{algorithm}{\protect\algorithmname}

 \theoremstyle{definition}
   \newtheorem{defn}{\protect\defnname}
  
  \theoremstyle{plain}
  \newtheorem{lemma}{\protect\lemmaname}
\theoremstyle{plain}
\newtheorem{thm}{\protect\theoremname}
\theoremstyle{plain}

\theoremstyle{plain}
\theoremstyle{plain}

\theoremstyle{plain}
\theoremstyle{plain}
\newtheorem{prop}{\protect\propositionname}
\theoremstyle{plain}

\usepackage{dsfont}
\hypersetup{pdftitle={Gibbs flow for approximate transport with applications to Bayesian
computation},pdfauthor={Heng et al},linkcolor=RoyalBlue,citecolor=RoyalBlue}
\usepackage[dvipsnames,svgnames,x11names,hyperref]{xcolor}

\@ifundefined{showcaptionsetup}{}{%
 \PassOptionsToPackage{caption=false}{subfig}}
\usepackage{subfig}
\makeatother

\providecommand{\defnname}{Definition}
\providecommand{\examplename}{Example}
\providecommand{\lemmaname}{Lemma}
\providecommand{\theoremname}{Theorem}
\providecommand{\assumptionname}{Assumption}
\providecommand{\propertyname}{Property}
\providecommand{\propositionname}{Proposition}

\begin{document}

\title{Gibbs flow for approximate transport with applications to Bayesian
computation}

\author{Jeremy Heng\thanks{ESSEC Business School; heng@essec.edu},  
Arnaud Doucet\thanks{University of Oxford and the Alan Turing Institute; doucet@stats.ox.ac.uk}, 
and Yvo Pokern\thanks{University College London; y.pokern@ucl.ac.uk} }
\date{}

\maketitle
\begin{abstract}
Let $\pi_{0}$ and $\pi_{1}$ be two distributions on the Borel space
{\normalsize{}$(\mathbb{R}^{d},\mathcal{B}(\mathbb{R}^{d}))$}. Any
measurable function $T:\mathbb{R}^{d}\rightarrow\mathbb{R}^{d}$ such
that $Y=T(X)\sim\pi_{1}$ if $X\sim\pi_{0}$ is called a transport
map from $\pi_{0}$ to $\pi_{1}$. For any $\pi_{0}$ and $\pi_{1}$,
if one could obtain an analytical expression for a transport map from
$\pi_{0}$ to $\pi_{1}$, then this could be straightforwardly applied
to sample from any distribution. One would map draws from an easy-to-sample
distribution $\pi_{0}$ to the target distribution $\pi_{1}$ using
this transport map. Although it is usually impossible to obtain an
explicit transport map for complex target distributions, we show here
how to build a tractable approximation of a novel transport map. This
is achieved by moving samples from $\pi_{0}$ using an ordinary differential
equation with a velocity field that depends on the full conditional
distributions of the target. Even when this ordinary differential
equation is time-discretized and the full conditional distributions
are numerically approximated, the resulting distribution of mapped
samples can be efficiently evaluated and used as a proposal within
sequential Monte Carlo samplers. 
We demonstrate significant gains over state-of-the-art sequential
Monte Carlo samplers at a fixed computational complexity on a variety
of applications.
\end{abstract}
\textbf{\small{}Keywords}{\small{}: Mass transport; Markov chain Monte
Carlo; Normalizing constants; Path Sampling; Sequential Monte Carlo.}{\small \par}

\section{Introduction\label{sec:Introduction}}

The use of the Bayesian formalism of inference is ubiquitous in many
areas of science. For statistical models of practical interest, implementation
usually relies on Monte Carlo methods to sample from the posterior
distribution which might be high dimensional and exhibit complex dependencies.
Most available Monte Carlo algorithms rely on proposal distributions
and the efficiency of these techniques is crucially dependent on whether
these proposals are able to capture important features of the target.
In this paper, we leverage ideas from the mass transport literature
to develop a new methodology to build efficient proposal distributions
which can be used within sequential Monte Carlo (SMC) samplers \cite{neal2001,chopin_2002,delmoraletal2006}.

Given initial and target distributions $\pi_{0}$ and $\pi_{1}$ defined
on $(\mathbb{R}^{d},\mathcal{B}(\mathbb{R}^{d}))$, which in a Bayesian
context may be interpreted as the prior and posterior, a transport
map is a measurable function $T:\mathbb{R}^{d}\rightarrow\mathbb{R}^{d}$
such that $Y=T(X)\sim\pi_{1}$ if $X\sim\pi_{0}$. The transport map
terminology arises from the fact that one can view $T$ as transporting
the probability mass represented by $\pi_{0}$ to the probability
mass represented by $\pi_{1}$. We will use the notation $\pi_{1}=(T)_{\#}\pi_{0}$
since $\pi_{1}$ is the push-forward measure of $\pi_{0}$ by $T$.
Characterizing the existence of transport maps has generated a large
literature in mathematics; see \cite{villani2008} for a recent review.
In particular, much work has been dedicated to the $L^{2}$ Monge-Kantorovich
problem, where one seeks the optimal transport map $T$ minimizing
the expected cost $\mathbb{E}|T(X)-X|^{2}$.

For the purposes of Monte Carlo simulation, any analytically tractable
transport map would allow us to map samples from $\pi_{0}$ to $\pi_{1}$.
However, even without imposing any optimality condition, such transport
maps have only been identified in simple scenarios; e.g. when both
$\pi_{0}$ and $\pi_{1}$ are Gaussian \cite[Remark 2.30]{peyre_cuturi_2017}.
To obtain an approximate transport map, \cite{meng2002,moselhymarzouk2012,parno2016}
proposed to minimize some measure of discrepancy between $(T_{\beta})_{\#}\pi_{0}$
and $\pi_{1}$, over a set of maps parametrized by a finite-dimensional
parameter $\beta$, e.g. a linear combination of some basis functions.
However, it can be difficult to identify an appropriate subspace of
candidate maps, and the resulting optimization problem is generally
non-convex unless stringent assumptions are made \cite{kim_etal_2013,parno2018}
and high dimensional in the absence of conditional independence structure
in the target $\pi_{1}$ \cite{spantini_etal_2018}. In this article,
we circumvent these difficulties by considering a different approach
to build approximate transport maps. 

The transport maps we will consider are derived from a fluid dynamics
interpretation of mass transport. Consider a curve of distributions
$\{\pi_{t}\}_{t\in(0,1)}$ connecting $\pi_{0}$ to $\pi_{1}$; e.g.
the geometric path $\pi_{t}\propto\pi_{0}^{1-\lambda\left(t\right)}\pi_{1}^{\lambda\left(t\right)}$
where $\lambda:\left[0,1\right]\rightarrow\left[0,1\right]$ is an
increasing smooth function satisfying $\lambda(0)=0$ and $\lambda(1)=1$.
The use of bridging distributions between distant $\pi_{0}$ and $\pi_{1}$
is at the core of many state-of-the-art Monte Carlo methods such as
path sampling \cite{gelmanmeng1998,oates2016} and annealed importance
sampling \cite{Crooks1998,jarzynski1997,neal2001,chopin_2002}. If
we view probability mass as an infinite ensemble of fluid particles,
the main idea is to move these particles deterministically, using
an ordinary differential equation (ODE) with a carefully designed
velocity field, so as to mimic the time evolution of $\pi_{t}$ over
the time interval $t\in[0,1]$. Loosely speaking, we may think of
the movement of particles under such a velocity field as implicitly
defining flow transport maps $\{T_{t}\}_{t\in[0,1]}$ satisfying $\pi_{t}=(T_{t})_{\#}\pi_{0}$
for each $t\in[0,1]$. 

The idea of constructing transport maps using flows originates from
\cite{moser1965}; see also \cite{Greene1979,dacorognamoser1990,benamou_brenier_2000}
for other early contributions. This approach has since been adopted
in a range of application domains ranging from engineering to physics
\cite{barronluo2007,CrisanXiong2010,daumhuangparticle2012,meyn2013,jarzynski2008}.
Noting that, for a given curve of distributions, there could be multiple
velocity fields achieving the flow transport, various optimality criteria
have been introduced to identify a unique solution \cite{moser1965,reichtransport2011,meyn2013};
e.g. \cite{reichtransport2011} proposed seeking the velocity field
minimizing kinetic energy. In these contributions, the optimal velocity
field is given by the solution of an elliptic partial differential
equation (PDE). However, when using a full grid, PDE solvers suffer
from the curse of dimensionality \cite{Dahmen2014,novak2009} which
could render them impractical. Sparse grid methods may be capable
of dealing with sufficiently high dimensions but they come with their
own set of approximations, e.g. tensor approximations \cite{chkifa2015,Dahmen2014}.
Using techniques from differential geometry, \cite{betancourt} constructed
a flow transport using contact Hamiltonian flows that also determines
$\lambda$ adaptively, but the velocity field depends on intractable
integrals on $\mathbb{R}^{d}$ which would have to be numerically
approximated. 

An alternative approach involves building analytically tractable approximations
of intractable flow transport maps. For example, in a Bayesian filtering
context where $\pi_{0}$ is a Gaussian prior distribution on unknown
states $X$, and the likelihood is also Gaussian distributed with
mean vector $\phi(X)$ and a known covariance matrix, \cite{bunchgodsill2014}
proposed linearizing $\phi$ locally to exploit analytical tractability
of Gaussian flows \cite{bergemannreich,reichMixture,cotter_reich}. 
This article also proposes approximate flow transport maps that are
analytically tractable, but the details of our construction are markedly
different. Our approach does not require any distributional assumptions
on $\pi_{0}$ and $\pi_{1}$, instead it is based on approximating
a novel flow that takes reference to the conditional distributions
$\pi_{t}(x_{1}|x_{2},\ldots,x_{d})$, $\pi_{t}(x_{2}|x_{3},\ldots,x_{d})$,
...,$\pi_{t}(x_{d-1}|x_{d})$ and the marginal distribution $\pi_{t}(x_{d})$
where $x_{i}\in\mathbb{R}$ for $i=1,\ldots,d$. As these distributions
are typically intractable, we propose a tractable approximation which
moves particles using a velocity field designed to track the full
conditional distributions $\{\pi_{t}(x_{i}|x_{-i})\}_{i=1,\ldots,p}$,
where $x_{i}\in\mathbb{R}^{d_{i}}$ and $x_{-i}=(x_{1},...,x_{i-1},x_{i+1},\ldots,x_{p})$.
We shall refer to the latter as the Gibbs flow in reference to the
Gibbs sampler. Contrary to existing transport-based methods, Gibbs
flow does not require selecting a parametric class of maps, solving
a non-convex optimization problem, approximating the solution of a
PDE or approximating $d$-dimensional integrals. Analogous to Gibbs
samplers, its implementation allows one to leverage any conditional
independence structure in the target $\pi_{1}$ and analytical tractability
of any full conditional distribution to move the corresponding component.
For components with intractable Gibbs flow, we will show that by further
blocking these components into one-dimensional components, the resulting
Gibbs velocity field only involves one-dimensional integrals w.r.t.
the corresponding full conditional distributions that can be efficiently
approximated using most quadrature routines. We will also introduce
a novel time discretization scheme reminiscent of the systematic scan
Gibbs sampler to numerically integrate the Gibbs flow. Although other
numerical integrators can also be considered, our scheme allows efficient
computation of the distribution of resulting mapped samples in high
dimensions, which is crucial when employing such distributions as
proposals within SMC samplers. Our approach only requires a computational
cost of $O(\sum_{i=1}^{p}d_{i}^{3})$ at each time step without requiring
additional approximations to reduce the computational complexity \cite{Han_Liu_2017}.
We establish various theoretical properties of the Gibbs flow and
demonstrate significant gains over state-of-the-art methods at a fixed
computational complexity on a variety of applications. 

The rest of the paper is organized as follows. In Section \ref{sec:Transport-with-flows},
we introduce the construction of transport maps using flows in a Bayesian
context. We present a novel flow transport, the Gibbs flow approximation
and its properties in Section \ref{sec:Solving-the-flow-transport-problem}.
We then discuss how the Gibbs flow can be numerically implemented
and employed as proposal distributions within SMC samplers in Section
\ref{sec:Gibbs-flow-samplers}. Lastly, in Section \ref{sec:Applications},
we illustrate the proposed methodology on a mixture model, a variance
component model, and a log-Gaussian Cox point process model. The proof
of all results are given in the Appendix. An R package
is available at \url{github.com/jeremyhengjm/GibbsFlow} to reproduce
all numerical results. 

\section{Transport with flows\label{sec:Transport-with-flows}}

\subsection{A curve from prior to posterior \label{subsec:A-path-from-prior-to-posterior}}

Let $\pi_{0}({\rm d}x)$ be a prior distribution on the Borel space
$(\mathbb{R}^{d},\mathcal{B}(\mathbb{R}^{d}))$ and $L:\mathbb{R}^{d}\rightarrow\mathbb{R}_{+}$
denote a likelihood function. To simplify presentation, we shall assume
that $\pi_{0}(\mathbf{{\rm d}}x)$ is absolutely continuous w.r.t.
the Lebesgue measure on $\mathbb{R}^{d}$, with an everywhere positive
density $x\mapsto\pi_{0}(x)$, and that $x\mapsto L(x)$ is also positive
everywhere and satisfies $\lim_{|x|\rightarrow\infty}L(x)=0$. We
will defer a discussion of improper priors to Section \ref{subsec:Variance-component-models}
and suppress all notational dependencies on observations. From Bayes'
rule, the resulting posterior distribution $\pi(\mathbf{{\rm d}}x)$
admits the density 
\begin{align}
 & \pi(x)=\frac{\pi_{0}(x)L(x)}{Z},\label{eq:targetposterior}
\end{align}
where $Z=\int_{\mathbb{R}^{d}}\pi_{0}(u)L(u)\:{\rm d}u$ denotes the
marginal likelihood. Henceforth we shall additionally assume that
$\pi_{0},L\in C^{1}(\mathbb{R}^{d},\mathbb{R}_{+})$, where $C^{k}(A,B)$
denotes the set of functions from $A$ to $B$ which are $k$-times
continuously differentiable.

We introduce a curve of distributions $\{\pi_{t}\}_{t\in[0,1]}$ smoothly
bridging the prior $\pi_{0}$ to the posterior $\pi_{1}=\pi$ by gradually
introducing the likelihood using a strictly increasing $C^{2}$-function
$\lambda:[0,1]\rightarrow[0,1]$ such that $\lambda(0)=0$ and $\lambda(1)=1$:
\begin{align}
\pi_{t}(x)=\frac{\gamma_{t}(x)}{Z(t)},\qquad & \gamma_{t}(x)=\pi_{0}(x)L(x)^{\lambda(t)},\label{eqn:tempering}
\end{align}
where $Z(t)=\int_{\mathbb{R}^{d}}\gamma_{t}(u)\,{\rm d}u$. The function
$\lambda$ is commonly known as inverse temperature in the context
of simulated annealing for optimization problems \cite{Kirkpatrick_Gelatt_Vecchi_1983}.
By differentiating (\ref{eqn:tempering}) w.r.t. the time variable
$t$, we obtain its time evolution along the curve 
\begin{align}
\partial_{t}\pi_{t}(x)=\lambda'(t)\left(\log L(x)-I_{t}\right)\pi_{t}(x),\label{eqn:derivdensity}
\end{align}
where $\lambda':[0,1]\rightarrow\mathbb{R}_{+}$ denotes the time
derivative of $\lambda$ and 
\begin{align}
I_{t}=\frac{1}{\lambda'(t)}\frac{{\rm d}}{{\rm d}t}\log Z(t)=\frac{\frac{{\rm d}}{{\rm d}t}\int_{\mathbb{R}^{d}}\pi_{0}(u)L(u)^{\lambda(t)}\,{\rm d}u}{\lambda'(t)Z(t)}=\mathbb{E}_{\pi_{t}}[\log L(X_{t})]\label{eqn:expectloglike}
\end{align}
is assumed to be finite for all $t\in[0,1]$. Under our assumptions,
the family of models $\{\pi_{t}\}_{t\in[0,1]}$ is regular so interchanging
the order of differentiation w.r.t. the time variable and integration
w.r.t. the spatial variable in the last equality of (\ref{eqn:expectloglike})
is valid. By integrating (\ref{eqn:expectloglike}) on the time interval
$[0,1]$, we recover the well-known path sampling identity \cite{gelmanmeng1998,oates2016}:
\begin{align}
\log Z=\int_{0}^{1}\lambda'(t)\,I_{t}\,{\rm d}t.
\end{align}

Equation (\ref{eqn:derivdensity}) reveals that the expected log-likelihood
$I_{t}$ plays the role of a reference value which controls the evolution
of the density $\pi_{t}(x)$, i.e. in logarithmic scale, the local
behaviour around a point $x\in\mathbb{R}^{d}$ is such that there
is an increase or decrease in density if $\log L(x)>I_{t}$ or $\log L(x)<I_{t}$,
respectively. In the following, we will see that this difference,
when integrated w.r.t. $\pi_{t}(x)$, provides us with the right direction
to move particles at time $t$. The factors $\lambda'(t)$ and $\pi_{t}(x)$
in (\ref{eqn:derivdensity}) are also intuitive as the change in density
must be proportional how quickly we introduce the likelihood and how
much probability mass there is locally. It will be apparent later
that these factors dictate the speed of particles. We note that the
contact Hamiltonian flow proposed in \cite{betancourt} also depends
on the term $\log L(x)-I_{t}$ which the author therein approximates
using Monte Carlo methods. 

\subsection{Particle dynamics, Liouville's equation and flow transport problem\label{subsec:Ordinary-differential-equation}}

Consider a particle trajectory $\{X_{t}\}_{t\in[0,1]}$ in $\mathbb{R}^{d}$,
initialized at time $t=0$ with a random draw $X_{0}\sim\pi_{0}$,
and evolved deterministically according to the following ODE 
\begin{align}
\frac{{\rm d}}{{\rm d}t}x(t)=f(t,x(t)),\quad t\in[0,1],\label{eqn:ODE}
\end{align}
with velocity field $f=(f_{1},\ldots,f_{d}):[0,1]\times\mathbb{R}^{d}\rightarrow\mathbb{R}^{d}$.
Under appropriate regularity conditions on $f$ which will be detailed
later, this ODE admits a unique solution $x(t;X_{0})$ for all $t\in[0,1]$.
Therefore we can define the flow map $T_{t}:\mathbb{R}^{d}\rightarrow\mathbb{R}^{d}$
as 
\begin{equation}
X_{t}=T_{t}(X_{0})=x(t;X_{0})\label{eq:flow_maps}
\end{equation}
which associates the initial position of the particle to its position
at time $t\in[0,1]$. It can be shown that flow maps are $C^{1}$-diffeomorphisms,
i.e. for each $t\in[0,1]$, $T_{t}$ is invertible and both $T_{t}$
and its inverse $T_{t}^{-1}:\mathbb{R}^{d}\rightarrow\mathbb{R}^{d}$
are continuously differentiable. These properties render flow maps
ideal candidates as transport maps. 

Additionally, if we denote the marginal distribution of $X_{t}$ by
$\tilde{\pi}_{t}=(T_{t})_{\#}\pi_{0}$, the curve of distributions
$\{\tilde{\pi}_{t}\}_{t\in[0,1]}$ satisfies, under regularity conditions,
the Liouville PDE \cite[eq. (3.5.13), p. 54]{Gardiner2002} also known
as the continuity equation \cite[eq. (8.1.1), p. 169]{ambrosio2005}:
\begin{align}
\partial_{t}\tilde{\pi}_{t}(x)=-\sum_{i=1}^{d}\partial_{x_{i}}(\tilde{\pi}_{t}(x)f_{i}(t,x))=-\nabla\cdot & (\tilde{\pi}_{t}(x)f(t,x))\label{eqn:liouville_initial}
\end{align}
for $(t,x)\in(0,1)\times\mathbb{R}^{d}$. Notationally, $\partial_{t}\varphi(t,x)$
and $\partial_{x_{i}}\varphi(t,x)$ denote the partial derivatives
of $\varphi\in C^{1}([0,1]\times\mathbb{R}^{d},\mathbb{R})$ w.r.t.
$t$ and $x_{i}$, respectively, and the divergence operator is defined
as $\nabla\cdot\varphi(x)=\sum_{i=1}^{d}\partial_{x_{i}}\varphi_{i}(x)$
for any $\varphi=(\varphi_{1},\ldots,\varphi_{d})\in C^{1}(\mathbb{R}^{d},\mathbb{R}^{d})$. 
The Liouville PDE can be seen as the Fokker\textendash Planck equation
in the case of zero diffusivity; an informal but intuitive derivation
of this PDE is given in Appendix \ref{append:liouville}. 

We can now describe the flow transport problem as identifying a velocity
field $f$ such that the curve of target distributions $\{\pi_{t}\}_{t\in[0,1]}$
in (\ref{eqn:tempering}) is the solution of Liouville equation (\ref{eqn:liouville_initial}),
i.e. we seek a $f$ that satisfies 
\begin{equation}
\partial_{t}\pi_{t}(x)=-\nabla\cdot(\pi_{t}(x)f(t,x))\label{eq:liouville}
\end{equation}
for $(t,x)\in(0,1)\times\mathbb{R}^{d}$. If such a velocity field
$f$ is regular enough that the resulting ODE (\ref{eqn:ODE}) admits
a unique solution for all $t\in[0,1]$ and initial positions $X_{0}\sim\pi_{0}$,
then this allows us to construct the flow maps (\ref{eq:flow_maps})
that satisfy $\pi_{t}=(T_{t})_{\#}\pi_{0}$ for all $t\in[0,1]$.
As a consequence, we can obtain samples from $\pi_{1}=\pi$ by taking
$X_{1}=T_{1}(X_{0})$. The following result presents sufficient conditions
on velocity fields $f$ that satisfy (\ref{eq:liouville}) to ensure
the validity of this approach. 
\begin{thm}
\label{thm:ambrosio}Suppose $f:[0,1]\times\mathbb{R}^{d}\rightarrow\mathbb{R}^{d}$
is a velocity field that satisfies Liouville equation (\ref{eq:liouville})
and the following conditions:
\begin{description}
\item [{A1.}] (continuously differentiable) $f\in C^{1}([0,1]\times\mathbb{R}^{d},\mathbb{R}^{d})$;
\item [{A2.}] (space-time integrability) $\int_{0}^{1}\int_{\mathbb{R}^{d}}|f(t,x)|\pi_{t}(x)\,\mathrm{d}x\,\mathrm{d}t<\infty$.
\end{description}
\noindent \begin{flushleft}
Then for $\pi_{0}$-almost everywhere $X_{0}\in\mathbb{R}^{d}$, there
exists a unique solution $x(t,X_{0})$ to the ODE (\ref{eqn:ODE})
for all $t\in[0,1]$. Therefore the flow maps $\{T_{t}\}_{t\in[0,1]}$
defined by (\ref{eq:flow_maps}) are flow transports, i.e. $\pi_{t}=(T_{t})_{\#}\pi_{0}$
for all $t\in[0,1]$.
\par\end{flushleft}

\end{thm}
Theorem \ref{thm:ambrosio} is a summary of results in \cite{ambrosio2005}
written for our purposes; see Appendix \ref{append:flowtransport}
for more details. With Theorem \ref{thm:ambrosio} in place, we can
now formally define the flow transport problem as identifying a velocity
field that satisfies Liouville's equation (\ref{eq:liouville}) and
Assumptions A1-A2. Although these assumptions are only sufficient
conditions, we stress that pathologies can occur when these regularity
conditions do not hold. This is illustrated in Appendix
\ref{subsec:divergence}, where we exhibit a velocity field that solves
(\ref{eq:liouville}) and prove that it yields divergent particle
trajectories. 

\section{A novel flow transport and Gibbs flow approximation\label{sec:Solving-the-flow-transport-problem}}

As alluded to in the introduction, the flow transport problem is typically
underdetermined. Although various optimality criteria could be employed
to attain unicity, they lead to velocity fields that are implicitly
defined by solutions of elliptic PDEs. In this section, we begin by
presenting an explicit solution to the flow transport problem before
introducing the Gibbs flow approximation. 

\subsection{A flow transport solution on $\mathbb{R}$\label{subsec:Flow-transport-problem-onrealline}}

We first discuss the one-dimensional case before considering the multivariate
case. In this case, there is a well-known solution to the flow transport
problem; see e.g. \cite{barronluo2007}. We will also establish that
this coincides with the minimal kinetic energy solution considered
in \cite{reichtransport2011,reichMixture}. 
\begin{prop}
\label{prop:1Dcase} Define the velocity field $f:[0,1]\times\mathbb{R}\rightarrow\mathbb{R}$
as
\begin{align}
f(t,x)=\frac{-\int_{-\infty}^{x}\partial_{t}\pi_{t}(u)\,{\rm d}u}{\pi_{t}(x)}\label{eqn:drift1D}
\end{align}
and assume that there exists an $\epsilon>0$ such that $x\mapsto|f(t,x)|\pi_{t}(x)=O\left(|x|^{-1-\epsilon}\right)$
as $|x|\rightarrow\infty$ with a constant that is independent of
$t\in[0,1]$. Then the velocity field (\ref{eqn:drift1D}) solves
the flow transport problem on $\mathbb{R}$ and is additionally the
minimal kinetic energy solution, i.e. for each $t\in[0,1]$ 
\begin{align}
f(t,\cdot)=\arg\min_{\varphi\in\mathcal{L}(\pi_{t})}\frac{1}{2}\int_{\mathbb{R}^{d}}\varphi^{2}(x)\pi_{t}(x)\,{\rm d}x,
\end{align}
where $\mathcal{L}(\pi_{t})=\left\{ \varphi:\mathbb{R}\rightarrow\mathbb{R}:\int_{\mathbb{R}}\varphi(x)^{2}\pi_{t}(x)\,\mathrm{d}x<\infty,\varphi(x)\mbox{ satisfies }\eqref{eq:liouville}\mbox{ for all }x\in\mathbb{R}\mbox{ at }t\in[0,1]\right\} $.
\end{prop}
To build intuition, we can rewrite (\ref{eqn:drift1D}) using (\ref{eqn:derivdensity})
as
\begin{align}
f(t,x)=\frac{\lambda'(t)I_{t}\left(F_{t}(x)-I_{t}^{x}/I_{t}\right)}{\pi_{t}(x)},\label{eqn:drift1Dbayes}
\end{align}
where $I_{t}^{x}=\int_{-\infty}^{x}\log L(u)\pi_{t}(u)\,{\rm d}u$
and $F_{t}(x)=\int_{-\infty}^{x}\pi_{t}(u)\,{\rm d}u$ is the cumulative
distribution function (CDF) of $\pi_{t}$. The velocity field (\ref{eqn:drift1Dbayes})
may be likened to driving a vehicle. The denominator corresponds to
the \emph{accelerator}, since, e.g., particles in the tails of $\pi_{t}$
need to \emph{speed up} to meet the changing schedule of intermediate
distributions. Also, it is intuitive that particle speeds are proportional
to the rate $\lambda'(t)$ at which we introduce the likelihood. The
numerator amounts to the \emph{steering wheel}: a particle's direction
of travel is given by the relative difference between its \emph{current
location} $x$, described by the term $F_{t}(x)$, and \emph{where
the particle needs to go}, prescribed by the term $I_{t}^{x}/I_{t}\in[0,1]$
which contains information from the likelihood. 

\subsection{A novel flow transport on $\mathbb{R}^{d}$, $d\ge1$\label{subsec:Bokanowski-Grebert-flow}}

It is tempting to extend (\ref{eqn:drift1D}) to the multivariate
case by simply introducing the velocity field $\bar{f}=(\bar{f}_{1},\ldots,\bar{f}_{d}):[0,1]\times\mathbb{R}^{d}\rightarrow\mathbb{R}^{d}$
given for $i=1,\ldots,d$ by 
\begin{align}
\bar{f}_{i}(t,x)=\frac{-\alpha_{i}\int_{-\infty}^{x_{i}}\partial_{t}\pi_{t}(u_{i},x_{-i})\,{\rm d}u_{i}}{\pi_{t}(x)},\label{eqn:antiDerivative}
\end{align}
where $\alpha_{i}\in\mathbb{R}$ and the integrand of (\ref{eqn:antiDerivative})
is to be understood as $\partial_{t}\pi_{t}(x_{1},\ldots,x_{i-1},u_{i},x_{i+1},\ldots,x_{d})$.
This velocity field was previously mentioned in \cite{barronluo2007}
and it can be shown to satisfy Liouville's equation (\ref{eq:liouville})
whenever $\sum_{i=1}^{d}\alpha_{i}=1$. However, we show in Appendix \ref{subsec:divergence} that (\ref{eqn:antiDerivative})
does not solve the flow transport problem as an ODE with velocity
field $\bar{f}$ would yield divergent particle trajectories even
on a simple Gaussian example. The main reason for this pathology is
the tail behaviour of $\bar{f}$. 

We now give our solution to the flow transport problem in the multivariate
case which recovers Proposition \ref{prop:1Dcase} when $d=1$. We
will write $x_{i:j}=(x_{i},\ldots,x_{j})\in\mathbb{R}^{j-i+1}$ and
denote the marginal distribution of $\pi_{t}$ in the $i=1,\ldots,d$
component by $\pi_{t}(x_{i})$ and its CDF by $F_{t}(x_{i})=\int_{-\infty}^{x_{i}}\pi_{t}(u_{i})\:\mathrm{d}u_{i}$. 
\begin{prop}
\label{prop:generalCase}Define the velocity field $f:[0,1]\times\mathbb{R}^{d}\rightarrow\mathbb{R}^{d}$
as 
\begin{align}
f_{i}(t,x)=-\Bigg(\prod_{j=1}^{i-1}\pi_{t}(x_{j}) & \int_{-\infty}^{x_{i}}\int_{\mathbb{R}^{i-1}}\partial_{t}\pi_{t}(u_{1:i-1},u_{i},x_{i+1:d})\,{\rm d}u_{1:i-1}{\rm d}u_{i}\nonumber \\
 & -\prod_{j=1}^{i-1}\pi_{t}(x_{j})F_{t}(x_{i})\int_{\mathbb{R}^{i}}\partial_{t}\pi_{t}(u_{1:i},x_{i+1:d})\,{\rm d}u_{1:i}\Bigg)\Bigg/\pi_{t}(x)\label{eqn:BGdrift1}
\end{align}
for $i=1,\ldots,d-1$ (use the convention $\prod_{1}^{0}=1$) and
\begin{align}
f_{d}(t,x)=-\Bigg(\prod_{j=1}^{d-1}\pi_{t}(x_{j})\int_{-\infty}^{x_{d}}\int_{\mathbb{R}^{d-1}}\partial_{t}\pi_{t}(u_{1:d-1},u_{d})\,{\rm d}u_{1:d-1}{\rm d}u_{d}\Bigg)\Bigg/\pi_{t}(x).\label{eqn:BGdrift2}
\end{align}
If there exists an $\epsilon>0$ such that $\sup_{\left\{ x\in\mathbb{R}^{d}:|x|=r\right\} }|f(t,x)|\pi_{t}(x)=O\left(r^{-d-\epsilon}\right)$
as $r\rightarrow\infty$ with a constant that is independent of $t\in[0,1]$,
then the velocity field (\ref{eqn:BGdrift1})-(\ref{eqn:BGdrift2})
solves the flow transport problem on $\mathbb{R}^{d}$.
\end{prop}
Our construction is a generalization of a method proposed by \cite{bokanowskigrebert1996}
to build a compactly supported three-dimensional velocity field solving
a flow transport problem in the context of molecular quantum chemistry.
When the target distributions factorize into independent one-dimensional
components, i.e. $\pi_{t}(x)=\prod_{i=1}^{d}\pi_{t}(x_{i})$, we establish
in Appendix \ref{append:solvingflow} that the velocity
field in (\ref{eqn:BGdrift1})-(\ref{eqn:BGdrift2}) would simply
reduce to 
\begin{equation}
f_{i}(t,x_{i})=\frac{-\int_{-\infty}^{x_{i}}\partial_{t}\pi_{t}(u_{i})\,{\rm d}u_{i}}{\pi_{t}(x_{i})},\quad i=1,\ldots,d,\label{eq:velocity_factorize}
\end{equation}
which is the solution of the one-dimensional flow transport problem
for each marginal distribution given by Proposition \ref{prop:1Dcase}.
As the integrals in (\ref{eqn:BGdrift1})-(\ref{eqn:BGdrift2}) can
be seen as expectations w.r.t. the conditional distributions $\pi_{t}(x_{1}|x_{2},\ldots,x_{d})$,
$\pi_{t}(x_{2}|x_{3},\ldots,x_{d})$, ..., $\pi_{t}(x_{d-1}|x_{d})$
and the marginal distribution $\pi_{t}(x_{d})$, we see that the flow
transport is achieved by taking reference to these conditionals. We
refer the reader to Appendix \ref{sec:gaussian_curve}
for an illustration of flow transport solutions when the curve of
distributions (\ref{eqn:tempering}) lies in the Gaussian family.

\subsection{Gibbs flow approximation }

Despite the explicit form of the flow transport solution in Proposition
\ref{prop:generalCase}, evaluating the velocity field (\ref{eqn:BGdrift1})-(\ref{eqn:BGdrift2})
would require computing integrals of dimension up to $d$. For computational
tractability, we propose an approximate flow transport that takes
reference to the full conditional distributions $\{\pi_{t}(x_{i}|x_{-i})\}_{i=1,\ldots,p}$,
where $x_{i}\in\mathbb{R}^{d_{i}}$ and $\sum_{i=1}^{p}d_{i}=d$.
For component $i=1,\ldots,p$, the time evolution of its full conditional
distribution is given by 
\begin{equation}
\partial_{t}\pi_{t}(x_{i}|x_{-i})=\lambda'(t)(\log L(x)-I_{t}(x_{-i}))\pi_{t}(x_{i}|x_{-i})\label{eq:evolution_conditionals}
\end{equation}
where $I_{t}(x_{-i})=\int_{\mathbb{R}^{d_{i}}}\log L(u_{i},x_{-i})\pi_{t}(u_{i}|x_{-i})\,\mathrm{d}u_{i}$.
We will design Gibbs velocity fields $\tilde{f}=(\tilde{f}_{1},\ldots,\tilde{f}_{p}):[0,1]\times\mathbb{R}^{d}\rightarrow\mathbb{R}^{d}$
that track changes in the full conditionals (\ref{eq:evolution_conditionals})
by seeking solutions to the following coupled system of Liouville
equations 
\begin{equation}
\partial_{t}\pi_{t}(x_{i}|x_{-i})=-\nabla\cdot(\pi_{t}(x_{i}|x_{-i})\tilde{f}_{i}(t,x)),\quad i=1,\ldots,p,\label{eqn:systemLiouville}
\end{equation}
for $t\in(0,1)$ and $x=(x_{1},\ldots,x_{p})\in\mathbb{R}^{d}$. Note
that the Liouville equation for each full conditional distribution
in (\ref{eqn:systemLiouville}) is defined on $(0,1)\times\mathbb{R}^{d_{i}}$,
so the divergence operator only acts on the variables $x_{i}\in\mathbb{R}^{d_{i}}$. 

For one-dimensional components, i.e. the case $d_{i}=1$, the velocity
field 
\begin{equation}
\tilde{f}_{i}(t,x)=\frac{-\int_{-\infty}^{x_{i}}\partial_{t}\pi_{t}(u_{i}|x_{-i})\,{\rm d}u_{i}}{\pi_{t}(x_{i}|x_{-i})},\label{eqn:gibbsAnti}
\end{equation}
which only involves one-dimensional integrals, can be shown to satisfy
(\ref{eqn:systemLiouville}) for the $i^{th}$-component, using similar
arguments as in Proposition \ref{prop:1Dcase}. For components with
dimension $d_{i}>1$, one could exploit analytical tractability of
full conditional distributions when they lie in the exponential family
to determine Gibbs velocity fields, or further block these components
into one-dimensional components and employ (\ref{eqn:gibbsAnti}).
We will illustrate how to systematically determine Gibbs velocity
fields on specific applications in Section \ref{sec:Applications},
and will assume for now that we have such a velocity field $\tilde{f}$
satisfying (\ref{eqn:systemLiouville}). The following result presents
sufficient conditions on a Gibbs velocity field $\tilde{f}$ to ensure
that, with initial position $X_{0}\sim\pi_{0},$ the ODE
\begin{equation}
\frac{{\rm d}}{{\rm d}t}x(t)=\tilde{f}(t,x(t))\label{eq:gibbs_ODE}
\end{equation}
 admits a unique solution for all $t\in[0,1]$.
\begin{prop}
\label{prop:gibbsFlow}Suppose $\tilde{f}:[0,1]\times\mathbb{R}^{d}\rightarrow\mathbb{R}^{d}$
is a velocity field that satisfies the system of Liouville equations
(\ref{eqn:systemLiouville}) and the following conditions:
\begin{description}
\item [{A3.}] (continuously differentiable) $\tilde{f}\in C^{1}([0,1]\times\mathbb{R}^{d},\mathbb{R}^{d})$;
\item [{A4.}] (tail behaviour) there exists $V\in C^{1}(\mathbb{R}^{d},\mathbb{R})$
satisfying $\lim_{|x|\rightarrow\infty}V(x)=\infty$ and $R>0$ such
that $\left<\nabla V(x),\tilde{f}(t,x)\right>\leq0$ for all $|x|>R$
and $t\in[0,1]$, where $\left<\cdot,\cdot\right>$ denotes the dot
product.
\end{description}
Then for $\pi_{0}$-almost everywhere $X_{0}\in\mathbb{R}^{d}$, there
exists a unique solution $x(t,X_{0})$ to the ODE (\ref{eq:gibbs_ODE})
for all $t\in[0,1]$.
\end{prop}
Assumption A3 imposes some regularity on the Gibbs velocity field
and Assumption A4 requires existence of a Lyapunov function $V\in C^{1}(\mathbb{R}^{d},\mathbb{R})$
so that a particle has non-increasing values of $V$ if it lies in
the tails. In some cases, one can choose $V(x)=|x|^{2}$ as a Lyapunov
function; we establish this in the $d=1$ case in Appendix
\ref{sec:gibbsflow_append}. Under the conclusions of Proposition
\ref{prop:gibbsFlow}, we can define the Gibbs flow map $\tilde{T}_{t}:\mathbb{R}^{d}\rightarrow\mathbb{R}^{d}$
as $X_{t}=\tilde{T}_{t}(X_{0})=x(t;X_{0})$ for each $t\in[0,1]$.
Since the system (\ref{eqn:systemLiouville}) is only a (tractable)
approximation of the desired Liouville equation (\ref{eq:liouville}),
the marginal distribution of $X_{t}$ under the Gibbs flow, $\tilde{\pi}_{t}=(\tilde{T}_{t})_{\#}\pi_{0}$,
will in general not be equal to the target distribution $\pi_{t}$. 
We now provide a characterization of this error in terms of the following
time-dependent local error:
\begin{align}
\varepsilon_{t}(x) & =\left|\partial_{t}\pi_{t}(x)+\nabla\cdot(\pi_{t}(x)\tilde{f}(t,x))\right|=\left|\partial_{t}\pi_{t}(x)-\sum_{i=1}^{p}\partial_{t}\pi_{t}(x_{i}|x_{-i})\pi_{t}(x_{-i})\right|\label{eqn:gibbsMimic}
\end{align}
which measures how well the Gibbs velocity field mimics the desired
change in density (\ref{eqn:derivdensity}). The sum over all components
in (\ref{eqn:gibbsMimic}) reveals the nature of the Gibbs flow approximation:
information about how much probability mass is changing in a particular
component is not communicated to other components. In other words,
computational tractability is gained at the expense of breaking down
a global problem in $d$ dimensions to $p$ many lower dimensional
problems. For any function $\varphi:\mathbb{R}^{d}\rightarrow\mathbb{R}$,
we will write $\|\varphi\|_{L^{2}}^{2}=\int_{\mathbb{R}^{d}}\varphi^{2}(x)\,{\rm d}x$
if $\varphi$ is $L^{2}$-integrable, and $\|\varphi\|_{\infty}=\sup_{x\in\mathbb{R}^{d}}|\varphi(x)|$
if $\varphi$ is bounded.
\begin{prop}
\label{prop:errorUpperBound}Suppose $\tilde{f}:[0,1]\times\mathbb{R}^{d}\rightarrow\mathbb{R}^{d}$
is a velocity field that satisfies the system of Liouville equations
(\ref{eqn:systemLiouville}), Assumptions A3-A4 and
\begin{description}
\item [{A5.}] (tail decay) there exists $\epsilon>0$ such that 
\[
\sup_{\left\{ x\in\mathbb{R}^{d}:|x|=r\right\} }|\tilde{f}(t,x)|\pi_{t}(x)=O\left(r^{-d-\epsilon}\right)\quad\mbox{and}\quad\sup_{\left\{ x\in\mathbb{R}^{d}:|x|=r\right\} }|\tilde{f}(t,x)|\tilde{\pi}_{t}(x)=O\left(r^{-d-\epsilon}\right)
\]
as $r\rightarrow\infty$ with constants that are independent of $t\in[0,1]$. 
\end{description}
Then the Gibbs flow approximation error is characterized by the following
inequality 
\begin{equation}
\|\tilde{\pi}_{t}-\pi_{t}\|_{L^{2}}^{2}\leq C(t)\int_{0}^{t}\|\varepsilon_{s}\|_{L^{2}}^{2}\,{\rm d}s\label{eqn:errorUpperBound}
\end{equation}
for $t\in[0,1]$, where $C(t)=t\exp\left(1+\int_{0}^{t}\|\nabla\cdot\tilde{f}(s,\cdot)\|_{\infty}\,{\rm d}s\right)$.
\end{prop}
The upper bound (\ref{eqn:errorUpperBound}) is tight in the sense
that it is equal to zero when the target distributions have independent
components, i.e. $\pi_{t}(x)=\prod_{i=1}^{p}\pi_{t}(x_{i})$. When
the latter is not the case, we observe that the bound deteriorates
with time, which is to be expected as errors can accumulate. To mitigate
accumulation of errors, we will combine Gibbs flow with Markov chain
Monte Carlo moves in Section \ref{subsec:Combining-Gibbs-flow-MCMC}.
Rewriting (\ref{eqn:gibbsMimic}) using (\ref{eqn:derivdensity})
and (\ref{eq:evolution_conditionals}) reveals that the inverse temperature
$\lambda(t)$ should be chosen such that its derivative $\lambda'(t)$
is small at those time instances when the integrated local error $\|\varepsilon_{t}\|_{L^{2}}^{2}$
is large, as this would reduce the magnitude of the resulting $L^{2}$-error
in (\ref{eqn:errorUpperBound}). We refer the reader to \cite{gelmanmeng1998,oates2016,zhou2016}
for other works on how to select $\lambda(t)$. For simplicity, all
simulations in Section \ref{sec:Applications} and the Appendix will employ a quadratic inverse temperature function, i.e.
$\lambda(t)=t^{2}$. Lastly, like with any Gibbs sampler, we expect
the use of any appropriate model specific reparameterization to also
reduce the $L^{2}$-error in (\ref{eqn:errorUpperBound}). 

\section{Gibbs flow samplers\label{sec:Gibbs-flow-samplers}}

\subsection{Numerical implementation\label{subsec:Numerical-implementation}}

Given a target distribution of interest, we advocate exploiting any
analytical tractability of full conditional distributions to determine
a Gibbs velocity field (e.g. Section \ref{subsec:Variance-component-models}).
For components without such tractability, a generic strategy would
be to further block these components into one-dimensional components
and rely on (\ref{eqn:gibbsAnti}). We first note that (\ref{eqn:gibbsAnti})
can be computed solely using one-dimensional integrals as the intractable
normalizing constant $Z(t)$ cancels in the expression: 
\begin{equation}
\tilde{f}_{i}(t,x)=\frac{\lambda'(t)\left\{ F_{t}(x_{i}|x_{-i})\int_{-\infty}^{\infty}\log L(u_{i},x_{-i})\gamma_{t}(u_{i},x_{-i})\,\mathrm{d}u_{i}-\int_{-\infty}^{x_{i}}\log L(u_{i},x_{-i})\gamma_{t}(u_{i},x_{-i})\,\mathrm{d}u_{i}\right\} }{\gamma_{t}(x)}\label{eq:rewrite_gibbs_velocity_1d}
\end{equation}
and the CDF of $\pi_{t}(x_{i}|x_{-i})$ can be rewritten as 
\begin{equation}
F_{t}(x_{i}|x_{-i})=\frac{\int_{-\infty}^{x_{i}}\gamma_{t}(u_{i},x_{-i})\:\mathrm{d}u_{i}}{\int_{-\infty}^{\infty}\gamma_{t}(v_{i},x_{-i})\:\mathrm{d}v_{i}}.\label{eq:rewrite_CDF_1d}
\end{equation}

The one-dimensional integrals in (\ref{eq:rewrite_gibbs_velocity_1d})-(\ref{eq:rewrite_CDF_1d})
are integrals of the form $\int_{D}\phi(u_{i},x_{-i})\,{\rm d}u_{i}$
for some integrand $\phi$ and domain $D\subseteq\mathbb{R}$. Here
we consider the class of composite Newton-Cotes quadrature rules
\begin{align}
\int_{D}\phi(u_{i},x_{-i})\,{\rm d}u_{i}\approx\sum_{r=1}^{R}\omega_{r}\phi(v_{r},x_{-i}),\label{eqn:newtonCotes}
\end{align}
where $\{\omega_{r}\}_{r=1,\ldots,R}$ are quadrature weights which
depend on the degree of the approximation and $\{v_{r}\}_{r=1,\ldots,R}$
are $R\in\mathbb{N}$ many equispaced quadrature points in $D$ \cite[p. 34]{Iserles2009}.
We take (\ref{eqn:newtonCotes}) to be of the closed type, i.e. $v_{1}$
and $v_{R}$ take the endpoints of $D$\footnote{Unbounded domains are treated with suitable truncation.},
as this choice will be convenient for domains of the type $D=(-\infty,x_{i}]$
for $x_{i}<\infty$. The composite quadrature rule (\ref{eqn:newtonCotes})
is derived by integrating Lagrange interpolation polynomials on subintervals;
the degree of which dictates the accuracy of the approximation on
each subinterval. We shall denote the resulting approximation of $\tilde{f}_{i}$
by $\hat{f}_{i}$; for components $i=1,\ldots,p$ with analytically
tractable Gibbs velocity field $\tilde{f}_{i}$, we set $\hat{f}_{i}=\tilde{f}_{i}$.

We now consider how to approximate a particle trajectory driven by
the ODE 
\begin{equation}
\frac{\mathrm{d}}{\mathrm{d}t}x(t)=\hat{f}(t,x(t)),\quad t\in[0,1],\label{eq:gibbs_velocity_approximated}
\end{equation}
with initial condition $X_{0}\sim\pi_{0}$. We will introduce a novel
numerical integration scheme that is reminiscent of the systematic
Gibbs scan. In the following, we will show that our proposed scheme,
in contrast to standard numerical integrators, allows efficient computation
of marginal distributions in high dimensions. For simplicity, we discretize
the time interval $[0,1]$ into a regular grid $t_{m}=mh,m=0,\ldots,M$
with a constant step size $h=1/M$; non-constant step sizes can also
be employed. To evolve a particle with position $X_{m-1}=(X_{m-1,1},\ldots,X_{m-1,p})\in\mathbb{R}^{d}$
at time $t_{m-1}$ on the subinterval $[t_{m-1},t_{m}]$, we consider
\[
\frac{\mathrm{d}}{\mathrm{d}t}x_{1}(t)=\hat{f}_{1}(t,x_{1}(t),x_{-1}),\quad t\in[t_{m-1},t_{m}],
\]
for the first component, with the other components fixed as $x_{-1}=(X_{m-1,2},\ldots,X_{m-1,p})=X_{m-1,2:p}$.
If the solution $x_{1}(t),t\in[t_{m-1},t_{m}]$ is analytically tractable,
we set $X_{m,1}=x_{1}(t_{m})$; otherwise we will rely on the Euler
discretization 
\begin{equation}
X_{m,1}=X_{m-1,1}+h\,\hat{f}_{1}(t_{m-1},X_{m-1,1},X_{m-1,2:p}).\label{eq:euler_discretization}
\end{equation}
Similarly, we update the second component by considering 
\begin{equation}
\frac{\mathrm{d}}{\mathrm{d}t}x_{2}(t)=\hat{f}_{2}(t,x_{2}(t),x_{-2}),\quad t\in[t_{m-1},t_{m}],\label{eq:conditional_ODE}
\end{equation}
with $x_{-2}=(X_{m,1},X_{m-1,3:p})$, and setting $X_{m,2}=x_{2}(t_{m})$
if the solution is available or 
\[
X_{m,2}=X_{m-1,2}+h\,\hat{f}_{2}(t_{m-1},X_{m,1},X_{m-1,2:p})
\]
otherwise. We then iteratively update all other components in a systematic
manner to obtain $X_{m}=(X_{m,1},\ldots,X_{m,p})\in\mathbb{R}^{d}$. 

In summary, the above procedure defines the maps 
\[
(X_{m,1:i},X_{m-1,(i+1):p})=\Psi_{m,i}(X_{m,1:i-1},X_{m-1,i:p}),\quad i=1,\ldots,p,
\]
(with $X_{m,1:0}=\emptyset$ and $X_{m-1,(p+1):p}=\emptyset$) which
update one component at a time. By iterating over all components,
the composition 
\begin{equation}
X_{m}=\Phi_{m}(X_{m-1})=\Psi_{m,p}\circ\cdots\circ\Psi_{m,1}(X_{m-1})\label{eq:novel_integrator}
\end{equation}
defines our numerical integration scheme. The flow maps $T_{t_{m}}:\mathbb{R}^{d}\rightarrow\mathbb{R}^{d}$
induced by this scheme 
\begin{equation}
X_{m}=T_{t_{m}}(X_{0})=\Phi_{m}\circ\cdots\circ\Phi_{1}(X_{0}),\quad m=0,\ldots,M,\label{eq:novel_integrator_flow}
\end{equation}
can be shown to be a first order approximation of the flow maps $\{\hat{T}_{t}\}_{t\in[0,1]}$
defined by (\ref{eq:gibbs_velocity_approximated}) (see Appendix \ref{sec:Numerical-integration-Gibbsflow}), i.e. $|T_{t_{m}}(X_{0})-\hat{T}_{t_{m}}(X_{0})|=O(h)$
for all $m=0,\ldots,M$ if the step size $h$ is sufficiently small\footnote{This error result holds even if Euler discretizations are employed
for some or all components.}. 

\subsection{Distribution of approximate Gibbs flow samples\label{subsec:Distribution-of-approximate}}

We now detail how to compute the marginal distributions of $X_{m},m=0,\ldots,M$
under the numerically approximated Gibbs flow (\ref{eq:novel_integrator_flow}).
This allows us to utilize these distributions as proposal distributions
within a sequential importance sampler. 

Under the assumptions of Proposition \ref{prop:gibbsFlow}, the Gibbs
flow maps $\{\tilde{T}_{t}\}_{t\in[0,1]}$ are $C^{1}$-diffeomorphisms
by construction. Hence their approximation (\ref{eq:novel_integrator_flow})
will be injective if the step size $h$ is sufficiently small and
quadrature approximations (if employed) are accurate enough - see
\cite{bunchgodsill2014,LiuWang2016} for similar arguments. Under
these conditions, it follows from a change of variables that the density
of $q_{t_{m}}=(T_{t_{m}})_{\#}\pi_{0}$, is 
\begin{equation}
q_{t_{m}}(X_{m})=\pi_{0}(X_{0})|\det(\nabla T_{t_{m}}(X_{0}))|^{-1}\label{eq:gibbsflow_proposals}
\end{equation}
where $X_{0}=T_{t_{m}}^{-1}(X_{m})$ is given by the inverse map,
$|\det(\nabla T_{t_{m}}(X_{0}))|$ denotes the absolute value of the
determinant of the Jacobian matrix of $T_{t_{m}}$. In numerical implementations,
monotonicity may be monitored by checking for any sign changes in
the Jacobian determinant. From (\ref{eq:novel_integrator_flow}),
the latter can be computed as 
\begin{align*}
\det(\nabla T_{t_{m}}(X_{0})) & =\prod_{k=1}^{m}\det(\nabla\Phi_{k}(X_{k-1})).
\end{align*}
Using the structure of our numerical integration scheme (\ref{eq:novel_integrator}),
the computational cost of computing 
\[
\det(\nabla\Phi_{k}(X_{k-1}))=\prod_{i=1}^{p}\det(\nabla\Psi_{k,i}(X_{k,1:i-1},X_{k-1,i:p}))
\]
is at most $O(\sum_{i=1}^{p}d_{i}^{3})$. This cost may be even lower
in statistical models with conditional independence structure as Gibbs
velocity fields will inherent such structures yielding sparse Jacobian
matrices (e.g. Section \ref{subsec:Variance-component-models}).

In the case of (\ref{eqn:gibbsAnti}) for one-dimensional components
and the Euler discretization (\ref{eq:euler_discretization}), computing
\[
\det(\nabla\Psi_{k,i}(X_{k,1:i-1},X_{k-1,i:p}))=\det(1+h\,\partial_{x_{i}}\hat{f}_{i}(t_{k-1},X_{k,1:i-1},X_{k-1,i:p}))
\]
requires the partial derivative of the approximate Gibbs velocity
field $\partial_{x_{i}}\hat{f}_{i}(t,x)$. It turns out that we can
compute $\partial_{x_{i}}\hat{f}_{i}(t,x)$ by simply replacing integrals
in the partial derivative of the Gibbs velocity field 
\[
\partial_{x_{i}}\tilde{f}_{i}(t,x)=\lambda'(t)\left\{ \frac{\int_{-\infty}^{\infty}\log L(u_{i},x_{-i})\gamma_{t}(u_{i},x_{-i})\:\mathrm{d}u_{i}}{\int_{-\infty}^{\infty}\gamma_{t}(u_{i},x_{-i})\:\mathrm{d}u_{i}}\right\} -\tilde{f}_{i}(t,x)\partial_{x_{i}}\log\gamma_{t}(x)
\]
with approximations based on the same quadrature rule. This follows
from the following argument which allows one to compute the partial
derivative w.r.t. $x_{i}$ of approximations of integrals of the form
$\int_{-\infty}^{x_{i}}\phi(u_{i},x_{-i})\,\mathrm{d}u_{i}$. Denote
by $\hat{\phi}$ the underlying Lagrange interpolant giving rise to
the quadrature rule (\ref{eqn:newtonCotes}). By the first fundamental
theorem of calculus and the closed property of (\ref{eqn:newtonCotes})
\begin{align}
\partial_{x_{i}}\sum_{r=1}^{R}\omega_{r}\phi(v_{r},x_{-i})=\partial_{x_{i}}\int_{-\infty}^{x_{i}}\hat{\phi}(u_{i},x_{-i})\,{\rm d}u_{i}=\hat{\phi}(x_{i},x_{-i})=\phi(x_{i},x_{-i}).
\end{align}

To illustrate the computational savings our proposed numerical integrator
(\ref{eq:novel_integrator}) offers over standard integrators like
the forward Euler method 
\begin{equation}
X_{m}=\Phi_{m}(X_{m-1})=X_{m-1}+h\:\hat{f}(t_{m-1},X_{m-1}),\label{eq:forward_euler}
\end{equation}
we consider the case of solely one-dimensional components, i.e. $d_{i}=1$
for all $i=1,\ldots,p$. In the absence of any sparsity, computing
the Jacobian determinant of the mapping in (\ref{eq:forward_euler})
would cost at most $O(d^{3})$; in contrast the cost associated to
(\ref{eq:novel_integrator}) is only $O(d)$. 

Given $N\in\mathbb{N}$ independent samples $X_{m}^{n},n=1,\ldots,N$
from (\ref{eq:novel_integrator_flow}), the above discussion allows
us to employ the marginal distribution $q_{t_{m}}$ in (\ref{eq:gibbsflow_proposals})
as a proposal distribution within an importance sampling approximation
of $\pi_{t_{m}}$. The importance weights $w_{m}(X_{m}^{n})=\gamma_{t_{m}}(X_{m}^{n})/q_{t_{m}}(X_{m}^{n})$
can be computed recursively using 
\[
w_{m}(X_{m}^{n})=w_{m-1}(X_{m-1}^{n})\frac{\gamma_{t_{m}}(X_{m}^{n})}{\gamma_{t_{m-1}}(X_{m-1}^{n})|\det(\nabla\Phi_{m}(X_{m-1}^{n}))|^{-1}},\quad m=1,\ldots,M,
\]
with $w_{0}(X_{0}^{n})=1$. An algorithmic description of the resulting
sequential importance sampler is detailed in Algorithm \ref{alg:GF-SIS}.
Using the output, we can approximate expectations of the form $\int_{\mathbb{R}^{d}}\phi(x)\pi(x)\:\mathrm{d}x$
with the weighted sum $\sum_{n=1}^{N}\phi(X_{M}^{n})W_{M}^{n}$, and
estimate the marginal likelihood $Z=\int_{\mathbb{R}^{d}}\pi_{0}(x)L(x)\,\mathrm{d}x$
unbiasedly with $\hat{Z}_{M}$. The adequacy of the importance sampling
approximation based on the Gibbs flow can be monitored using the effective
sample size (ESS) introduced in \cite{Kong1994}. This quantity takes
values between $1$ and $N$, and will be equal to $N$ if samples
are distributed according to the target distribution. 

\begin{algorithm}[H]
\protect\caption{Gibbs flow sequential importance sampler (GF-SIS)~\label{alg:GF-SIS}}

\textbf{Input}: prior $\pi_{0}$, likelihood $L$, inverse temperature
$\lambda$, step size $h$, and Gibbs velocity field $\tilde{f}$.

For time step $m=0$

\qquad{}For $n=1,\ldots,N$

\qquad{}\qquad{}(a) sample $X_{0}^{n}=(X_{0,1}^{n},\ldots,X_{0,p}^{n})\sim\pi_{0}$;

\qquad{}\qquad{}(b) set $w_{0}^{n}=1$ and $W_{0}^{n}=N^{-1}$;

\qquad{}(c) set $\mathrm{ESS}_{0}=N$ and $\hat{Z}_{0}=1$.

\textcompwordmark{}

For time step $m=1,\ldots,M$ 

\qquad{}For $n=1,\ldots,N$

\qquad{}\qquad{}For $i=1,\ldots,p$

\qquad{}\qquad{}\qquad{}(d) set $(X_{m,1:i}^{n},X_{m-1,(i+1):p}^{n})=\Psi_{m,i}(X_{m,1:i-1}^{n},X_{m-1,i:p}^{n})$
using Section \ref{subsec:Numerical-implementation};

\qquad{}\qquad{}\qquad{}(e) compute $J_{m,i}^{n}=\det(\nabla\Psi_{m,i}(X_{m,1:i-1}^{n},X_{m-1,i:p}^{n}))$
using Section \ref{subsec:Distribution-of-approximate};

\qquad{}\qquad{}(f) set $X_{m}^{n}=(X_{m,1}^{n},\ldots,X_{m,p}^{n})$
and $J_{m}^{n}=\prod_{i=1}^{p}J_{m,i}^{n}$;

\qquad{}\qquad{}(g) compute unnormalized weights 
\[
w_{m}^{n}=w_{m-1}^{n}\frac{\gamma_{t_{m}}(X_{m}^{n})}{\gamma_{t_{m-1}}(X_{m-1}^{n})|J_{m}^{n}|^{-1}};
\]

\qquad{}\qquad{}(h) compute normalized weights $W_{m}^{n}=w_{m}^{n}/\sum_{\ell=1}^{N}w_{m}^{\ell}$;

\qquad{}(i) compute effective sample size $\mathrm{ESS}_{m}=\left\{ \sum_{n=1}^{N}(W_{m}^{n})^{2}\right\} ^{-1}$;

\qquad{}(j) compute normalizing constant estimator $\hat{Z}_{m}=N^{-1}\sum_{n=1}^{N}w_{m}^{n}$.

\textbf{Output}: samples $\{X_{M}^{n}\}_{n=1,\ldots,N}$, normalized
weights $\{W_{M}^{n}\}_{n=1,\ldots,N}$ and normalizing constant estimator
$\hat{Z}_{M}$.
\end{algorithm}

\subsection{Combining Gibbs flow with Markov chain Monte Carlo \label{subsec:Combining-Gibbs-flow-MCMC}}

State-of-the-art methods based on annealed importance sampling (AIS)
simulate $N\in\mathbb{N}$ inhomogeneous Markov chains $X_{0}^{n}\sim\pi_{0}$
and $X_{m}^{n}\sim K_{m}(X_{m-1}^{n},\cdot),$ for $m=1,\ldots,M$
and $n=1,\ldots,N$, where $K_{m}$ is a $\pi_{t_{m}}$-invariant
Markov chain Monte Carlo (MCMC) kernel. For each $m=1,\ldots,M$,
although the marginal distribution of $\{X_{m}^{n}\}_{n=1,\ldots,N}$
is typically intractable, one can still use these samples within an
importance sampling approximation of $\pi_{t_{m}}$, by associating
sample $n=1,\ldots,N$ with the importance weight 
\begin{equation}
w_{m}(X_{0:m-1}^{n})=w_{m-1}(X_{0:m-2}^{n})\frac{\gamma_{t_{m}}(X_{m-1}^{n})}{\gamma_{t_{m-1}}(X_{m-1}^{n})},\quad m=1,\ldots,M,\label{eq:AIS_weights}
\end{equation}
with $w_{0}(X_{0:-1}^{n})=1$ and $X_{0:m-1}^{n}=(X_{0}^{n},\ldots,X_{m-1}^{n})$.
The choice of bridging distributions $\{\pi_{t_{m}}\}_{m=0,\ldots,M}$
and MCMC kernels $\{K_{m}\}_{m=1,\ldots,M}$ can have a large impact
on algorithmic performance; if these kernels mix slowly and/or the
intermediate distributions are too distant, the variance of the importance
weights (\ref{eq:AIS_weights}) can be very high. 

To improve the performance of AIS, references \cite{jarzynski2008,Vaikuntanatha2011}
suggested adding deterministic maps $\Phi_{m}$ which attempt to ``push''
samples from $\pi_{t_{m-1}}$ to $\pi_{t_{m}}$, but the authors did
not propose a generic methodology to construct such transport maps.
In our context, we will rely on numerical approximation of the Gibbs
flow, as described in Section \ref{subsec:Numerical-implementation},
to build these maps. Practically, for $n=1,\ldots,N$, we initialize
by sampling $X_{0}^{n}\sim\pi_{0}$ and setting $\tilde{X}_{0}^{n}=X_{0}^{n}$.
For $m=1,\ldots,M$, we then iterate by setting $X_{m}^{n}=\Phi_{m}(\tilde{X}_{m-1}^{n})$,
as defined in (\ref{eq:novel_integrator}), and sampling $\tilde{X}_{m}^{n}\sim K_{m}(X_{m}^{n},\cdot)$
from a $\pi_{t_{m}}$-invariant MCMC kernel. Like in AIS, we can also
use the samples $\{\tilde{X}_{m}^{n}\}_{n=1,\ldots,N}$ within an
importance sampling approximation of $\pi_{t_{m}}$. The importance
weights are given by 
\[
w_{m}(X_{0:m}^{n},\tilde{X}_{0:m}^{n})=w_{m-1}(X_{0:m-1}^{n},\tilde{X}_{0:m-1}^{n})\frac{\gamma_{t_{m}}(X_{m}^{n})}{\gamma_{t_{m-1}}(\tilde{X}_{m-1}^{n})|\det(\nabla\Phi_{m}(\tilde{X}_{m-1}^{n}))|^{-1}},
\]
for $m=1,\ldots,M$ with $w_{0}(X_{0}^{n},\tilde{X}_{0}^{n})=1$.
We provide an algorithmic description of the resulting annealed importance
sampler in Algorithm \ref{alg:GF-AIS}. From the output, expectations
$\int_{\mathbb{R}^{d}}\phi(x)\pi(x)\:\mathrm{d}x$ can be approximated
by the weighted sum $\sum_{n=1}^{N}W_{M}^{n}\phi(\tilde{X}_{M}^{n})$
and the marginal likelihood by the unbiased estimator $\hat{Z}_{M}$.
Although resampling is not considered in Algorithms \ref{alg:GF-SIS}-\ref{alg:GF-AIS}
to simplify our exposition, any resampling scheme can also be employed
with minor modifications; this is detailed in Appendix
\ref{sec:extra_algorithms} for completeness. 

\begin{algorithm}[H]
\protect\caption{Gibbs flow annealed importance sampler (GF-AIS)~\label{alg:GF-AIS}}

\textbf{Input}: prior $\pi_{0}$, likelihood $L$, inverse temperature
$\lambda$, step size $h$, Gibbs velocity field $\tilde{f}$, MCMC
kernels $\{K_{m}\}_{m=1,\ldots,M}$. 

For time step $m=0$

\qquad{}For $n=1,\ldots,N$

\qquad{}\qquad{}(a) sample $X_{0}^{n}=(X_{0,1}^{n},\ldots,X_{0,p}^{n})\sim\pi_{0}$
and set $\tilde{X}_{0}^{n}=X_{0}^{n}$;

\qquad{}\qquad{}(b) set $w_{0}^{n}=1$ and $W_{0}^{n}=N^{-1}$;

\qquad{}(c) set $\mathrm{ESS}_{0}=N$ and $\hat{Z}_{0}=1$.

\textcompwordmark{}

For time step $m=1,\ldots,M$ 

\qquad{}For $n=1,\ldots,N$

\qquad{}\qquad{}For $i=1,\ldots,p$

\qquad{}\qquad{}\qquad{}(d) set $(X_{m,1:i}^{n},\tilde{X}_{m-1,(i+1):p}^{n})=\Psi_{m,i}(X_{m,1:i-1}^{n},\tilde{X}_{m-1,i:p}^{n})$
using Section \ref{subsec:Numerical-implementation};

\qquad{}\qquad{}\qquad{}(e) compute $J_{m,i}^{n}=\det(\nabla\Psi_{m,i}(X_{m,1:i-1}^{n},\tilde{X}_{m-1,i:p}^{n}))$
using Section \ref{subsec:Distribution-of-approximate};

\qquad{}\qquad{}(f) set $X_{m}^{n}=(X_{m,1}^{n},\ldots,X_{m,p}^{n})$
and $J_{m}^{n}=\prod_{i=1}^{p}J_{m,i}^{n}$;

\qquad{}\qquad{}(g) compute unnormalized weights 
\[
w_{m}^{n}=w_{m-1}^{n}\frac{\gamma_{t_{m}}(X_{m}^{n})}{\gamma_{t_{m-1}}(\tilde{X}_{m-1}^{n})|J_{m}^{n}|^{-1}};
\]

\qquad{}\qquad{}(h) compute normalized weights $W_{m}^{n}=w_{m}^{n}/\sum_{\ell=1}^{N}w_{m}^{\ell}$;

\qquad{}\qquad{}(i) sample $\tilde{X}_{m}^{n}\sim K_{m}(X_{m}^{n},\cdot)$
from $\pi_{t_{m}}$-invariant MCMC kernel;

\qquad{}(j) compute effective sample size $\mathrm{ESS}_{m}=\left\{ \sum_{n=1}^{N}(W_{m}^{n})^{2}\right\} ^{-1}$;

\qquad{}(k) compute normalizing constant estimator $\hat{Z}_{m}=N^{-1}\sum_{n=1}^{N}w_{m}^{n}$.

\textbf{Output}: samples $\{\tilde{X}_{M}^{n}\}_{n=1,\ldots,N}$,
normalized weights $\{W_{M}^{n}\}_{n=1,\ldots,N}$ and normalizing
constant estimator $\hat{Z}_{M}$.
\end{algorithm}

\section{Applications\label{sec:Applications}}

\subsection{Bayesian mixture modelling \label{subsec:Bayesian-mixture-modelling}}

We now investigate the performance of Gibbs flow samplers on a Bayesian
mixture model, where the posterior distribution of mixture means is
inferred. This is a canonical example of distributions with multiple
well-separated modes. 

Consider $J\in\mathbb{N}$ independent observations from a univariate
Gaussian mixture model with $d$ components, i.e. for $j=1,\ldots,J$
each observation is distributed according to $Y_{j}\sim\frac{1}{d}\sum_{i=1}^{d}\mathcal{N}(x_{i},\sigma_{i}^{2})$,
where $\mathcal{N}(\mu,\varsigma^{2})$ (and $y\mapsto\mathcal{N}(y;\mu,\varsigma^{2})$)
denotes the Gaussian distribution (and density) with mean $\mu$ and
variance $\varsigma^{2}$. Following \cite{Lee2010}, we set $d=4$,
$\sigma_{i}=\sigma=0.55$ for $i=1,\ldots,d$ and perform inference
only on the mean parameters $x=(x_{1},\ldots,x_{4})\in\mathbb{R}^{4}$.
We generate the data $\{y_{j}\}_{j=1,\ldots,J}$ using $J=100$ simulations
from the model with parameter value $x^{*}=(-3,0,3,6)$ and stratification
between components. We adopt a uniform prior distribution on the $d$-dimensional
hypercube $[-10,10]^{d}$. The curve of distributions in (\ref{eqn:tempering})
is 
\begin{align}
\pi_{t}(x)=\frac{\mathbb{I}_{[-10,10]^{d}}(x)L(x)^{\lambda(t)}}{20^{d}Z(t)},\quad t\in[0,1],\label{eq:mixtureGaussCurve}
\end{align}
where $\mathbb{I}_{[-10,10]^{d}}(x)=1$ if $x\in[-10,10]^{d}$ and
$0$ otherwise, and the likelihood is 
\begin{align}
L(x)=\frac{1}{d^{J}}\prod_{j=1}^{J}\sum_{i=1}^{d}\phi(y_{j};x_{i},\sigma^{2}).
\end{align}
It follows from exchangeability of the prior and non-identifiability
of mixture components that the posterior distribution (\ref{eq:targetposterior})
is invariant under ``label permutation''. Therefore $\pi_{1}=\pi$
admits $d!=24$ well-separated modes centered approximately around
all permutations of $x^{*}$. As it is known that simple MCMC and
importance sampling methods typically perform poorly for such problems
\cite{Celeux_etal_2000}, we will determine the quality of the Gibbs
flow approximation (\ref{eqn:systemLiouville}) by examining how well
it can explore all $24$ modes equally.

\begin{figure}[htbp]
\centering{}\includegraphics[scale=0.5]{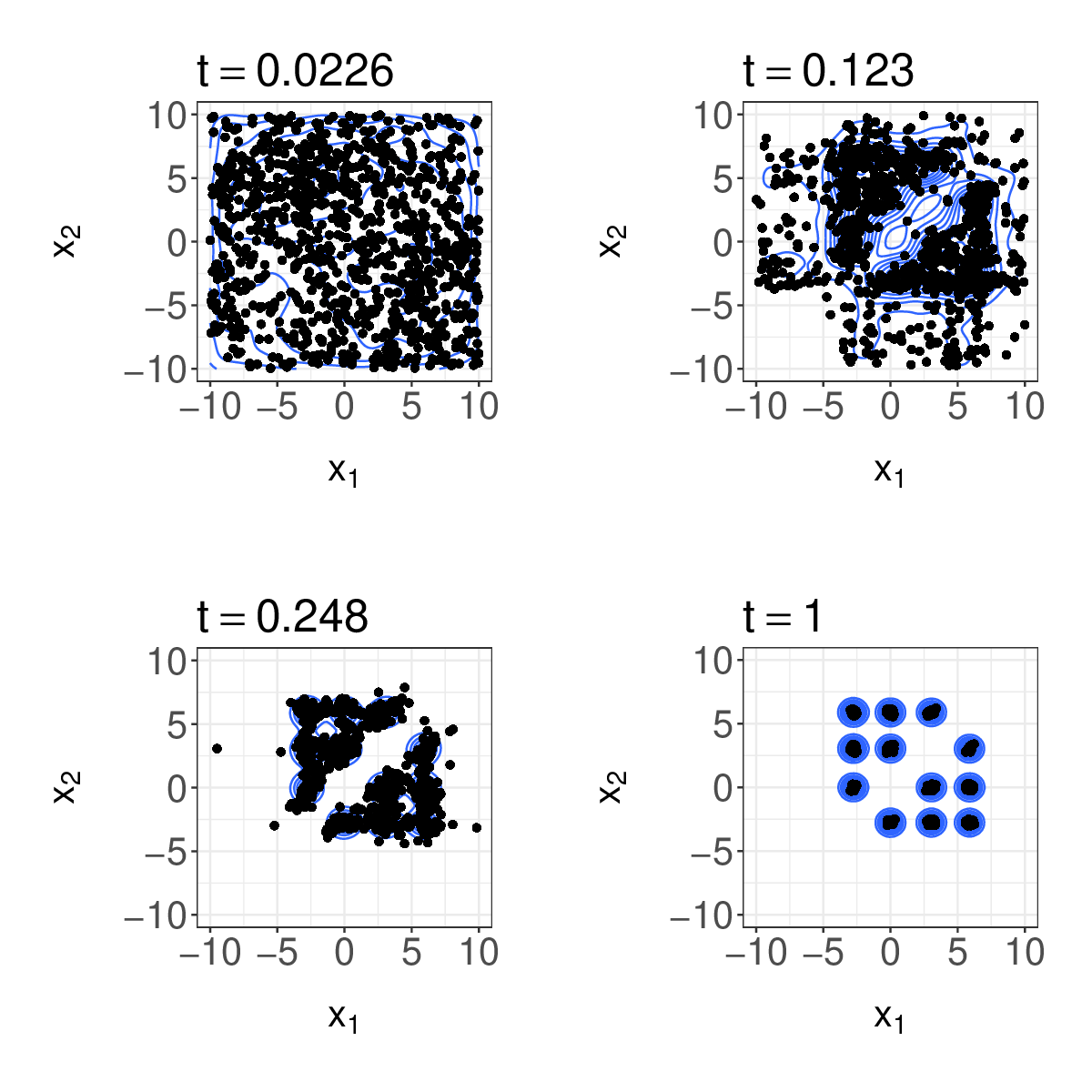} \caption{Time evolution of $N=1024$ prior samples in the first two dimensions
under the Gibbs flow (\emph{black dots}) for the Bayesian mixture
model in Section \ref{subsec:Bayesian-mixture-modelling}. For each
time instance, the superimposed (\emph{blue}) contours represent the
marginal of the target distribution obtained as a kernel density estimate
from the output of a SMC sampler.\label{fig:mixturemodel_evolution}}
\end{figure}

\begin{figure}[htbp]
\centering{}\includegraphics[scale=0.5]{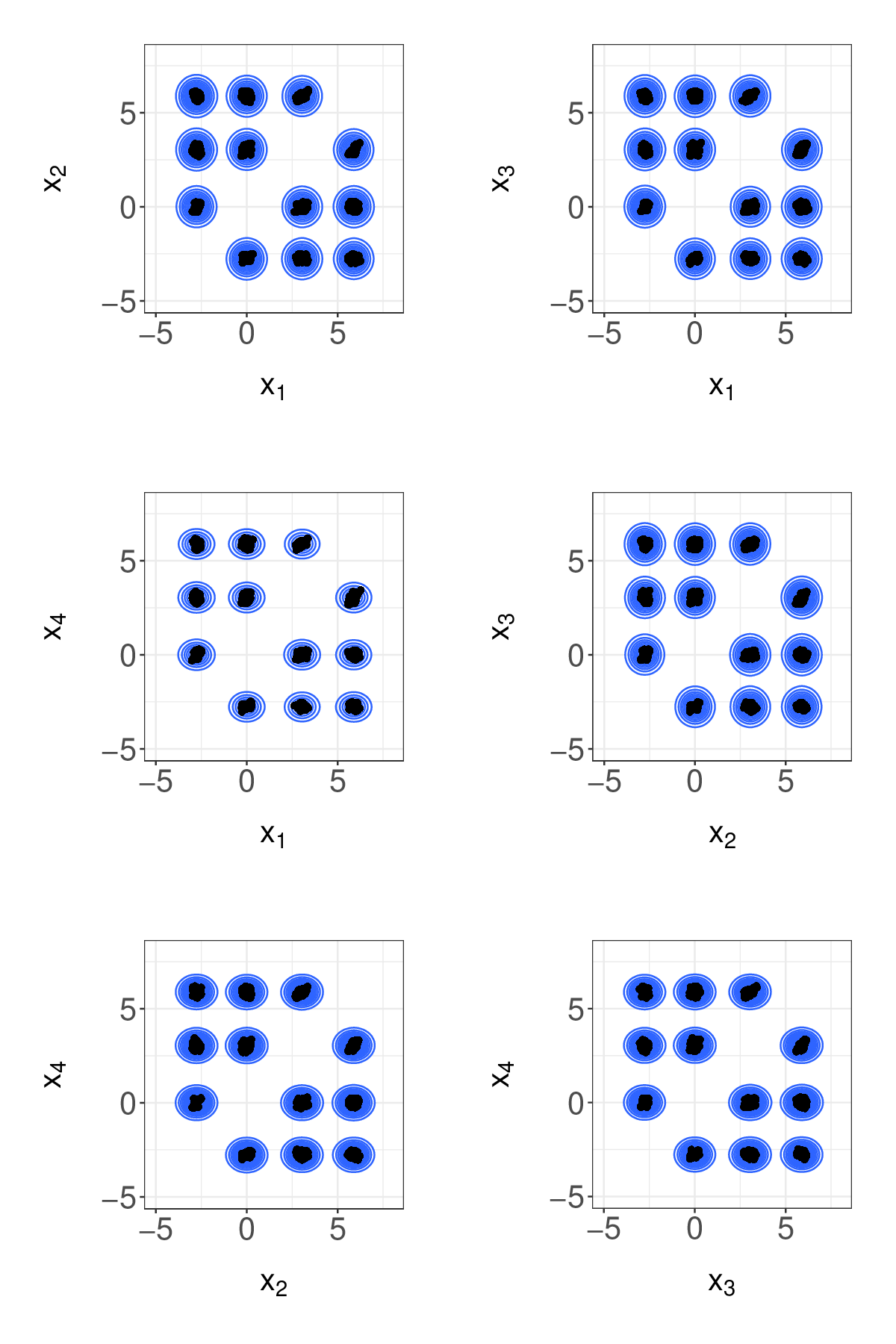} \caption{All pairs of marginal posterior distributions on $\mathbb{R}^{2}$
for the Bayesian mixture model in Section \ref{subsec:Bayesian-mixture-modelling}.\label{fig:mixturemodel_allpairs}}
\end{figure}

\begin{figure}[htbp]
\centering{}\includegraphics[scale=0.5]{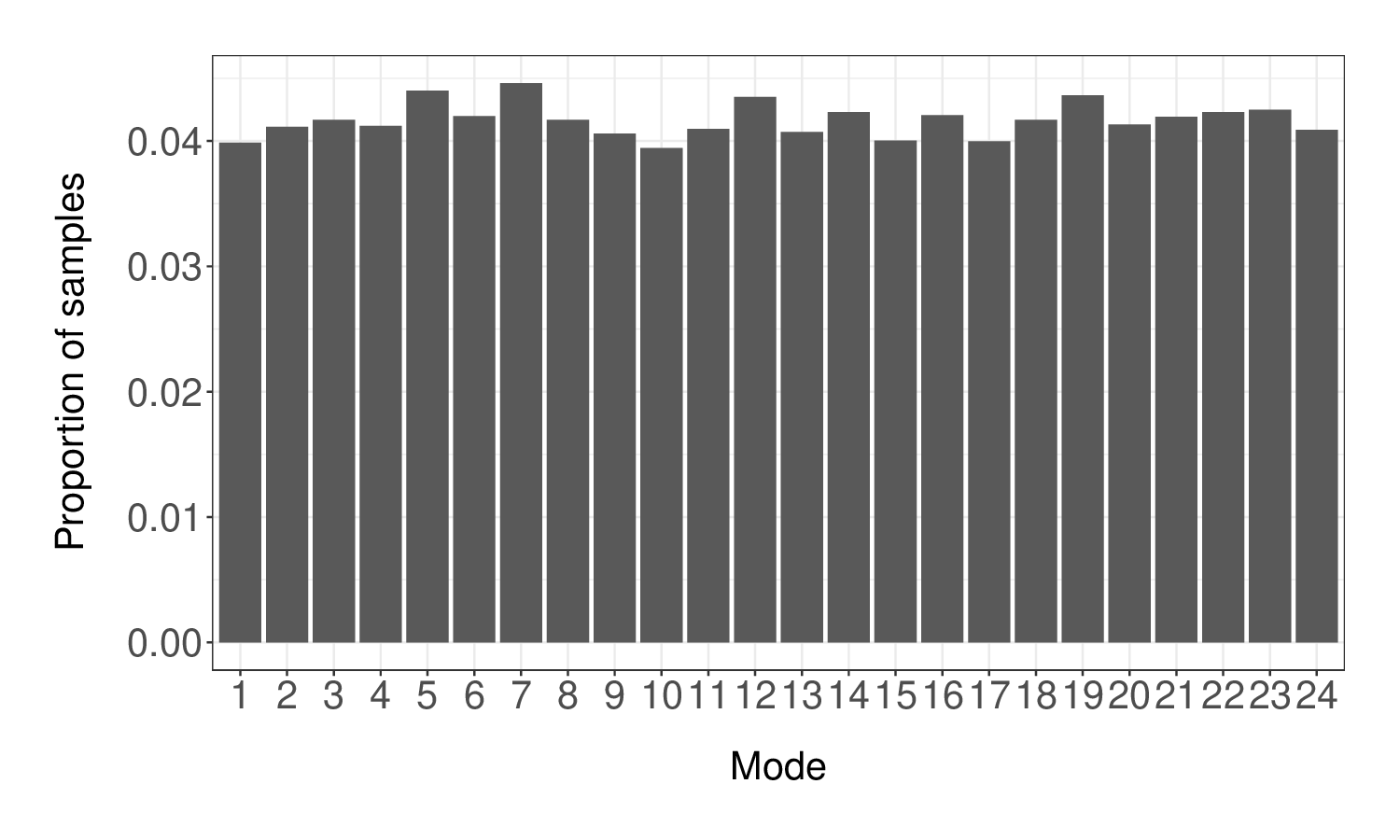}\caption{Proportion of Gibbs flow samples in each of the $24$ modes for the
Bayesian mixture model in Section \ref{subsec:Bayesian-mixture-modelling}.
\label{fig:mixturemodel_proportions}}
\end{figure}

\begin{figure}[htbp]
\centering{}\includegraphics[scale=0.5]{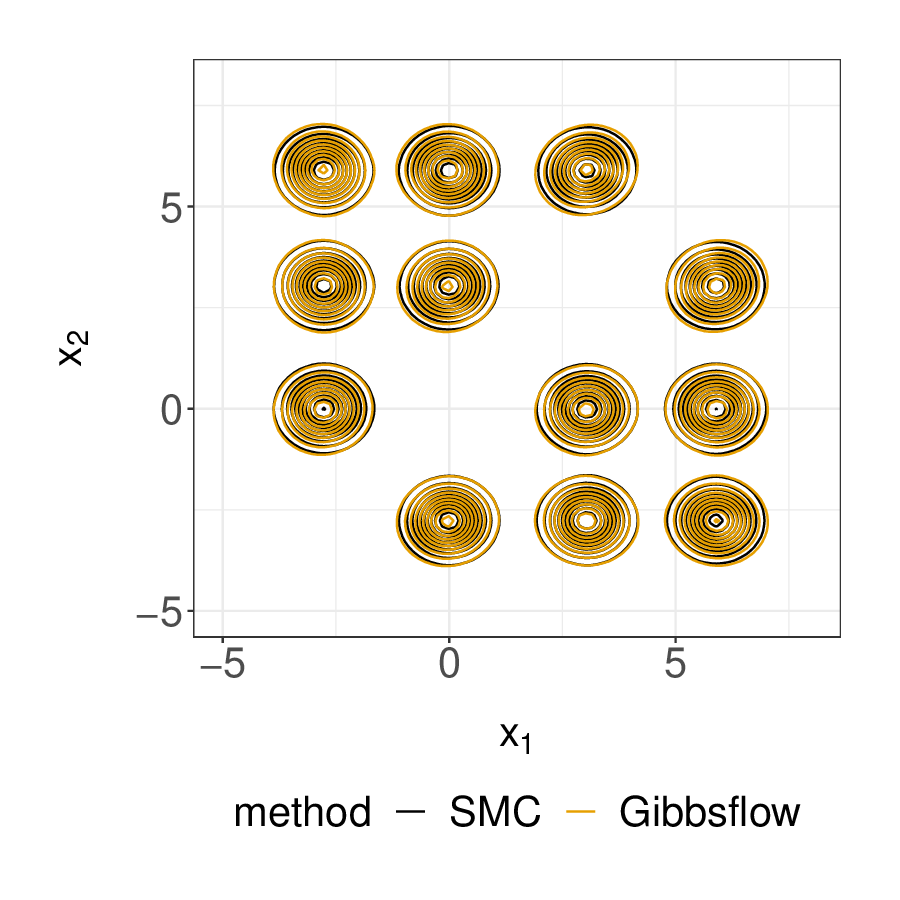}\includegraphics[scale=0.5]{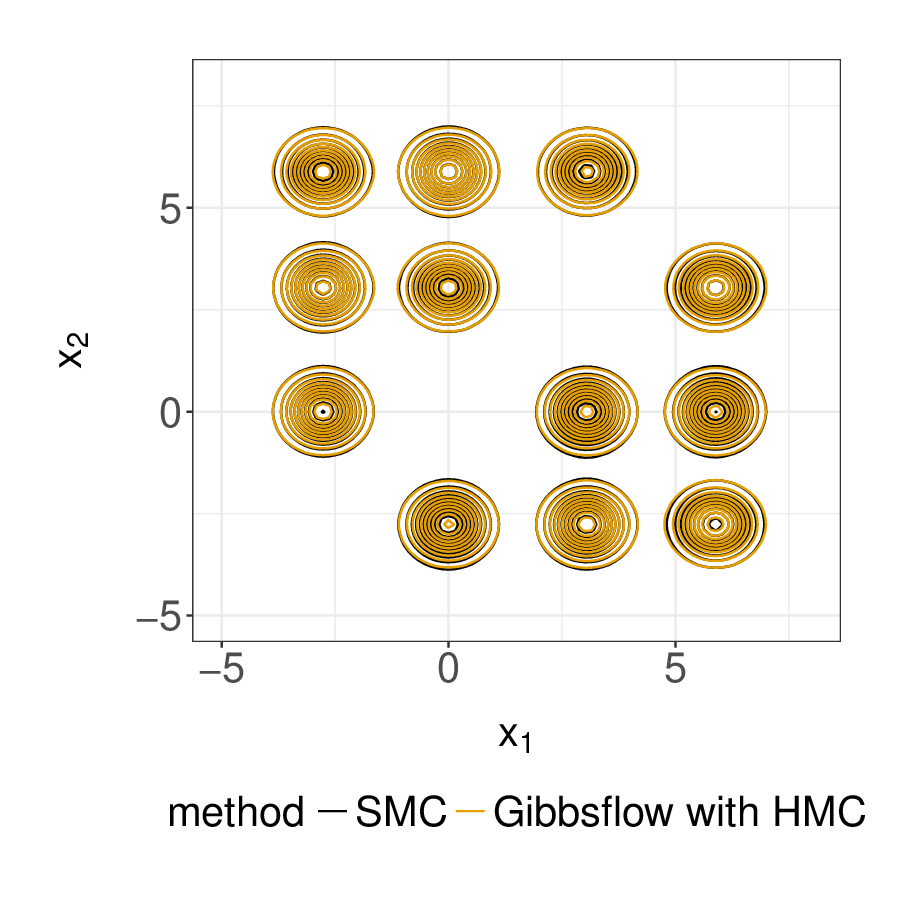}
\caption{Marginal posterior distribution (\emph{black}), marginal distribution
of Gibbs flow samples (\emph{left-orange}), and marginal distribution
of samples under the Gibbs flow and Hamiltonian Monte Carlo kernels
(\emph{right-orange}) for the Bayesian mixture model in Section \ref{subsec:Bayesian-mixture-modelling},
obtained using kernel density estimates from the output of a SMC sampler
and $N=16,384$ independent samples, respectively. \label{fig:mixturemodel_kdes}}
\end{figure}

As the full conditional distributions of the posterior are not in
the exponential family, we employ the Gibbs velocity field (\ref{eq:rewrite_gibbs_velocity_1d})-(\ref{eq:rewrite_CDF_1d})
for one-dimensional components. Using a composite trapezoidal rule with $R=100$ quadrature points
and the default ODE solver from the \texttt{deSolve} R package, we
compare the time evolution of $N=1024$ prior samples under the Gibbs
flow with the output of a standard SMC sampler with many particles
as the reference truth in Figure \ref{fig:mixturemodel_evolution}.
The performance of the Gibbs flow for this challenging problem is
striking as the samples reach all modes. 
This is also seen in Figure \ref{fig:mixturemodel_allpairs} that
shows all pairwise marginal posterior distributions on $\mathbb{R}^{2}$
(note that each of these marginals admits $12$ well-separated modes).
To corroborate these observations, we simulate another $N=16,384$
independent Gibbs flow samples and display the proportion of samples
in each of the $24$ modes in Figure \ref{fig:mixturemodel_proportions}.
The uniformity of these proportions is then tested using a Pearson's
Chi-squared goodness-of-fit test, which gives a p-value of $0.8522$.
Next, we examine how well the distribution of Gibbs flow samples matches
the posterior distribution in the left panel of Figure \ref{fig:mixturemodel_kdes}.
Although there is good agreement between these distributions, there
is still some discrepancy which is analyzed in Proposition \ref{prop:errorUpperBound}.
In the right panel of Figure \ref{fig:mixturemodel_kdes}, we show
that this difference can be reduced by combining the Gibbs flow with
Hamiltonian Monte Carlo (HMC) kernels\footnote{Here we apply a Hamiltonian Monte Carlo kernel between time intervals
of $h=0.0025$. We use a step size of $0.1$ for the leapfrog integrator
and an integration time of $1.0$.}, as discussed in Section \ref{subsec:Combining-Gibbs-flow-MCMC}.

\subsection{Variance component models\label{subsec:Variance-component-models}}

We now apply our proposed methodology to variance component models,
which is typical of problems in Bayesian statistics where one would
employ a Gibbs sampler \cite{gelfand_smith_1990,rosenthal_1996}.
Firstly, there are two hyperparameters with prior distributions $\sigma_{\theta}^{2}\sim\mathcal{IG}(\alpha_{0},\beta_{0})$
and $\mu\sim\mathcal{N}(\mu_{0},\sigma_{0}^{2})$, where $\mathcal{IG}(a,b)$
(and $s\mapsto\mathcal{IG}(s;a,b)$) denotes the inverse Gamma distribution
(and density) with shape parameter $a$ and scale parameter $b$.
Following \cite{rosenthal_1996}, we adopt an improper prior for $\sigma_{\theta}^{2}$
and a flat or vague prior for $\mu$. Given these hyperparameters,
there are $K\in\mathbb{N}$ location parameters $\theta=(\theta_{1},\ldots,\theta_{K})\in\mathbb{R}^{K}$
that are conditionally independent and distributed as $\theta_{i}\sim\mathcal{N}(\mu,\sigma_{\theta}^{2})$
for $i=1,\ldots,K$. With these parameters, $J\in\mathbb{N}$ observations
at each location $i=1,\ldots,K$ are modeled as conditionally independent
and distributed as $Y_{ij}\sim\mathcal{N}(\theta_{i},\sigma_{e}^{2})$
for $j=1,\ldots,J,$ where $\sigma_{e}^{2}$ is estimated empirically.
We will write $y=(y_{ij})\in\mathbb{R}^{K\times J}$ as the observed
dataset. 

In this application, the improper prior (with a possibly negative
value of $\alpha_{0}$) is 
\begin{equation}
p_{0}(\sigma_{\theta}^{2},\mu,\theta)=\mathcal{IG}(\sigma_{\theta}^{2};\alpha_{0},\beta_{0})\mathcal{N}(\mu;\mu_{0},\sigma_{0}^{2})\prod_{i=1}^{K}\mathcal{N}(\theta_{i};\mu,\sigma_{\theta}^{2})\label{eq:vc_improper_prior}
\end{equation}
and the likelihood function is $p(y|\sigma_{\theta}^{2},\mu,\theta)=\prod_{i=1}^{K}\prod_{j=1}^{J}\mathcal{N}(y_{ij};\theta_{i},\sigma_{e}^{2})$
for $(\sigma_{\theta}^{2},\mu,\theta)\in\mathbb{R}_{+}\times\mathbb{R}\times\mathbb{R}^{K}$.
To employ the methodology described in Section \ref{subsec:A-path-from-prior-to-posterior},
we set $(x_{1},x_{2},x_{3})=(\sigma_{\theta}^{2},\mu,\theta)$ as
the parameters to be inferred and consider the following ``artificial''
prior distribution
\begin{equation}
\pi_{0}(\sigma_{\theta}^{2},\mu,\theta)=\mathcal{IG}(\sigma_{\theta}^{2};\alpha_{1},\beta_{1})\mathcal{N}(\mu;\mu_{1},\sigma_{1}^{2})\prod_{i=1}^{K}\mathcal{N}(\theta_{i};\mu_{2},\sigma_{2}^{2})\label{eq:vcmodel_artificial_prior}
\end{equation}
to initialize our method, for some fixed $\alpha_{1}>0,\beta_{1}>0,\mu_{1}\in\mathbb{R},\sigma_{1}^{2}>0,\mu_{2}\in\mathbb{R},\sigma_{2}^{2}>0$.
The corresponding ``artificial'' likelihood function that would
yield the desired posterior $p(\sigma_{\theta}^{2},\mu,\theta|y)\propto p_{0}(\sigma_{\theta}^{2},\mu,\theta)p(y|\sigma_{\theta}^{2},\mu,\theta)$
is 
\[
L(\sigma_{\theta}^{2},\mu,\theta)=\frac{p_{0}(\sigma_{\theta}^{2},\mu,\theta)p(y|\sigma_{\theta}^{2},\mu,\theta)}{\pi_{0}(\sigma_{\theta}^{2},\mu,\theta)}.
\]
Given these choices, which are necessary to deal with the improper
prior (\ref{eq:vc_improper_prior}), we can then define the curve
of distributions $\{\pi_{t}\}_{t\in[0,1]}$ in (\ref{eqn:tempering}).

It can be shown that the full conditional distributions of $\pi_{t},t\in[0,1]$
are 
\begin{align}
 & \pi_{t}(\sigma_{\theta}^{2}|\mu,\theta)=\mathcal{IG}(\sigma_{\theta}^{2};\alpha(t),\beta(t|\mu,\theta)),\quad\pi_{t}(\mu|\sigma_{\theta}^{2},\theta)=\mathcal{N}(\mu;\nu(t|\sigma_{\theta}^{2},\theta),\varsigma^{2}(t|\sigma_{\theta}^{2})),\nonumber \\
 & \pi_{t}(\theta|\sigma_{\theta}^{2},\mu)=\prod_{i=1}^{K}\mathcal{N}(\theta_{i};\xi_{i}(t|\sigma_{\theta}^{2},\mu,y),\tau^{2}(t|\sigma_{\theta}^{2})),\label{eq:vcmodel_fullconditionals}
\end{align}
where the summary statistics $\alpha,\beta,\nu,\varsigma^{2},\xi_{1},\ldots,\xi_{K},\tau^{2}$
are given in Appendix \ref{sec:Expressions-for-variance-component}.
Since these full conditionals lie in the exponential family, we can
exploit such analytical tractability to determine a Gibbs velocity
field. For the parameter $\sigma_{\theta}^{2}$, we use (\ref{eqn:gibbsAnti})
which reduces to 
\begin{equation}
\tilde{f}_{1}(t,\sigma_{\theta}^{2},\mu,\theta)=\frac{-\int_{0}^{\sigma_{\theta}^{2}}\left\{ \kappa(t|\mu,\theta)-\alpha'(t)\log(u_{1})-\beta'(t|\mu,\theta)u_{1}^{-1}\right\} \mathcal{IG}(u_{1};\alpha(t),\beta(t|\mu,\theta))\,\mathrm{d}u_{1}}{\mathcal{IG}(\sigma_{\theta}^{2};\alpha(t),\beta(t|\mu,\theta))}\label{eq:vcmodel_sigma_gibbs}
\end{equation}
where $\alpha'(t)$ and $\beta'(t|\mu,\theta)$ denote the time derivatives
of $\alpha(t)$ and $\beta(t|\mu,\theta)$ respectively, 
\[
\kappa(t|\mu,\theta)=\alpha'(t)\psi(\alpha(t))-\alpha'(t)\log(\beta(t|\mu,\theta))-\alpha(t)\beta^{-1}(t|\mu,\theta)\beta'(t|\mu,\theta)
\]
and $\psi$ is the digamma function\footnote{We evaluate this function using the \texttt{digamma} function in the
R \texttt{base} package.}. For parameters $\mu$ and $\theta$ which have Gaussian full conditional
distributions, the corresponding components of the Gibbs velocity
field are more explicit 
\begin{align}
 & \tilde{f}_{2}(t,\sigma_{\theta}^{2},\mu,\theta)=\frac{\varsigma'(t|\sigma_{\theta}^{2})}{\sigma_{1}}(\mu-\mu_{1})+\nu'(t|\sigma_{\theta}^{2},\theta),\label{eq:vcmodel_mu_gibbs}\\
 & \tilde{f}_{3}(t,\sigma_{\theta}^{2},\mu,\theta)=\frac{\tau'(t|\sigma_{\theta}^{2})}{\sigma_{2}}(\theta-\mu_{2})+\xi'(t|\sigma_{\theta}^{2},\mu,y),\label{eq:vcmodel_theta_gibbs}
\end{align}
where $\varsigma',\nu',\tau'$ and $\xi'=(\xi_{1}',\ldots,\xi_{K}')$
denote the time derivatives of $\varsigma,\nu,\tau$ and $\xi=(\xi_{1},\ldots,\xi_{K})$
respectively. To approximate the Gibbs flow, we update the parameter
$\sigma_{\theta}^{2}$ at time step $m=1,\ldots,M$ using the Euler
discretization (\ref{eq:euler_discretization}), which defines the
map 
\[
\Psi_{m,1}(\sigma_{\theta}^{2},\mu,\theta)=(\sigma_{\theta}^{2}+h\hat{f}_{1}(t_{m-1},\sigma_{\theta}^{2},\mu,\theta),\mu,\theta),
\]
where $\hat{f}_{1}$ denotes an approximation of (\ref{eq:vcmodel_sigma_gibbs})
using a composite trapezoidal rule with $R=50$ quadrature points.
To update the parameter $\mu$ or $\theta$ conditionally on other
parameters, since the solution of (\ref{eq:conditional_ODE}) under
the linear velocity (\ref{eq:vcmodel_mu_gibbs}) or (\ref{eq:vcmodel_theta_gibbs})
is tractable, we have 
\begin{align*}
 & \Psi_{m,2}(\sigma_{\theta}^{2},\mu,\theta)=\left(\sigma_{\theta}^{2},\frac{\varsigma(t_{m}|\sigma_{\theta}^{2})}{\varsigma(t_{m-1}|\sigma_{\theta}^{2})}(\mu-\nu(t_{m-1}|\sigma_{\theta}^{2},\theta))+\nu(t_{m}|\sigma_{\theta}^{2},\theta),\theta\right),\\
 & \Psi_{m,3}(\sigma_{\theta}^{2},\mu,\theta)=\left(\sigma_{\theta}^{2},\mu,\frac{\tau(t_{m}|\sigma_{\theta}^{2})}{\tau(t_{m-1}|\sigma_{\theta}^{2})}(\theta-\xi(t_{m-1}|\sigma_{\theta}^{2},\mu,y))+\xi(t_{m}|\sigma_{\theta}^{2},\mu,y)\right),
\end{align*}
for $m=1,\ldots,M$. In contrast to the generic expressions in (\ref{eq:rewrite_gibbs_velocity_1d})-(\ref{eq:rewrite_CDF_1d}),
approximating the Gibbs flow for this model only requires computing
summary statistics and evaluating inverse Gamma densities. 

\begin{figure}[htbp]
\centering{}\includegraphics[scale=0.5]{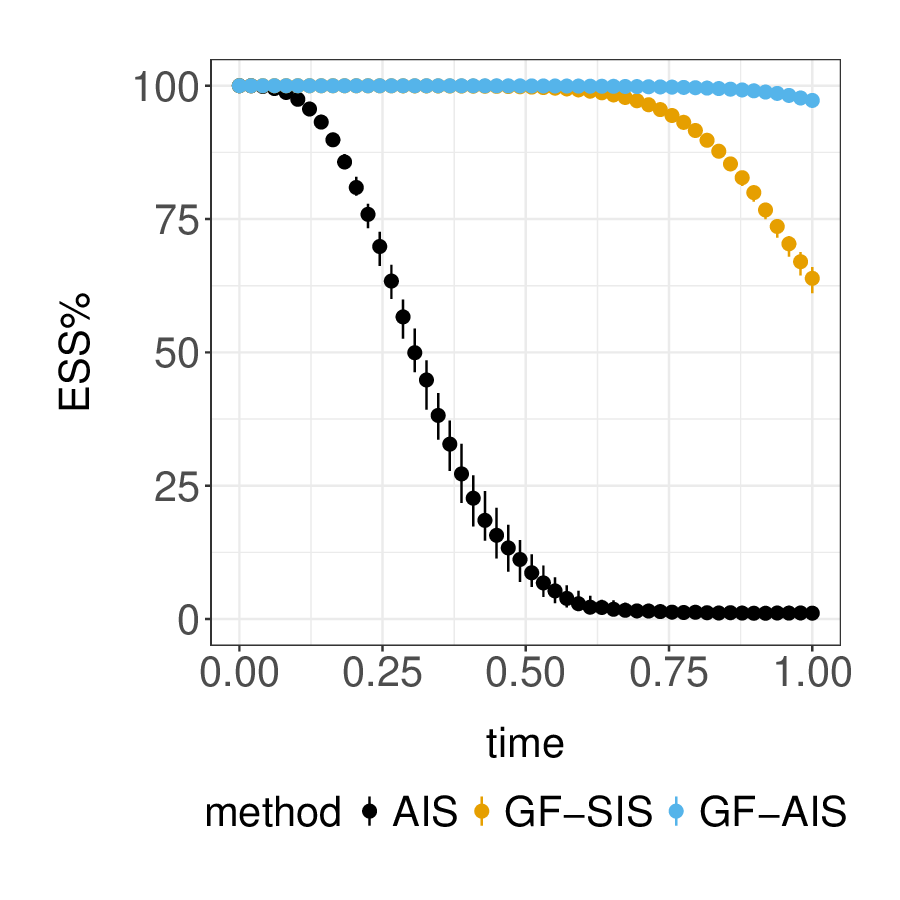}\includegraphics[scale=0.5]{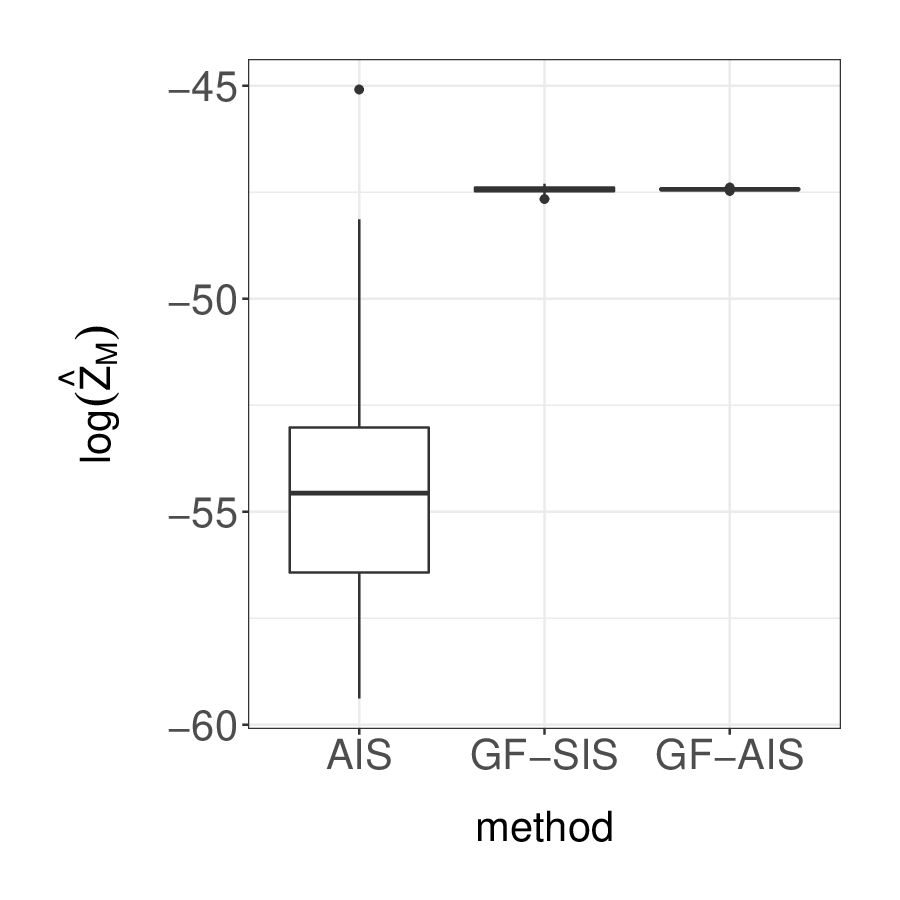}
\caption{Boxplots of effective sample size percentage (\emph{left}) and log-marginal
likelihood estimates (\emph{right}) when fitting variance component
model on the baseball dataset, obtained with $100$ independent repetitions
of AIS, GF-SIS (Algorithm \ref{alg:GF-SIS}) and GF-AIS (Algorithm
\ref{alg:GF-AIS}). \label{fig:vcmodel_baseball}}
\end{figure}

We consider a dataset of $K=18$ baseball players\textquoteright{}
batting averages ($J=1$) taken from \cite[Table 1]{morris1983}.
In this case, the number of parameters to be inferred is $d=K+2=20$.
Following \cite{rosenthal_1996}, we adopt the empirical estimate
$\sigma_{e}^{2}=4.34\times10^{-3}$ and a prior specification corresponding
to $\alpha_{0}=-1,\beta_{0}=2,\mu_{0}=0,\sigma_{0}=10$. We initialize
the Gibbs flow using an ``artificial'' prior distribution (\ref{eq:vcmodel_artificial_prior})
with $\alpha_{1}=\beta_{1}=4,\mu_{1}=\mu_{2}=0,\sigma_{1}=\sigma_{2}=0.1$. 
In Figure \ref{fig:vcmodel_baseball}, we display the performance
of the resulting GF-SIS (Algorithm \ref{alg:GF-SIS}) using $N=128$
samples and $M=50$ time steps. 
To improve performance with fixed $N$ and $M$, we combine approximate
Gibbs flow with HMC kernels\footnote{We apply a Hamiltonian Monte Carlo kernel at each time step. To achieve
suitable acceptance probabilities, we use a step size of $0.05$ for
the leapfrog integrator and an integration time of $0.5$. } within GF-AIS (Algorithm \ref{alg:GF-AIS}): this increases the ESS\%
from $63\%$ to $97\%$ on average, and reduces the variance of the
log-marginal likelihood estimator by a factor of $22$, at the expense
of $4$ times the compute time of GF-SIS. As competing algorithm,
we consider AIS with the same $N$, $M$ and HMC kernels as GF-AIS,
but we increase the number of HMC iterations at each time step to
match the computational time of GF-AIS, so as to ensure a fair comparison.
Based on $100$ independent repetitions of all three algorithms, the
sample variance of log-marginal likelihood estimates relative to AIS
was observed to be $1255$ and $27,928$ times smaller for GF-SIS
and GF-AIS respectively. 

\begin{figure}[htbp]
\centering{}\includegraphics[scale=0.5]{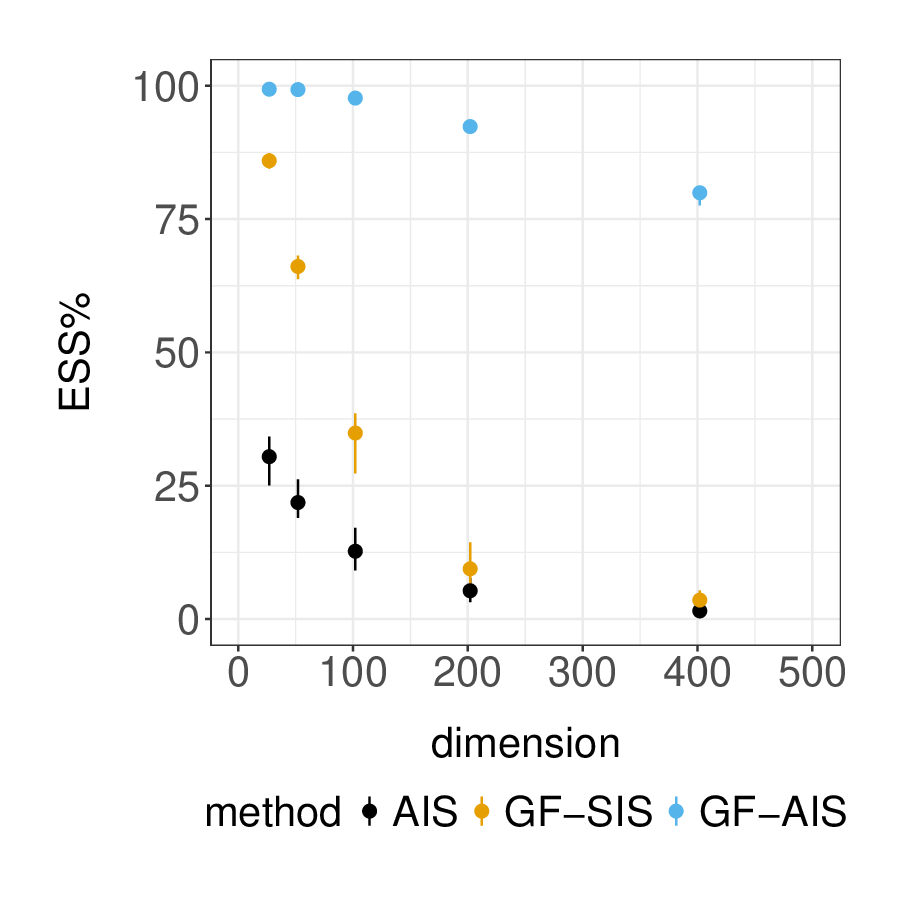}\includegraphics[scale=0.5]{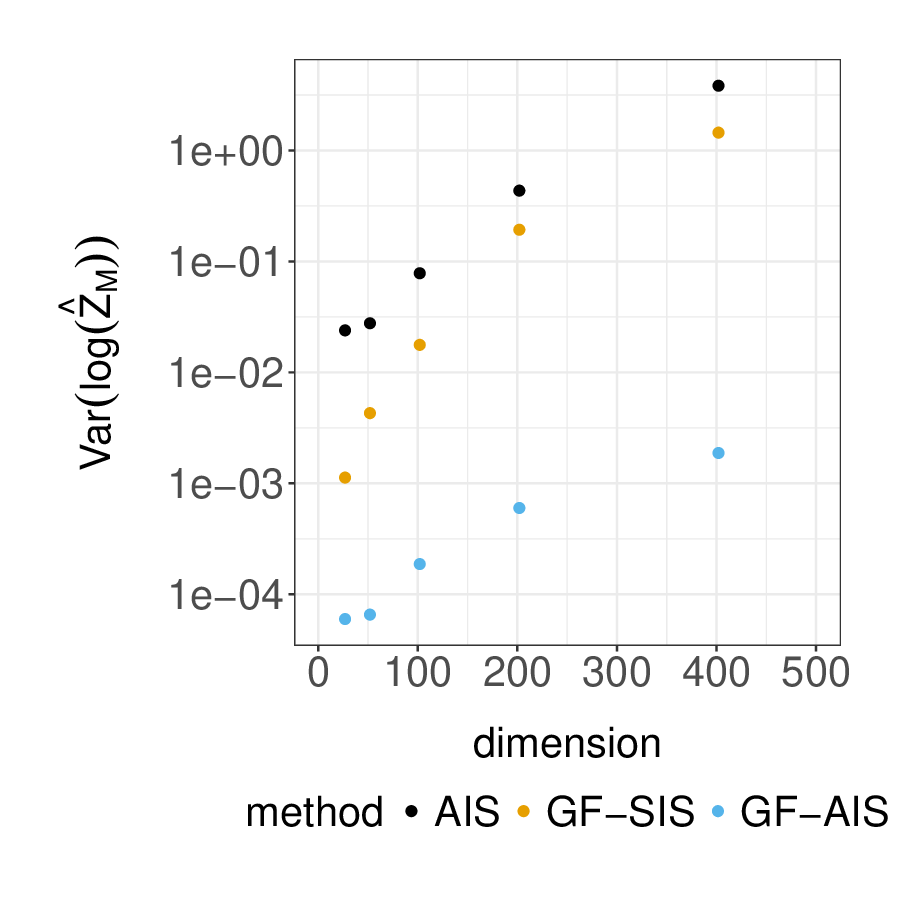}
\caption{Boxplots of terminal effective sample size percentage (\emph{left})
and variance of log-marginal likelihood estimates (\emph{right}) when
fitting variance component model on simulated data in various dimensions,
obtained with $100$ independent repetitions of AIS, GF-SIS (Algorithm
\ref{alg:GF-SIS}) and GF-AIS (Algorithm \ref{alg:GF-AIS}). \label{fig:vcmodel_simulated}}
\end{figure}

We then investigate how the performance of these algorithms behaves
with dimension on simulated data. The model specification and algorithmic
settings remain the same as we scale $d\in\{27,52,102,202,402\}$,
with the exception of increasing time steps $M\in\{125,250,500,750,1000\}$
linearly with $d$ and decreasing the step size of the leapfrog integrator
in HMC to achieve stable acceptance probabilities\footnote{For $d\in\{27,52,102,202,402\}$, we use $10$ steps of the leapfrog
integrator with step size $\{0.0125,0.0100,0.0075,0.0050,0.0025\}$
respectively. }. Like before, we select the number of HMC iterations in AIS to match
the compute time of GF-AIS; both algorithms require approximately
$\{5,7,14,15,16\}$ times more compute time than GF-SIS as $d$ varies.
Figure \ref{fig:vcmodel_simulated} summarizes how the performance
of these algorithms scale with dimension. Relative to standard AIS,
GF-SIS performed better in all of the observed dimensions despite
costing less compute time, and GF-AIS offers much better ESS\% and
variance reduction of several orders at a fixed computational cost.

\subsection{Log-Gaussian Cox point processes}

Lastly, we present an application of our methodology on a model from
spatial statistics. In particular, we consider Bayesian inference
for log-Gaussian Cox point processes on a dataset\footnote{The dataset can be found in the R package \texttt{spatstat} as \texttt{finpines}.}
concerning the locations of $126$ Scots pine saplings in a natural
forest in Finland \cite{moller1998,christensen2005,girolami2011}.
The actual square plot of $10\times10$ square metres is standardized
to the unit square and discretized into a $J\times J$ regular grid.
Given a latent intensity process $\Lambda_{j},j\in\{1,\ldots,J\}^{2}$,
the number of points in each cell $Y_{j}$ for $j\in\{1,\ldots,J\}^{2}$
are modeled as conditionally independent and Poisson distributed with
mean $a\Lambda_{j}$, where $a=J^{-2}$ is the area of each cell.
The prior distribution of the intensity is specified by the relation
$\Lambda_{j}=\exp(X_{j})$ for $j\in\{1,\ldots,J\}^{2}$, where $X=(X_{j})\in\mathbb{R}^{J\times J}$
is a Gaussian process with constant mean $\mu_{0}\in\mathbb{R}$ and
exponential covariance function $\Sigma_{0}(i,j)=\sigma^{2}\exp(-|i-j|/(J\beta))$
for $i,j\in\{1,\ldots,J\}^{2}$ and $\sigma^{2},\beta>0$. We will
adopt the parameter values $\sigma^{2}=1.91$, $\beta=1/33$ and $\mu_{0}=\log(126)-\sigma^{2}/2$
estimated by \cite{moller1998}. This application corresponds to working
in dimension $d=J^{2}$ with a prior distribution of $p_{0}(x)=\mathcal{N}(x;\mu_{0}1_{d},\Sigma_{0})$
where $1_{d}=(1,\ldots,1)\in\mathbb{R}^{d}$ and a likelihood function
of $p(y|x)=\prod_{j\in\{1,\ldots,J\}^{2}}p(y_{j}|x_{j})=\prod_{j\in\{1,\ldots,J\}^{2}}\exp(x_{j}y_{j}-a\exp(x_{j}))$,
where $y=(y_{j})\in\mathbb{N}^{J\times J}$ denotes the dataset. 

We will apply the methodology described in Section \ref{subsec:A-path-from-prior-to-posterior}
with initialization from the prior distribution $\pi_{0}=p_{0}$ or
a Gaussian approximation of the posterior distribution; given by either
a mean field variational Bayes (VB) approximation \cite{teng2017}
$\pi_{0}(x)=\prod_{j\in\{1,\ldots,J\}^{2}}\mathcal{N}(x_{j};\mu_{j},\sigma_{j}^{2})$,
or of the form $\pi_{0}(x)\propto p_{0}(x)\prod_{j\in\{1,\ldots,J\}^{2}}\mathcal{N}(x_{j};\mu_{j},\sigma_{j}^{2})$,
where $(\mu_{j},\sigma_{j}^{2}),j\in\{1,\ldots,J\}^{2}$ are fitted
using expectation-propagation (EP) \cite{minka2001}, as advocated
in \cite{chopin2017}. To accommodate these choices, we take as \textquotedblleft artificial\textquotedblright{}
likelihood function $L(x)=p_{0}(x)p(y|x)/\pi_{0}(x)$ to define the
curve of distributions $\{\pi_{t}\}_{t\in[0,1]}$ in (\ref{eqn:tempering}).
Although the full conditional distributions of $\pi_{t},t\in[0,1]$
are not in the exponential family, computation of the Gibbs velocity
field (\ref{eq:rewrite_gibbs_velocity_1d})-(\ref{eq:rewrite_CDF_1d})
for one-dimensional components can be greatly simplified by rewriting
\begin{equation}
\tilde{f}_{i}(t,x)=\frac{\lambda'(t)\left\{ F_{t}(x_{i}|x_{-i})\int_{-\infty}^{\infty}\log L_{i}(u_{i},x_{-i})\pi_{t}(u_{i}|x_{-i})\,\mathrm{d}u_{i}-\int_{-\infty}^{x_{i}}\log L_{i}(u_{i},x_{-i})\pi_{t}(u_{i}|x_{-i})\,\mathrm{d}u_{i}\right\} }{\pi_{t}(x_{i}|x_{-i})}\label{eq:cox_gibbs_velocity_1d}
\end{equation}
where $L_{i}(x)=p_{0}(x_{i}|x_{-i})p(y_{i}|x_{i})/\pi_{0}(x_{i}|x_{-i})$
and 
\[
\frac{\pi_{t}(u_{i}|x_{-i})}{\pi_{t}(x_{i}|x_{-i})}=\frac{\gamma_{t}(u_{i},x_{-i})}{\gamma_{t}(x_{i},x_{-i})}=\frac{\pi_{0}(u_{i}|x_{-i})^{1-\lambda(t)}p_{0}(u_{i}|x_{-i})^{\lambda(t)}p(y_{i}|u_{i})^{\lambda(t)}}{\pi_{0}(x_{i}|x_{-i})^{1-\lambda(t)}p_{0}(x_{i}|x_{-i})^{\lambda(t)}p(y_{i}|x_{i})^{\lambda(t)}}
\]
for $i=1,\ldots,d$, and noting that the full conditional distributions
$\{p_{0}(x_{i}|x_{-i})\}_{i=1,\ldots,d}$ and $\{\pi_{0}(x_{i}|x_{-i})\}_{i=1,\ldots,d}$
are univariate Gaussians that can be precomputed. We approximate (\ref{eq:cox_gibbs_velocity_1d})
using a composite trapezoidal rule with $R=40$ quadrature points,
and the Gibbs flow using the Euler discretization (\ref{eq:euler_discretization}),
which defines the map $\Psi_{m,i}$ for time step $m=1,\ldots,M$
and component $i=1,\ldots,d$. 

\begin{figure}[htbp]
\centering{}\includegraphics[scale=0.5]{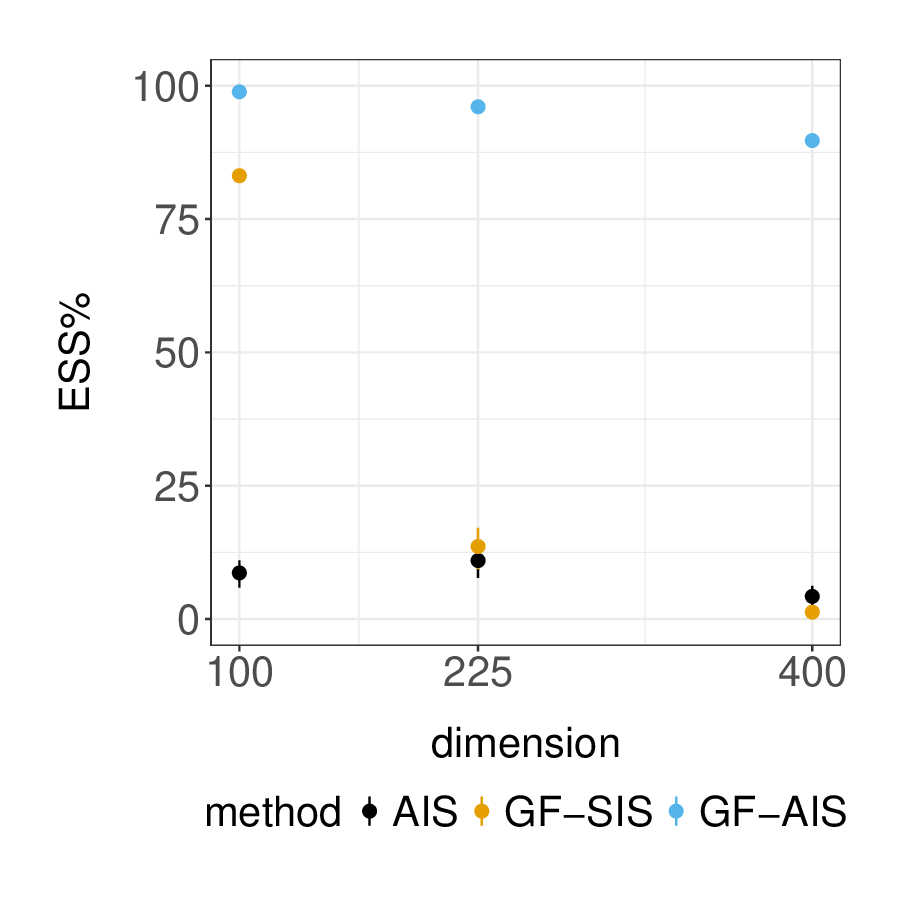}\includegraphics[scale=0.5]{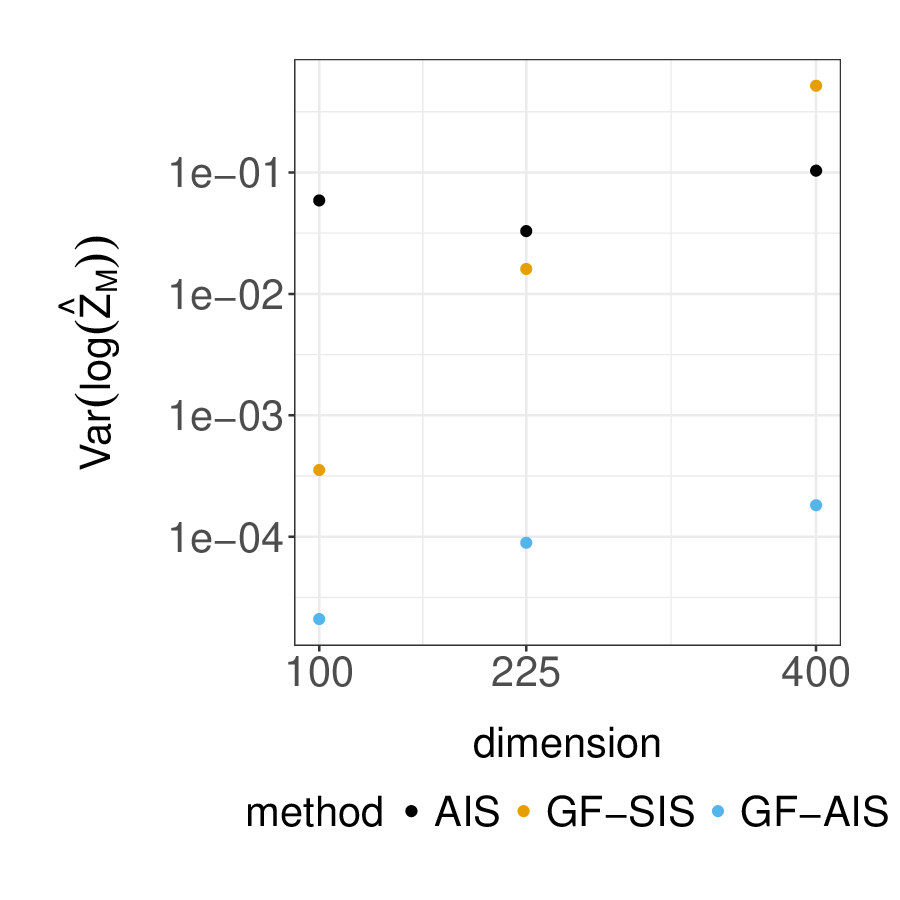}
\caption{Boxplots of terminal effective sample size percentage (\emph{left})
and variance of log-marginal likelihood estimates (\emph{right}) when
fitting log-Gaussian Cox point process model on Scots pine saplings
dataset with various spatial resolutions, obtained with $100$ independent
repetitions of AIS, GF-SIS (Algorithm \ref{alg:GF-SIS}) and GF-AIS
(Algorithm \ref{alg:GF-AIS}). \label{fig:coxprocess_scaling}}
\end{figure}

We first consider initialization from the prior $\pi_{0}=p_{0}$ and
vary the spatial resolution by taking $d\in\{10^{2},15^{2},20^{2}\}$.
Figure \ref{fig:coxprocess_scaling} displays the performance of GF-SIS
(Algorithm \ref{alg:GF-SIS}) using $N=512$ samples and as we increase
the time steps $M\in\{40,60,80\}$ with dimension correspondingly.
To obtain better performance for the same number of samples $N$ and
time steps $M$, we combine approximate Gibbs flow with Riemann manifold
Hamiltonian Monte Carlo (RM-HMC) kernels\footnote{We apply a Riemann manifold Hamiltonian Monte Carlo kernel at each
time step, with a leapfrog integrator step size of 0.25 and an integration
time of 2.5.} that employ the metric tensor $\Sigma_{0}^{-1}+a\exp(\mu_{0}+\sigma^{2}/2)I_{d}$
\cite{girolami2011}, where $I_{d}\in\mathbb{R}^{d\times d}$ denotes
the identity matrix. Although the resulting GF-AIS (Algorithm \ref{alg:GF-AIS})
requires approximately $\{25\%,50\%,100\%\}$ more compute time than
GF-SIS as dimension increases, it is apparent from Figure \ref{fig:coxprocess_scaling}
that it improves algorithmic performance by several orders of magnitude.
Like before, we compare GF-AIS to an AIS with the same $N$, $M$
and RM-HMC kernels, but to ensure a fair comparison, the number of
RM-HMC iterations at each time step is increased to match computational
time. The results summarized in Figure \ref{fig:coxprocess_scaling}
indicate that GF-AIS can offer very significant numerical gains over
standard AIS in all three dimensions considered. 

\begin{figure}[htbp]
\centering{}\includegraphics[scale=0.5]{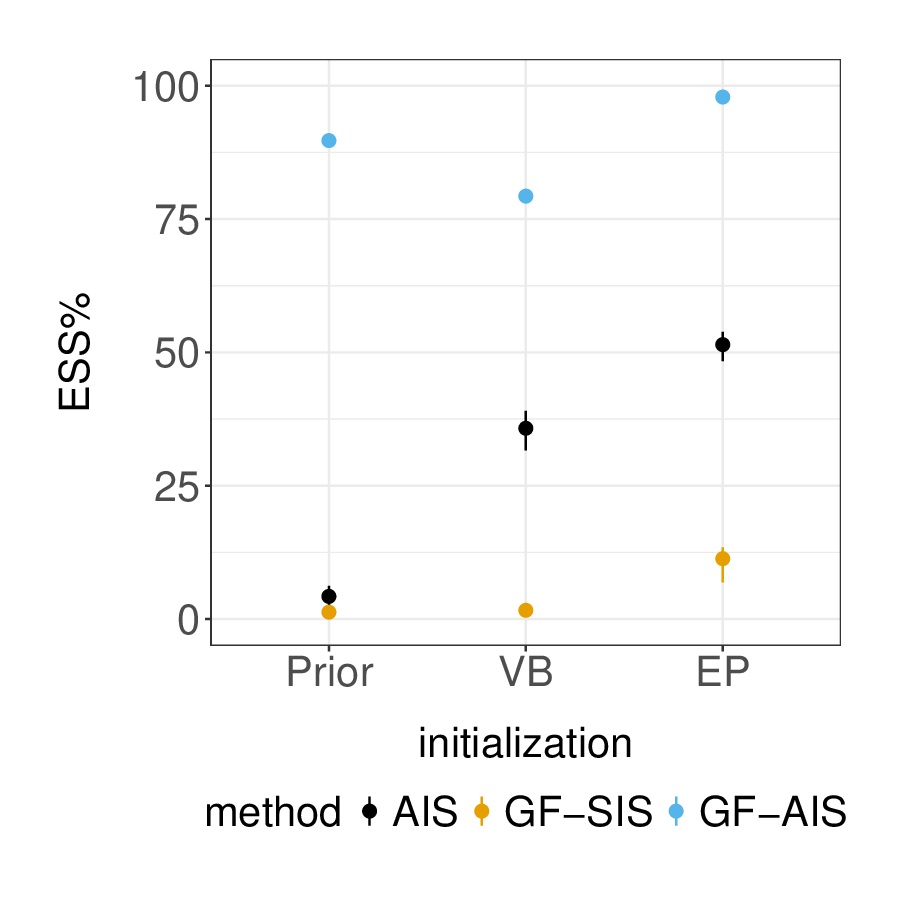}\includegraphics[scale=0.5]{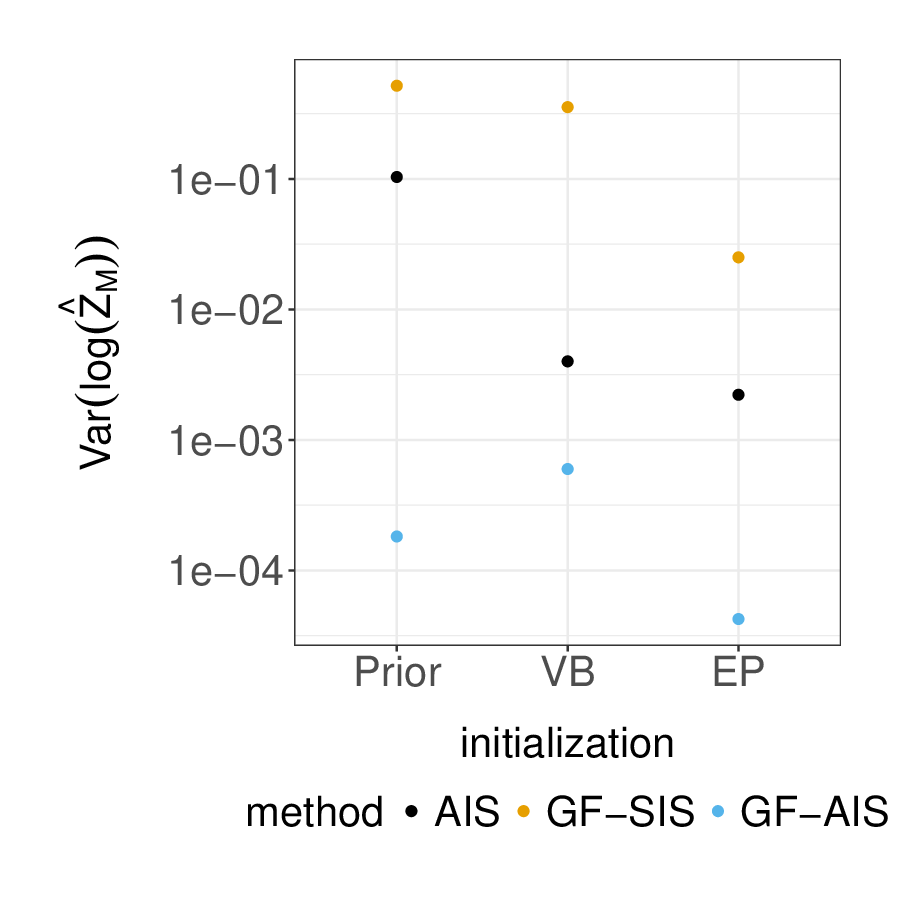}
\caption{Boxplots of terminal effective sample size percentage (\emph{left})
and variance of log-marginal likelihood estimates (\emph{right}) when
fitting log-Gaussian Cox point process model on Scots pine saplings
dataset with various initial distributions, obtained with $100$ independent
repetitions of AIS, GF-SIS (Algorithm \ref{alg:GF-SIS}) and GF-AIS
(Algorithm \ref{alg:GF-AIS}). The initial distributions considered
here are the prior, a variational Bayes (VB) and an expectation-propagation
(EP) approximation of the posterior distribution. \label{fig:coxprocess_initial}}
\end{figure}

Lastly, we investigate the impact of the initial distribution for
a spatial resolution of $d=20^{2}$. Figure \ref{fig:coxprocess_initial}
shows that initial distributions that are ``closer'' to the posterior
than the prior typically lead to better algorithmic performance. Interestingly,
due to the nature of the Gibbs flow approximation, we find that an
EP approximation of the posterior provides a better initialization
than a VB approximation. 

\appendix

\section{Informal derivation of Liouville's equation\label{append:liouville}}

Consider a $d$-dimensional hyper-rectangle $\Delta V(x)$, defined
formally as the Cartesian product of intervals $(x_{i},x_{i}+\Delta_{i})$
for $i=1,\ldots,d$ and some small $\Delta=(\Delta_{1},\ldots,\Delta_{d})\in\mathbb{R}_{+}^{d}$,
to be thought of as an infinitesimal control volume at a point $x\in\mathbb{R}^{d}$. 
%as depicted in Figure \ref{fig:conserveMass}. 

If we perceive particles as constituents of a fluid representing probability
mass, then the fluid flow driven by a velocity field $f$ will cause
the probability mass in $\Delta V(x)$ to change. Along the $i^{\mathrm{th}}$
axis, for sufficiently small $|\Delta|_{\infty}:=\max_{i=1,\ldots,d}\Delta_{i}$,
this change is given by the difference between the rate at which mass
flows into $\Delta V(x)$ 
\begin{align}
\tilde{\pi}_{t}(x)f_{i}(t,x)\prod_{j\neq i}\Delta_{j}+o(|\Delta|_{{\infty}}^{d})\label{eqn:massflowin}
\end{align}
and the rate at which mass flows out of $\Delta V(x)$ 
\begin{align}
\tilde{\pi}_{t}(x+\Delta_{i}e_{i})f_{i}(t,x+\Delta_{i}e_{i})\prod_{j\neq i}\Delta_{j}+o(|\Delta|_{\infty}^{d}),\label{eqn:massflowout}
\end{align}
where $\{e_{i}\}_{i=1}^{d}$ denote the canonical basis vectors in
$\mathbb{R}^{d}$. In fluid dynamics terminology, the leading terms
in (\ref{eqn:massflowin}) and (\ref{eqn:massflowout}) are simply
the density multiplied by the volume metric flow rate in and out of
the control volume respectively.

Summing over all axes yields the net rate at which probability mass
is accumulating in $\Delta V(x)$: 
\begin{align}
\sum_{i=1}^{d}\left(\tilde{\pi}_{t}(x)f_{i}(t,x)\prod_{j\neq i}\Delta_{j}-\tilde{\pi}_{t}(x+\Delta_{i}e_{i})f_{i}(t,x+\Delta_{i}e_{i})\prod_{j\neq i}\Delta_{j}\right)+o(|\Delta|_{\infty}^{d}).\label{eqn:changeinmass}
\end{align}
For probability mass to be conserved, (\ref{eqn:changeinmass}) has
to be equal to 
\begin{align}
\partial_{t}\tilde{\pi}_{t}(x)\prod_{i=1}^{d}\Delta_{i}+o(|\Delta|_{\infty}^{d}).\label{eqn:desiredchangeinmass}
\end{align}
Equating (\ref{eqn:changeinmass}) and (\ref{eqn:desiredchangeinmass})
and dividing by the volume $\prod_{i=1}^{d}\Delta_{i}$ of $\Delta V(x)$
gives 
\begin{align}
\partial_{t}\tilde{\pi}_{t}(x)=\sum_{i=1}^{d}\frac{\tilde{\pi}_{t}(x)f_{i}(t,x)-\tilde{\pi}_{t}(x+\Delta_{i}e_{i})f_{i}(t,x+\Delta_{i}e_{i})}{\Delta_{i}}+o(1).
\end{align}
Finally, taking the limit of $|\Delta|_{\infty}\rightarrow0$ gives
(\ref{eqn:liouville_initial}). 
%\begin{figure}[h!]
%\centering \includegraphics[scale=0.3]{conserveMass} \caption{Illustrating the conservation of mass argument in $\mathbb{R}^{2}$.}
%\label{fig:conserveMass} 
%\end{figure}

\section{Flow transport problem\label{append:flowtransport}}

In this section, we supply additional details behind the validity
of the flow transport problem. We first establish a preliminary result
about the curve of distributions (\ref{eqn:tempering}). 
\begin{lemma}
\label{lem:narrow_continuous}The curve of distributions $\{\pi_{t}\}_{t\in[0,1]}$
defined in (\ref{eqn:tempering}) is narrowly continuous, i.e. for
any bounded function $\varphi\in C^{0}(\mathbb{R}^{d},\mathbb{R})$
and any sequence $(t_{n})_{n\geq1}\subset[0,1]$ such that $t_{n}\rightarrow t_{*}$
we have
\begin{align}
\lim_{n\rightarrow\infty}\int_{\mathbb{R}^{d}}\varphi(x)\pi_{t_{n}}(x)\,\mathrm{d}x=\int_{\mathbb{R}^{d}}\varphi(x)\pi_{t_{*}}(x)\,\mathrm{d}x.\label{eq:narrow_continuous}
\end{align}
\end{lemma}
\begin{proof}
Using the dominated convergence theorem with dominating function
\[
\pi_{0}(x)L(x)^{\lambda(t)}\leq\pi_{0}(x)\sup_{t\in[0,1]}L(x)^{\lambda(t)}\leq\pi_{0}(x)(1+L(x))
\]
shows that $Z(t)$ is continuous on $[0,1]$. Together with continuity
of $\lambda(t)$, it then follows that $t\mapsto\pi_{t}(x)\in C^{0}([0,1],\mathbb{R}_{+})$
for each $x\in\mathbb{R}^{d}$. Hence for any bounded function $\varphi\in C^{0}(\mathbb{R}^{d},\mathbb{R})$
and any sequence $(t_{n})_{n\geq1}\subset[0,1]$ such that $t_{n}\rightarrow t_{*}$,
we have $\varphi(x)\pi_{t_{n}}(x)\rightarrow\varphi(x)\pi_{t_{*}}(x)$
pointwise. Note that
\begin{align}
|\varphi(x)\pi_{t_{n}}(x)|\leq\sup_{u\in\mathbb{R}^{d}}|\varphi(u)|\frac{\pi_{0}(x)\sup_{t\in[0,1]}L(x)^{\lambda(t)}}{\inf_{t\in[0,1]}Z(t)}.\label{eqn:boundNarrow}
\end{align}
Since $Z(t)$ is continuous on $[0,1]$, the infimum in (\ref{eqn:boundNarrow})
is attained and is strictly positive under positivity assumptions
made on $\pi_{0}(x)$ and $L(x)$. Hence the upper bound in (\ref{eqn:boundNarrow})
is integrable and by the dominated convergence theorem, (\ref{eq:narrow_continuous})
follows. 
\end{proof}
We now introduce the notion of weak solutions which is needed to establish
Theorem \ref{thm:ambrosio}.
\begin{defn}
A curve of distributions $\{\tilde{\pi}_{t}\}_{t\in[0,1]}$ is a weak
solution of (\ref{eqn:liouville_initial}) if 
\begin{align}
\int_{0}^{1}\int_{\mathbb{R}^{d}}\left(\partial_{t}\varphi(t,x)+\left<f(t,x),\nabla\varphi(t,x)\right>\right)\tilde{\pi}_{t}(x)\,\mathrm{d}x\,{\rm d}t=0
\end{align}
for all compactly supported $\varphi\in C^{\infty}((0,1)\times\mathbb{R}^{d},\mathbb{R})$,
where $\left<\cdot,\cdot\right>$ denotes the inner product in $\mathbb{R}^{d}$. 
\end{defn}
\begin{proof}[Proof of Theorem \ref{thm:ambrosio}]
If $f$ satisfies (\ref{eq:liouville}) for all $(t,x)\in(0,1)\times\mathbb{R}^{d}$,
then $\{\pi_{t}\}_{t\in[0,1]}$ is a weak solution of Liouville's
equation (\ref{eqn:liouville_initial}) and is narrowly continuous
by Lemma \ref{lem:narrow_continuous}. Therefore under Assumptions
A1-A2, the conclusions of Theorem \ref{thm:ambrosio} follow from
\cite[Propositions 8.1.7-8.1.8]{ambrosio2005}.
\end{proof}
It is worth noting that converse of Theorem \ref{thm:ambrosio} also
holds under the same conditions \cite[Lemma 8.1.6]{ambrosio2005}.
These two results describe an equivalence between the Eulerian perspective
characterized by Liouville's PDE (\ref{eqn:liouville_initial}) and
the Lagrangian perspective described in terms of particle trajectories
governed by the ODE (\ref{eqn:ODE}). It is possible to weaken Assumption
A1; see \cite{dipernaLions1989} for earlier work and \cite{ambrosio2004},
\cite[Theorem 8.2.1]{ambrosio2005} for recent advances. 

\section{Solving the flow transport problem\label{append:solvingflow}}

In this section, we first detail the proofs of Propositions \ref{prop:1Dcase}-\ref{prop:generalCase}
before giving additional remarks on our solution to the flow transport
problem. 
\begin{proof}[Proof of Proposition \ref{prop:1Dcase}]
Using continuity of $\pi_{0},L$ and positivity of $L$, an application
of the first fundamental theorem of calculus shows that $f$ satisfies
(\ref{eq:liouville}). The assumptions on $\pi_{0}$ and $L$ imply
$f\in C^{1}([0,1]\times\mathbb{R},\mathbb{R})$ hence Assumption A1
of Theorem \ref{thm:ambrosio} holds. The integrability Assumption
A2 in Theorem \ref{thm:ambrosio} follows from the prescribed tail
behaviour of $x\mapsto|f(t,x)|\pi_{t}(x)$ uniformly over $t\in[0,1]$.
Therefore the assumptions of Theorem \ref{thm:ambrosio} hold and
$f$ solves the flow transport problem on $\mathbb{R}$. To see that
(\ref{eqn:drift1D}) is indeed the minimal kinetic energy solution,
we note that the optimality condition in \cite{reichtransport2011,reichMixture}
requires existence of a function $\varphi:[0,1]\times\mathbb{R}\rightarrow\mathbb{R}$
such that $f(t,x)=\nabla\varphi(t,x)$. This follows as a consequence
of working on $\mathbb{R}$ since we may set $\varphi(t,x)=\int_{-\infty}^{x}f(t,u)\,\mathrm{d}u<\infty$.
\end{proof}
\begin{proof}[Proof of Proposition \ref{prop:generalCase}]
The arguments are similar to those used in Proposition \ref{prop:1Dcase}.
By straightforward verification $f$ satisfies (\ref{eq:liouville}):
\begin{align*}
 & -\sum_{i=1}^{d}\partial_{x_{i}}(\pi_{t}(x)f_{i}(t,x))\\
 & =\sum_{i=1}^{d-1}\partial_{x_{i}}\Bigg(\prod_{j=1}^{i-1}\pi_{t}(x_{j})\int_{\mathbb{R}^{i-1}}\int_{-\infty}^{x_{i}}\partial_{t}\pi_{t}(u_{1:i-1},u_{i},x_{i+1:d})\,{\rm d}u_{1:i-1}{\rm d}u_{i}\\
 & \hspace{2cm}-\prod_{j=1}^{i-1}\pi_{t}(x_{j})F_{t}(x_{i})\int_{\mathbb{R}^{i}}\partial_{t}\pi_{t}(u_{1:i},x_{i+1:d})\,{\rm d}u_{1:i}\Bigg)\\
 & \hspace{2cm}+\partial_{x_{d}}\Bigg(\prod_{j=1}^{d-1}\pi_{t}(x_{j})\int_{\mathbb{R}^{d-1}}\int_{-\infty}^{x_{d}}\partial_{t}\pi_{t}(u_{1:d-1},u_{d})\,{\rm d}u_{1:d-1}{\rm d}u_{d}\Bigg)\\
 & =\sum_{i=1}^{d-1}\Bigg(\prod_{j=1}^{i-1}\pi_{t}(x_{j})\int_{\mathbb{R}^{i-1}}\partial_{t}\pi_{t}(u_{1:i-1},x_{i},x_{i+1:d})\,{\rm d}u_{1:i-1}\\
 & \hspace{2cm}-\prod_{j=1}^{i}\pi_{t}(x_{j})\int_{\mathbb{R}^{i}}\partial_{t}\pi_{t}(u_{1:i},x_{i+1:d})\,{\rm d}u_{1:i}\Bigg)\\
 & \hspace{2cm}+\prod_{j=1}^{d-1}\pi_{t}(x_{j})\int_{\mathbb{R}^{d-1}}\partial_{t}\pi_{t}(u_{1:d-1},x_{d})\,{\rm d}u_{1:d-1}\\
 & =\partial_{t}\pi_{t}(x).
\end{align*}
The penultimate line applies the first fundamental theorem of calculus
and the final equality comes from the telescopic sum. The assumptions
on $\pi_{0}$ and $L$ imply $f\in C^{1}([0,1]\times\mathbb{R}^{d},\mathbb{R}^{d})$
hence Assumption A1 of Theorem \ref{thm:ambrosio} holds. The integrability
Assumption A2 in Theorem \ref{thm:ambrosio} follows from the prescribed
tail behaviour of $x\mapsto|f(t,x)|\pi_{t}(x)$ uniformly over $t\in[0,1]$.
Therefore the assumptions of Theorem \ref{thm:ambrosio} hold and
$f$ solves the flow transport problem on $\mathbb{R}^{d}$. 
\end{proof}

We note that the velocity field $f$ defined in (\ref{eqn:BGdrift1})-(\ref{eqn:BGdrift2})
satisfies $|f(t,x)|\pi_{t}(x)\rightarrow0$ as $|x|\rightarrow\infty$
for each $t\in[0,1]$ by construction. The tail behaviour prescribed
in Proposition \ref{prop:generalCase} assumes that this decay happens
fast enough to ensure the integrability Assumption A2 in Theorem \ref{thm:ambrosio}. 

Suppose that the curve of distributions (\ref{eqn:tempering}) factorize
into independent one-dimensional components, i.e. 
\[
\pi_{t}(x)=\prod_{i=1}^{d}\pi_{t}(x_{i})=\prod_{i=1}^{d}\frac{\pi_{0}(x_{i})L_{i}(x_{i})^{\lambda(t)}}{Z_{i}(t)},
\]
where for $i=1,\ldots,d$, $\pi_{0}(x_{i})$ denotes the $i^{th}$
marginal distribution of $\pi_{0}$, $L_{i}:\mathbb{R}\rightarrow\mathbb{R}_{+}$
the corresponding likelihood function and $Z_{i}(t)=\int_{\mathbb{R}}\pi_{0}(u_{i})L_{i}(u_{i})^{\lambda(t)}\mathrm{d}u_{i}$.
In this case, the time evolution along the curve for each marginal
distribution is given by 
\[
\partial_{t}\pi_{t}(x_{i})=\lambda'(t)(\log L_{i}(x_{i})-I_{t}^{(i)})\pi_{t}(x_{i})
\]
where $I_{t}^{(i)}=\int_{\mathbb{R}}\log L_{i}(u_{i})\pi_{t}(u_{i})\,\mathrm{d}u_{i}$.
We now show that the velocity field in (\ref{eqn:BGdrift1})-(\ref{eqn:BGdrift2})
would reduce to (\ref{eq:velocity_factorize}), which is the minimal
kinetic energy solution of the following system of uncoupled Liouville
PDEs 
\[
\partial_{t}\pi_{t}(x_{i})=-\partial_{x_{i}}(\pi_{t}(x_{i})f_{i}(t,x_{i})),\quad i=1,\ldots,d,
\]
for $(t,x_{i})\in(0,1)\times\mathbb{R}$, given by Proposition \ref{prop:1Dcase}
for each marginal distribution. From (\ref{eqn:BGdrift1}), for $i=1,\ldots,d-1$
\begin{align}
f_{i}(t,x) & =\frac{\lambda'(t)}{\prod_{l=1}^{d}\pi_{t}(x_{l})}\Bigg(\prod_{j=1}^{i-1}\pi_{t}(x_{j})\int_{-\infty}^{x_{i}}\int_{\mathbb{R}^{i-1}}\Bigg(I_{t}-\sum_{l=1}^{i}\log L_{l}(u_{l})-\sum_{k=i+1}^{d}\log L_{k}(x_{k})\Bigg)\nonumber \\
 & \hspace{1.5cm}\times\prod_{j=1}^{i}\pi_{t}(u_{j})\prod_{k=i+1}^{d}\pi_{t}(x_{k})\,{\rm d}u_{1:i-1}{\rm d}u_{i}\nonumber \\
 & \hspace{1.5cm}-\prod_{j=1}^{i-1}\pi_{t}(x_{j})\int_{-\infty}^{x_{i}}\pi_{t}(u_{i})\,{\rm d}u_{i}\int_{\mathbb{R}^{i}}\Bigg(I_{t}-\sum_{l=1}^{i}\log L_{l}(u_{l})-\sum_{k=i+1}^{d}\log L_{k}(x_{k})\Bigg)\nonumber \\
 & \hspace{1.5cm}\times\prod_{j=1}^{i}\pi_{t}(u_{j})\prod_{k=i+1}^{d}\pi_{t}(x_{k})\,{\rm d}u_{1:i}\Bigg)\nonumber \\
 & =\frac{\lambda'(t)}{\pi_{t}(x_{i})}\Bigg(\int_{-\infty}^{x_{i}}\pi_{t}(u_{i})\,{\rm d}u_{i}\Bigg(I_{t}-\sum_{l=1}^{i-1}I_{t}^{(l)}-\sum_{k=i+1}^{d}\log L_{k}(x_{k})\Bigg)-\int_{-\infty}^{x_{i}}\log L_{i}(u_{i})\pi_{t}(u_{i})\,{\rm d}u_{i}\nonumber \\
 & \hspace{1.5cm}-\int_{-\infty}^{x_{i}}\pi_{t}(u_{i})\,{\rm d}u_{i}\Bigg(I_{t}-\sum_{l=1}^{i}I_{t}^{(l)}-\sum_{k=i+1}^{d}\log L_{k}(x_{k})\Bigg)\Bigg)\nonumber \\
 & =\frac{\lambda'(t)}{\pi_{t}(x_{i})}\Bigg(\int_{-\infty}^{x_{i}}(I_{t}^{(i)}-\log L_{i}(u_{i}))\pi_{t}(u_{i})\,{\rm d}u_{i}\Bigg),
\end{align}
and from (\ref{eqn:BGdrift2}) 
\begin{align}
f_{d}(t,x) & =\frac{\lambda'(t)}{\prod_{l=1}^{d}\pi_{t}(x_{l})}\Bigg(\prod_{j=1}^{d-1}\pi_{t}(x_{j})\int_{-\infty}^{x_{d}}\int_{\mathbb{R}^{d-1}}\sum_{l=1}^{d}(I_{t}^{(l)}-\log L_{l}(u_{l}))\prod_{k=1}^{d}\pi_{t}(u_{k})\,{\rm d}u_{1:d-1}{\rm d}u_{d}\Bigg)\nonumber \\
 & =\frac{\lambda'(t)}{\pi_{t}(x_{d})}\Bigg(\int_{-\infty}^{x_{d}}\pi_{t}(u_{d})\,{\rm d}u_{d}\Bigg(\sum_{l=1}^{d}I_{t}^{(l)}-\sum_{l=1}^{d-1}I_{t}^{(l)}\Bigg)-\int_{-\infty}^{x_{d}}\log L_{d}(u_{d})\pi_{t}(u_{d})\,{\rm d}u_{d}\Bigg)\nonumber \\
 & =\frac{\lambda'(t)}{\pi_{t}(x_{d})}\Bigg(\int_{-\infty}^{x_{d}}(I_{t}^{(d)}-\log L_{d}(u_{d}))\pi_{t}(u_{d})\,{\rm d}u_{d}\Bigg).
\end{align}

\section{Gibbs flow approximation\label{sec:gibbsflow_append}}

This section concerns properties of the Gibbs flow approximation.
We first give the proof of Proposition \ref{prop:gibbsFlow}.
\begin{proof}[Proof of Proposition \ref{prop:gibbsFlow}]
By Assumption A3, $\tilde{f}\in C^{1}([0,1]\times\mathbb{R}^{d},\mathbb{R}^{d})$
which implies that it is locally Lipschitz. With local Lipschitzness,
we need to establish that the solution $x(t;X_{0})$ of (\ref{eq:gibbs_ODE}),
with initial condition $X_{0}\sim\pi_{0}$, is bounded whenever it
exists to complete the proof. Boundedness will be obtained by showing
that $V$ is a Lyapunov function. Define $\alpha(R)=\max_{\{x\in\mathbb{R}^{d}:|x|\leq R\}}V(x)$
for $R>0$ and note that $\alpha(R)\rightarrow\infty$ as $R\rightarrow\infty$
under our assumption. Using Assumption A4, there exists $R_{1}>0$
such that 
\begin{align}
\frac{\mathrm{d}}{\mathrm{d}t}V(x(t))=\left<\nabla V(x(t)),\tilde{f}(t,x(t))\right>\leq0,\label{eq:lyapunov_gibbs}
\end{align}
for all $x(t)\in\mathbb{R}^{d}$ such that $|x(t)|\geq R_{1}$. It
follows that $|x(t;X_{0})|\leq\max\{R_{1},R_{2}(X_{0})\}<\infty$
where $R_{2}(X_{0})=\sup\{R>0:\alpha(R)\leq V(X_{0})\}$.
\end{proof}
We now show that Assumption A4 can be verified for the Gibbs velocity
field in the $d=1$ case by choosing $V(x)=|x|^{2}$. Clearly, $V\in C^{1}(\mathbb{R}^{d},\mathbb{R})$
and $V(x)\rightarrow\infty$ as $|x|\rightarrow\infty$. By assumption,
$\log L(x)\rightarrow-\infty$ as $|x|\rightarrow\infty$ so there
exists $R>0$ such that $\log L(x)<I_{t}$ for $|x|>R$. It follows
from (\ref{eqn:derivdensity}) that 
\[
-x\int_{-\infty}^{x}\partial_{t}\pi_{t}(u)\,\mathrm{d}u=x\int_{x}^{\infty}\partial_{t}\pi_{t}(u)\,\mathrm{d}u<0
\]
and therefore 
\[
\frac{\mathrm{d}}{\mathrm{d}t}V(x(t))=2x(t)\tilde{f}(t,x(t))<0
\]
for $|x|>R$. Next we detail the proof of Proposition \ref{prop:errorUpperBound}. 
\begin{proof}[Proof of Proposition \ref{prop:errorUpperBound}]
Let $f$ denote the velocity field (\ref{eqn:BGdrift1})-(\ref{eqn:BGdrift2})
in Proposition \ref{prop:generalCase} and recall that it satisfies
the Liouville equation (\ref{eq:liouville}). 
Define $\Delta:[0,1]\times\mathbb{R}^{d}\rightarrow\mathbb{R}$ as
the difference $\Delta_{t}(x)=\pi_{t}(x)-\tilde{\pi}_{t}(x)$ for
$(t,x)\in[0,1]\times\mathbb{R}^{d}$. By taking the difference between
(\ref{eq:liouville}) and 
\[
\partial\tilde{\pi}_{t}(x)=-\nabla\cdot(\tilde{\pi}_{t}(x)\tilde{f}(t,x))
\]
and introducing a cross term, we obtain 
\begin{align*}
\partial_{t}\Delta_{t}(x)=-\nabla\cdot(\pi_{t}(x)(f(t,x)-\tilde{f}(t,x))+\Delta_{t}(x)\tilde{f}(t,x)).
\end{align*}
Multiplying throughout by $\Delta_{t}(x)$ and applying chain rule
yields 
\begin{align*}
\frac{1}{2}\partial_{t}\Delta_{t}^{2}(x)=-(\nabla\cdot\tilde{f}(t,x))\Delta_{t}^{2}(x)-\frac{1}{2}\left<\tilde{f}(t,x),\nabla\Delta_{t}^{2}(x)\right>-\nabla\cdot(\pi_{t}(x)(f(t,x)-\tilde{f}(t,x)))\Delta_{t}(x).
\end{align*}
We then integrate by parts to obtain 
\begin{align*}
\partial_{t}\|\Delta_{t}\|_{L^{2}}^{2}=-\int_{\mathbb{R}^{d}}(\nabla\cdot\tilde{f}(t,x))\Delta_{t}^{2}(x)\,\mathrm{d}x-2\int_{\mathbb{R}^{d}}\nabla\cdot(\pi_{t}(x)(f(t,x)-\tilde{f}(t,x)))\Delta_{t}(x)\,\mathrm{d}x,
\end{align*}
noting that the boundary term vanishes by Assumption A5. Using Young's
inequality gives 
\begin{align}
\partial_{t}\|\Delta_{t}\|_{L^{2}}^{2} & \leq\left|\int_{\mathbb{R}^{d}}(\nabla\cdot\tilde{f}(t,x))\Delta_{t}^{2}(x)\,\mathrm{d}x\right|+2\left|\int_{\mathbb{R}^{d}}\nabla\cdot(\pi_{t}(x)(f(t,x)-\tilde{f}(t,x)))\Delta_{t}(x)\,\mathrm{d}x\right|\label{eqn:holderYoung}\\
 & \leq\|\nabla\cdot\tilde{f}\left(t,\cdot\right)\|_{\infty}\|\Delta_{t}\|_{L^{2}}^{2}+\delta^{-1}\|\Delta_{t}\|_{L^{2}}^{2}+\delta\|\varepsilon_{t}\|_{L^{2}}^{2}\nonumber 
\end{align}
for any $\delta>0$. Since $\tilde{\pi}_{0}=\pi_{0}$, integrating
both sides of (\ref{eqn:holderYoung}) on $[0,t]$ yields 
\begin{align*}
\|\Delta_{t}\|_{L^{2}}^{2}\leq\delta\int_{0}^{t}\|\varepsilon_{s}\|_{L^{2}}^{2}\,{\rm d}s+\int_{0}^{t}\left(\|\nabla\cdot\tilde{f}\left(s,\cdot\right)\|_{\infty}+\delta^{-1}\right)\|\Delta_{s}\|_{L^{2}}^{2}\,{\rm d}s.
\end{align*}
Now applying Gronwall's lemma on the time interval $[0,t]$ combined
with the fact that $t\mapsto\delta\int_{0}^{t}\|\varepsilon_{s}\|_{L^{2}}^{2}\,{\rm d}s$
is non-decreasing 
\begin{align*}
\|\Delta_{t}\|_{L^{2}}^{2}\leq\delta\int_{0}^{t}\|\varepsilon_{s}\|_{L^{2}}^{2}\,{\rm d}s\thinspace\cdot\thinspace\exp\left(t\delta^{-1}+\int_{0}^{t}\|\nabla\cdot\tilde{f}(s,\cdot)\|_{\infty}\,{\rm d}s\right).
\end{align*}
Lastly, minimizing this upper bound w.r.t. $\delta$ gives (\ref{eqn:errorUpperBound}). 
\end{proof}

\section{Numerical integration of the Gibbs flow \label{sec:Numerical-integration-Gibbsflow}}

In the following, we will show that the numerical integration scheme
(\ref{eq:novel_integrator}) is a first order method, i.e. the global
error 
\[
|e_{m}|=|X_{m}-x(t_{m})|=|T_{t_{m}}(X_{0})-\hat{T}_{t_{m}}(X_{0})|=O(h)
\]
for all $m=0,\ldots,M$, if the step size $h$ is sufficiently small.
For ease of presentation, we will consider $p=2$ components; extension
to the case $p>2$ is straightforward but less instructive. We consider
the case where the maps $\Phi_{m}=\Psi_{m,2}\circ\Psi_{m,1}$ are
defined by Euler discretizations, i.e. 
\begin{equation}
\Psi_{m,1}(x_{1},x_{2})=\left(\begin{array}{c}
x_{1}\\
x_{2}
\end{array}\right)+\left(\begin{array}{c}
h\hat{f}_{1}(t,x_{1},x_{2})\\
0
\end{array}\right),\quad\Psi_{m,2}(x_{1},x_{2})=\left(\begin{array}{c}
x_{1}\\
x_{2}
\end{array}\right)+\left(\begin{array}{c}
0\\
h\hat{f}_{2}(t,x_{1},x_{2})
\end{array}\right).\label{eq:two_euler_discretization}
\end{equation}
By a Taylor expansion, the case where $\Psi_{m,1}$ and $\Psi_{m,2}$
are defined by analytically tractable flows (when the other component
is fixed) differs from (\ref{eq:two_euler_discretization}) by an
$O(h^{2})$ term; therefore, it will be apparent that there is no
loss of generality in considering Euler discretizations. We will assume
that the (approximated) Gibbs velocity field $\hat{f}$ is Lipschitz
continuous, i.e. $|\hat{f}(t,x)-\hat{f}(t,u)|\leq\ell_{f}|x-u|$ for
all $(t,x),(t,u)\in[0,1]\times\mathbb{R}^{d}$.

For each step size $h>0$, we define the function $F_{h}:[0,1]\times\mathbb{R}^{d}\rightarrow\mathbb{R}^{d}$
as
\[
F_{h}(t,x_{1},x_{2})=\left(\begin{array}{c}
\hat{f}_{1}(t,x_{1},x_{2})\\
\hat{f}_{2}(t,x_{1}+h\hat{f}_{1}(t,x_{1},x_{2}),x_{2})
\end{array}\right),
\]
which allows us to express the numerical integration scheme as a one-step
method 
\begin{equation}
X_{m}=\Phi_{m}(X_{m-1})=X_{m-1}+hF_{h}(t_{m-1},X_{m-1,1},X_{m-1,2}).\label{eq:one-step-method}
\end{equation}
We note that $F_{h}$ is also Lipschitz continuous with a Lipschitz
constant $L_{F}$ that depends on $\ell_{f}$. We first examine the
local truncation error 
\begin{equation}
|\varepsilon_{m}|=\left|\frac{x(t_{m})-x(t_{m-1})}{h}-F_{h}(t_{m-1},x_{1}(t_{m-1}),x_{2}(t_{m-1}))\right|.\label{eq:local_truncation_error}
\end{equation}
Assuming that the solution defined by (\ref{eq:gibbs_velocity_approximated})
satisfies $x\in C^{2}([0,1],\mathbb{R}^{d})$, by Taylor's theorem
\[
x(t_{m})=x(t_{m-1})+h\hat{f}(t_{m-1},x(t_{m-1}))+\frac{1}{2}h^{2}x''(\xi_{m-1})
\]
for some $\xi_{m-1}\in(t_{m-1},t_{m})$, where $x''(t)=(\mathrm{d}^{2}/\mathrm{d}t^{2})x(t)$
denotes the second derivative of $x(t)$. Substituting this expansion
into (\ref{eq:local_truncation_error}), we have by Lipschitz continuity
of $\hat{f}$ and the form of (\ref{eq:two_euler_discretization})
that 
\begin{align*}
|\varepsilon_{m}| & =\left|\left(\begin{array}{c}
0\\
\hat{f}_{2}(t_{m-1},x(t_{m-1}))-\hat{f}_{2}(t_{m-1},\Psi_{m,1}(x(t_{m-1})))
\end{array}\right)+\frac{1}{2}hx''(\xi_{m-1})\right|\\
 & \leq\ell_{f}|x(t_{m-1})-\Psi_{m,1}(x(t_{m-1}))|+\frac{1}{2}h|x''(\xi_{m-1})|\\
 & \leq\ell_{f}h|\hat{f}_{1}(t_{m-1},x(t_{m-1}))|+\frac{1}{2}h|x''(\xi_{m-1})|.
\end{align*}
Defining $D_{1}=\sup_{t\in[0,1]}|x'(t)|$ and $D_{2}=\sup_{t\in[0,1]}|x''(t)|$,
the local truncation errors are bounded by 
\begin{equation}
|\varepsilon_{m}|\leq(\ell_{f}D_{1}+D_{2}/2)h.\label{eq:bound_local_errors}
\end{equation}

To relate local truncation errors to global errors, we rewrite (\ref{eq:local_truncation_error})
as

\[
x(t_{m})=x(t_{m-1})+hF_{h}(t_{m-1},x_{1}(t_{m-1}),x_{2}(t_{m-1}))+h\varepsilon_{m}
\]
and subtract (\ref{eq:one-step-method}) from this equation to obtain
\[
e_{m}=e_{m-1}+h\left(F_{h}(t_{m-1},x_{1}(t_{m-1}),x_{2}(t_{m-1}))-F_{h}(t_{m-1},X_{m-1,1},X_{m-1,2})\right)+h\varepsilon_{m}.
\]
By Lipschitz continuity of $F_{h}$ and the bound in (\ref{eq:bound_local_errors}),
we have 
\[
|e_{m}|\leq|e_{m-1}|+hL_{F}|e_{m-1}|+h^{2}(\ell_{f}D_{1}+D_{2}/2).
\]
Applying Gronwall's Lemma then shows that the global errors satisfy
\[
|e_{m}|\leq(\ell_{f}D_{1}+D_{2}/2)L_{F}^{-1}\left(\exp(t_{m}L_{F})-1\right)h.
\]

\section{Gibbs flow samplers with resampling\label{sec:extra_algorithms}}

In this section, we detail modifications of Algorithms \ref{alg:GF-SIS}-\ref{alg:GF-AIS}
to incorporate resampling at every time step. We will use the notation
$\mathcal{R}(W^{1},\ldots,W^{N})$ to denote a resampling operation
based on a vector of normalized weights $\{W^{n}\}_{n=1,\ldots,N}$,
i.e. $W^{n}\geq0$ for all $n=1,\ldots,N$ and $\sum_{n=1}^{N}W^{n}=1$.
In this case, Algorithm \ref{alg:GF-SISR} replaces Algorithm \ref{alg:GF-SIS}
and Algorithm \ref{alg:GF-SMC} replaces Algorithm \ref{alg:GF-AIS}.
The normalizing constant estimators $\hat{Z}_{M}$ returned by Algorithms
\ref{alg:GF-SISR}-\ref{alg:GF-SMC} are both unbiased estimators
of the marginal likelihood $Z=\int_{\mathbb{R}^{d}}\pi_{0}(x)L(x)\,\mathrm{d}x$.
To approximate expectations of the form $\int_{\mathbb{R}^{d}}\phi(x)\pi(x)\:\mathrm{d}x$,
we will use $N^{-1}\sum_{n=1}^{N}\phi(X_{M}^{A_{M}^{n}})$ from the
output of Algorithm \ref{alg:GF-SISR}, and $N^{-1}\sum_{n=1}^{N}\phi(\tilde{X}_{M}^{n})$
from the output of Algorithm \ref{alg:GF-SMC}.

\begin{algorithm}[H]
\protect\caption{Gibbs flow sequential importance sampling resampling (GF-SISR)~\label{alg:GF-SISR}}

\textbf{Input}: prior $\pi_{0}$, likelihood $L$, inverse temperature
$\lambda$, step size $h$, and Gibbs velocity field $\tilde{f}$.

For time step $m=0$

\qquad{}For $n=1,\ldots,N$

\qquad{}\qquad{}(a) sample $X_{0}^{n}=(X_{0,1}^{n},\ldots,X_{0,p}^{n})\sim\pi_{0}$;

\qquad{}\qquad{}(b) set $w_{0}^{n}=1$, $W_{0}^{n}=N^{-1}$ and
$A_{0}^{n}=n$;

\qquad{}(c) set $\mathrm{ESS}_{0}=N$ and $\hat{Z}_{0}=1$.

\textcompwordmark{}

For time step $m=1,\ldots,M$ 

\qquad{}For $n=1,\ldots,N$

\qquad{}\qquad{}For $i=1,\ldots,p$

\qquad{}\qquad{}\qquad{}(d) set $(X_{m,1:i}^{n},X_{m-1,(i+1):p}^{A_{m-1}^{n}})=\Psi_{m,i}(X_{m,1:i-1}^{n},X_{m-1,i:p}^{A_{m-1}^{n}})$
using Section \ref{subsec:Numerical-implementation};

\qquad{}\qquad{}\qquad{}(e) compute $J_{m,i}^{n}=\det(\nabla\Psi_{m,i}(X_{m,1:i-1}^{n},X_{m-1,i:p}^{A_{m-1}^{n}}))$
using Section \ref{subsec:Distribution-of-approximate};

\qquad{}\qquad{}(f) set $X_{m}^{n}=(X_{m,1}^{n},\ldots,X_{m,p}^{n})$
and $J_{m}^{n}=\prod_{i=1}^{p}J_{m,i}^{n}$;

\qquad{}\qquad{}(g) compute unnormalized weights 
\[
w_{m}^{n}=\frac{\gamma_{t_{m}}(X_{m}^{n})}{\gamma_{t_{m-1}}(X_{m-1}^{A_{m-1}^{n}})|J_{m}^{n}|^{-1}};
\]

\qquad{}\qquad{}(h) compute normalized weights $W_{m}^{n}=w_{m}^{n}/\sum_{\ell=1}^{N}w_{m}^{\ell}$;

\qquad{}\qquad{}(i) sample ancestor index $A_{m}^{n}\sim\mathcal{R}(W_{m}^{1},\ldots,W_{m}^{N})$;

\qquad{}(j) compute effective sample size $\mathrm{ESS}_{m}=\left\{ \sum_{n=1}^{N}(W_{m}^{n})^{2}\right\} ^{-1}$;

\qquad{}(k) compute normalizing constant estimator $\hat{Z}_{m}=\hat{Z}_{m-1}N^{-1}\sum_{n=1}^{N}w_{m}^{n}$.

\textbf{Output}: samples $\{X_{M}^{n}\}_{n=1,\ldots,N}$, ancestors
$\{A_{M}^{n}\}_{n=1,\ldots,N}$ and normalizing constant estimator
$\hat{Z}_{M}$.
\end{algorithm}

\begin{algorithm}[H]
\protect\caption{Gibbs flow sequential Monte Carlo sampler (GF-SMC)~\label{alg:GF-SMC}}

\textbf{Input}: prior $\pi_{0}$, likelihood $L$, inverse temperature
$\lambda$, step size $h$, Gibbs velocity field $\tilde{f}$, MCMC
kernels $\{K_{m}\}_{m=1,\ldots,M}$. 

For time step $m=0$

\qquad{}For $n=1,\ldots,N$

\qquad{}\qquad{}(a) sample $X_{0}^{n}=(X_{0,1}^{n},\ldots,X_{0,p}^{n})\sim\pi_{0}$
and set $\tilde{X}_{0}^{n}=X_{0}^{n}$;

\qquad{}\qquad{}(b) set $w_{0}^{n}=1$ and $W_{0}^{n}=N^{-1}$;

\qquad{}(c) set $\mathrm{ESS}_{0}=N$ and $\hat{Z}_{0}=1$.

\textcompwordmark{}

For time step $m=1,\ldots,M$ 

\qquad{}For $n=1,\ldots,N$

\qquad{}\qquad{}For $i=1,\ldots,p$

\qquad{}\qquad{}\qquad{}(d) set $(X_{m,1:i}^{n},\tilde{X}_{m-1,(i+1):p}^{n})=\Psi_{m,i}(X_{m,1:i-1}^{n},\tilde{X}_{m-1,i:p}^{n})$
using Section \ref{subsec:Numerical-implementation};

\qquad{}\qquad{}\qquad{}(e) compute $J_{m,i}^{n}=\det(\nabla\Psi_{m,i}(X_{m,1:i-1}^{n},\tilde{X}_{m-1,i:p}^{n}))$
using Section \ref{subsec:Distribution-of-approximate};

\qquad{}\qquad{}(f) set $X_{m}^{n}=(X_{m,1}^{n},\ldots,X_{m,p}^{n})$
and $J_{m}^{n}=\prod_{i=1}^{p}J_{m,i}^{n}$;

\qquad{}\qquad{}(g) compute unnormalized weights 
\[
w_{m}^{n}=w_{m-1}^{n}\frac{\gamma_{t_{m}}(X_{m}^{n})}{\gamma_{t_{m-1}}(\tilde{X}_{m-1}^{n})|J_{m}^{n}|^{-1}};
\]

\qquad{}\qquad{}(h) compute normalized weights $W_{m}^{n}=w_{m}^{n}/\sum_{\ell=1}^{N}w_{m}^{\ell}$;

\qquad{}\qquad{}(i) sample ancestor index $A_{m}^{n}\sim\mathcal{R}(W_{m}^{1},\ldots,W_{m}^{N})$;

\qquad{}\qquad{}(j) sample $\tilde{X}_{m}^{n}\sim K_{m}(X_{m}^{A_{m}^{n}},\cdot)$
from $\pi_{t_{m}}$-invariant MCMC kernel;

\qquad{}(k) compute effective sample size $\mathrm{ESS}_{m}=\left\{ \sum_{n=1}^{N}(W_{m}^{n})^{2}\right\} ^{-1}$;

\qquad{}(l) compute normalizing constant estimator $\hat{Z}_{m}=\hat{Z}_{m-1}N^{-1}\sum_{n=1}^{N}w_{m}^{n}$.

\textbf{Output}: samples $\{\tilde{X}_{M}^{n}\}_{n=1,\ldots,N}$ and
normalizing constant estimator $\hat{Z}_{M}$.
\end{algorithm}

\section{Flow transports for curve of Gaussian distributions \label{sec:gaussian_curve} }

In this section, we illustrate the flow transports introduced in Section
\ref{sec:Solving-the-flow-transport-problem} on a curve of Gaussian
distributions. 

\subsection{Curve of Gaussian distributions \label{subsec:Curve-of-Gaussian}}

Consider the prior distribution $\pi_{0}(x)=\mathcal{N}(x;\mu_{0},\Sigma_{0})$
and likelihood function 
\[
L(x;y)=\exp\left(-\frac{1}{2}\left\langle x-y,R^{-1}(x-y)\right\rangle \right)
\]
with symmetric positive definite $R\in\mathbb{R}^{d\times d}$ and
observation $y\in\mathbb{R}^{d}$. By conjugacy, the curve of distributions
$\{\pi_{t}\}_{t\in[0,1]}$ defined in (\ref{eqn:tempering}) lies
in the Gaussian family, i.e. $\pi_{t}(x)=\mathcal{N}(x;\mu_{t},\Sigma_{t})$
for $t\in[0,1]$ with 
\begin{align}
\Sigma_{t}^{-1}=\Sigma_{0}^{-1}+\lambda(t)R^{-1},\quad\mu_{t}=\Sigma_{t}\left(\Sigma_{0}^{-1}\mu_{0}+\lambda(t)R^{-1}y\right),\label{eqn:gaussianParameters}
\end{align}
and the expected log-likelihood (\ref{eqn:expectloglike}) is 
\begin{align}
I_{t}=-\frac{1}{2}\left(\textrm{Tr}(R^{-1}\Sigma_{t})+\left\langle \mu_{t}-y,R^{-1}(\mu_{t}-y)\right\rangle \right)\label{eqn:gaussianIT}
\end{align}
where $\mathrm{Tr}(A)$ denotes the trace of a square matrix $A$.
In this Gaussian setting, the minimum kinetic energy solution to the
flow transport problem \cite{bergemannreich,reichMixture,cotter_reich}
\begin{equation}
f^{*}(t,\cdot)=\arg\min_{\varphi\in\mathcal{L}(\pi_{t})}\frac{1}{2}\int_{\mathbb{R}^{d}}\left\langle \varphi(x),\Sigma_{t}^{-1}\varphi(x)\right\rangle \pi_{t}(x)\,\mathrm{d}x\label{eq:min_kinetic_energy}
\end{equation}
where
\[
\mathcal{L}(\pi_{t})=\{\varphi:\mathbb{R}^{d}\rightarrow\mathbb{R}^{d}:\int_{\mathbb{R}^{d}}\left\langle \varphi(x),\Sigma_{t}^{-1}\varphi(x)\right\rangle \pi_{t}(x)\,\mathrm{d}x<\infty,\,\varphi\mbox{ satisfies }\eqref{eq:liouville}\mbox{ for all }x\in\mathbb{R}^{d}\mbox{ at }t\in[0,1]\},
\]
is analytically tractable and is given by
\begin{align}
f(t,x)=-\frac{\lambda'(t)}{2}\Sigma_{t}R^{-1}(x+\mu_{t}-2y).\label{eqn:min_kinetic_gaussian}
\end{align}

\subsection{Univariate case\label{subsec:Scalar-case-normal}}

\begin{figure}[htbp]
\hspace{5cm}\includegraphics[scale=0.35]{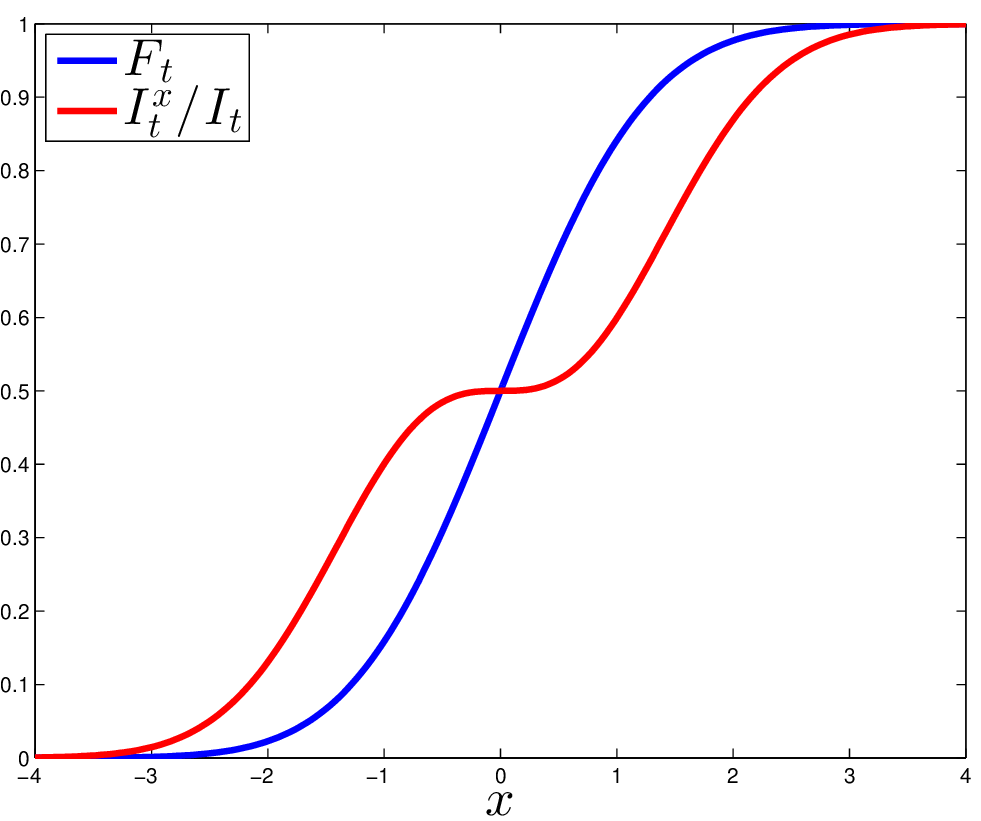} \caption{Illustrating steering property of (\ref{eqn:min_kinetic_gaussian})
on univariate Gaussian example with $\lambda(t)=t$ and $t=0$. }
\label{fig:1DflowIntuition} 
\end{figure}

As noted in Proposition \ref{prop:1Dcase}, the velocity field in
(\ref{eqn:drift1D}) corresponds exactly to (\ref{eq:min_kinetic_energy})
when $d=1$. For a more concrete example, we shall consider $\mu_{0}=0,\Sigma_{0}=1,y=0,R=1$.
The curve of distributions in Appendix \ref{subsec:Curve-of-Gaussian}
is given by $\pi_{t}(x)=\mathcal{N}\left(x;0,(1+\lambda(t))^{-1}\right)$,
so as time progresses, we expect particles to have a mean-reverting
behaviour towards the origin. This is indeed the case in (\ref{eqn:min_kinetic_gaussian})
which gives a linear mean-reverting drift $f(t,x)=-\lambda'(t)x\{2(1+\lambda(t))\}^{-1}$.
Figure \ref{fig:1DflowIntuition} also illustrates this behaviour
with the steering property mentioned in Section \ref{subsec:Flow-transport-problem-onrealline}:
since $I_{t}=-\{2(1+\lambda(t))\}^{-1}<0$ for all $t\in[0,1]$, reversion
to the stable stationary point at the origin dictates that $F_{t}(x)<I_{t}^{x}/I_{t}$
for $x<0$ and $F_{t}(x)>I_{t}^{x}/I_{t}$ for $x>0$. 

\subsection{Pathological case\label{subsec:divergence}}

Using the same arguments in the proof of Propositions \ref{prop:1Dcase}-\ref{prop:generalCase},
the velocity field $\bar{f}$ given in (\ref{eqn:antiDerivative})
satisfies Liouville equation (\ref{eq:liouville}) whenever $\sum_{i=1}^{d}\alpha_{i}=1$.
However, Theorem \ref{thm:ambrosio} does not apply as $|\bar{f}_{i}(t,x)|\pi_{t}(x)\nrightarrow0$
as $x_{i}\rightarrow\infty$ for any $i=1,\ldots,d$, so Assumption
A2 does not hold. On a simple Gaussian example detailed below, we
will establish that (\ref{eqn:antiDerivative}) does not solve the
flow transport problem by showing that an ODE with velocity field
$\bar{f}$ would yield divergent particle trajectories.

Consider $d=2$ and the curve of distributions in Appendix
\ref{subsec:Curve-of-Gaussian} with parameters $\mu_{0}=(0,0),\Sigma_{0}=R=I_{2}$
and $y=(0,0)$. This setup corresponds to independent components that
are marginally distributed according to the univariate Gaussian model
of Appendix \ref{subsec:Scalar-case-normal}. Hence
we would expect a particle under a valid flow transport to have a
mean-reverting behaviour towards the origin. The velocity field in
(\ref{eqn:antiDerivative}) has the form 
\begin{align}
\begin{pmatrix}\bar{f}_{1}(t,x)\\
\bar{f}_{2}(t,x)
\end{pmatrix}=\begin{pmatrix}\frac{\alpha_{1}\lambda'(t)}{2\pi_{t}(x_{1})}\left(\int_{-\infty}^{x_{1}}u_{1}^{2}\pi_{t}(u_{1})\,{\rm d}u_{1}+x_{2}^{2}F_{t}(x_{1})-\frac{F_{t}(x_{1})}{1+\lambda(t)}\right)\\
\frac{\alpha_{2}\lambda'(t)}{2\pi_{t}(x_{2})}\left(\int_{-\infty}^{x_{2}}u_{2}^{2}\pi_{t}(u_{2})\,{\rm d}u_{2}+x_{1}^{2}F_{t}(x_{2})-\frac{F_{t}(x_{2})}{1+\lambda(t)}\right)
\end{pmatrix}\label{eqn:antiD_gaussian}
\end{align}
for $x=(x_{1},x_{2})\in\mathbb{R}^{2}$ and $t\in[0,1]$, where $\pi_{t}(x)=\mathcal{N}\left(x;(0,0),(1+\lambda(t))^{-1}I_{2}\right)$
and $F_{t}(x_{i})$ denotes the marginal CDFs. We note that the two
components of the velocity field are coupled.

Now consider $\alpha_{1},\alpha_{2}>0$ with $\alpha_{1}+\alpha_{2}=1$.
We investigate the behaviour of particles in the upper-right quadrant
of the space. For each $t\in[0,1]$, define the sets $\mathcal{S}_{t}=\left\lbrace x\in\mathbb{R}^{2}:x_{1},x_{2}>1/\sqrt{1+\lambda(t)}\right\rbrace $
and $\mathcal{P}_{t}=\left\lbrace x\in\mathbb{R}^{2}:\bar{f}(t,x)>(0,0)\right\rbrace $;
noting that $\lambda'(t)>0$, it follows from (\ref{eqn:antiD_gaussian})
that $\mathcal{S}_{0}\subset\mathcal{S}_{t}\subset\mathcal{P}_{t}$
for any $t\in(0,1]$. Since $\pi_{0}(\mathcal{S}_{0})>0$, we can
conclude that there exist particle trajectories which only move farther
away from the origin with positive probability. Analytical tractability
in this simple example allows us to strengthen the previous statement
and show that these trajectories in fact blow up in finite time. We
start by seeking a lower bound on $\bar{f}$; by symmetry, it suffices
to consider only the first component. On the set $\mathcal{S}_{0}$,
we have $\int_{-\infty}^{x_{1}}u_{1}^{2}\pi_{t}(u_{1})\,{\rm d}u_{1}>\{2(1+\lambda(t))\}^{-1}\geq\frac{1}{4}$,
hence
\begin{align}
\bar{f}_{1}(t,x)\geq\frac{c}{4}\exp\left(\frac{1}{2}x_{1}^{2}\right)\geq\frac{c}{32}x_{1}^{4},\label{eqn:lowerBoundAnti-1}
\end{align}
where $c=\min_{i=1,2}\frac{\alpha_{i}\sqrt{\pi}}{2}\inf_{t\in[0,1]}\lambda'(t)>0$.
Now consider an uncoupled system of ODEs with velocity field 
\begin{align}
\begin{pmatrix}\hat{f}_{1}(t,x_{1})\\
\hat{f}_{2}(t,x_{2})
\end{pmatrix}=\begin{pmatrix}\frac{c}{32}x_{1}^{4}\\
\frac{c}{32}x_{2}^{4}
\end{pmatrix}\leq\begin{pmatrix}\bar{f}_{1}(t,x)\\
\bar{f}_{2}(t,x)
\end{pmatrix},
\end{align}
and note that its solution $x_{i}(t;x_{0,i})=1/\sqrt[3]{3\left(\frac{1}{3x_{0,i}^{3}}-\frac{c}{32}t\right)}$,
corresponding to an initial condition $x_{0}=(x{}_{0,1},x_{0,2})\in\mathbb{R}^{2}$,
diverges as $t\rightarrow\frac{32}{3cx_{0,i}^{3}}$. Define the set
$\mathcal{V}=\left\lbrace x\in\mathbb{R}^{2}:x_{1},x_{2}>\sqrt[3]{\frac{32}{3c}}\right\rbrace $.
Noting that $\hat{f}$ is locally Lipschitz and component-wise increasing,
the comparison theorem \cite[Theorem III.10.XII (b), p. 112]{walter1998}
implies that a particle starting in $\mathcal{S}_{0}\cap\mathcal{V}$
and evolving under (\ref{eqn:antiD_gaussian}) has a trajectory that
explodes before $t=1$. Since $\pi_{0}(\mathcal{S}_{0}\cap\mathcal{V})>0$,
we conclude the claim that there exist divergent particle trajectories
with positive probability. 

\subsection{Multivariate case\label{subsec:Multidimensional-flow-transport-normal}}

\begin{figure}[htbp]
\hspace{-1.8cm}\hspace{1.5cm}\includegraphics[scale=0.35]{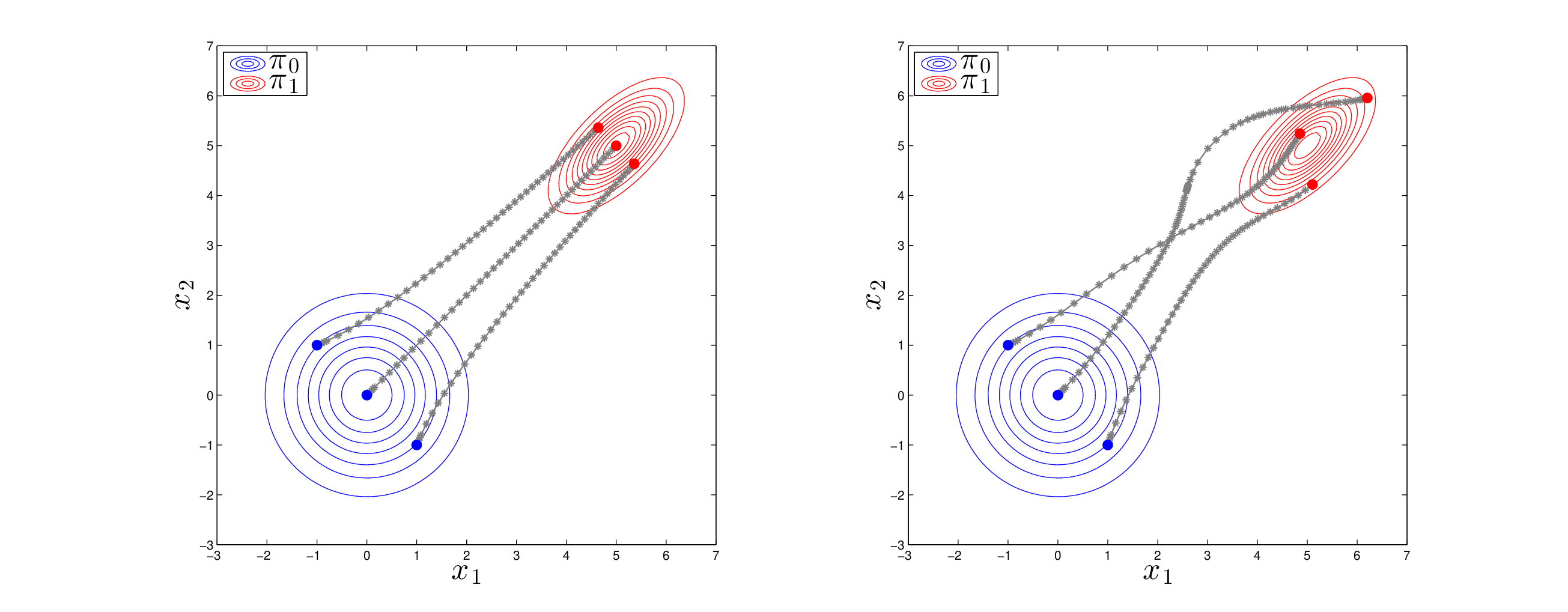}
\protect\caption{Bivariate Gaussian example. Three particle trajectories driven under
different velocity fields but with the same initial conditions in
both panels: (\emph{left}) minimal kinetic energy velocity field (\ref{eqn:min_kinetic_gaussian});
(\emph{right}) velocity field (\ref{eqn:BGdrift1})-(\ref{eqn:BGdrift2})
in Proposition \ref{prop:generalCase}. The asterisk symbols displayed
correspond to steps taken by an adaptive explicit fourth-order Runge-Kutta
numerical integrator.}

\label{fig:2DgaussianTraj} 
\end{figure}

Consider $d=2$ and the curve of distributions in Appendix
\ref{subsec:Curve-of-Gaussian} with parameters $\mu_{0}=(0,0),\Sigma_{0}=I_{2},R=\begin{pmatrix}1 & \rho\\
\rho & 1
\end{pmatrix},y=(14.25,14.25)$ and $\rho=0.85$. In this setting, as time progresses, the independent
prior distribution simultaneously gets deformed and translated. Figure
\ref{fig:2DgaussianTraj} illustrates that, on average, particles
driven by (\ref{eqn:min_kinetic_gaussian}) require less kinetic energy
than that of (\ref{eqn:BGdrift1})-(\ref{eqn:BGdrift2}). However,
in the general non-Gaussian case, obtaining the minimal kinetic energy
velocity field requires numerical resolution of an elliptic PDE. 

\subsection{Gibbs flow approximation }

\begin{figure}[htbp]
\hspace{-1.8cm}\hspace{1.6cm}\includegraphics[scale=0.35]{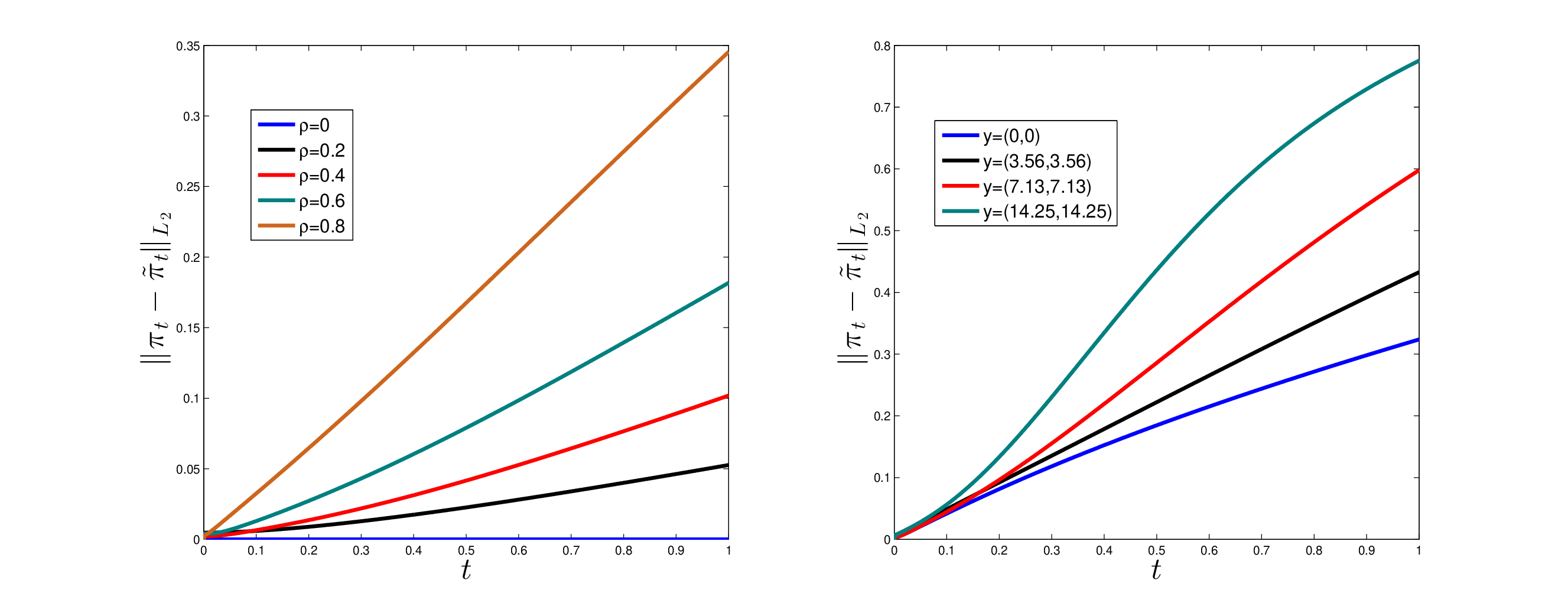}
\caption{Bivariate Gaussian example. Error in $L^{2}$-norm at varying degrees
of correlation $\rho$ (\emph{left}) and extremality of the observation
$y$ (\emph{right}).}

\label{fig:L2norm} 
\end{figure}

To illustrate the nature of the Gibbs flow approximation (\ref{eqn:gibbsAnti}),
we consider the setting in Appendix \ref{subsec:Multidimensional-flow-transport-normal}
and observe the $L^{2}$-error analyzed in Proposition \ref{prop:errorUpperBound}
at varying degrees of correlation, induced by the parameter $\rho$,
and extremality of the observation $y$. The left panel of Figure
\ref{fig:L2norm} shows that while performance degrades with $\rho$,
as expected from our construction, the approximation is able to exploit
any local independence structure in the target distributions, thus
keeping the error reasonably small for moderate degrees of correlation.
The right panel of Figure \ref{fig:L2norm} reveals the inadequacy
of the approximation when the overlap between the prior distribution
and the likelihood function decreases, which is also to be expected. 

\section{Expressions for variance component models\label{sec:Expressions-for-variance-component}}

This section provides additional details for the variance component
models of Section \ref{subsec:Variance-component-models}. The summary
statistics of the full conditional distributions in (\ref{eq:vcmodel_fullconditionals})
are given by 
\begin{align*}
 & \alpha(t)=\alpha_{1}+\lambda(t)\left\{ \alpha_{0}-\alpha_{1}+\frac{K}{2}\right\} ,\beta(t|\mu,\theta)=\beta_{1}+\lambda(t)(\beta_{0}-\beta_{1})+\frac{1}{2}\lambda(t)\sum_{i=1}^{K}(\theta_{i}-\mu)^{2},\\
 & \nu(t|\sigma_{\theta}^{2},\theta)=\frac{\sigma_{\theta}^{2}(1-\lambda(t))\mu_{1}+\lambda(t)\sigma_{1}^{2}\sum_{i=1}^{K}\theta_{i}}{\sigma_{\theta}^{2}(1-\lambda(t))+\sigma_{1}^{2}K\lambda(t)},\quad\varsigma^{2}(t|\sigma_{\theta}^{2})=\frac{\sigma_{1}^{2}\sigma_{\theta}^{2}}{\sigma_{\theta}^{2}(1-\lambda(t))+\sigma_{1}^{2}K\lambda(t)},\\
 & \xi_{i}(t|\sigma_{\theta}^{2},\mu,y)=\frac{\sigma_{e}^{2}m(t|\sigma_{\theta}^{2},\mu)+\lambda(t)s^{2}(t|\sigma_{\theta}^{2})\sum_{j=1}^{J}y_{ij}}{\sigma_{e}^{2}+\lambda(t)s^{2}(t|\sigma_{\theta}^{2})},\quad i=1,\ldots,K,\\
 & \tau^{2}(t|\sigma_{\theta}^{2})=\frac{s^{2}(t|\sigma_{\theta}^{2})\sigma_{e}^{2}}{\sigma_{e}^{2}+\lambda(t)s^{2}(t|\sigma_{\theta}^{2})},
\end{align*}
where 
\[
m(t|\sigma_{\theta}^{2},\mu)=\frac{\lambda(t)\mu\sigma_{2}^{2}+(1-\lambda(t))\mu_{2}\sigma_{\theta}^{2}}{\lambda(t)\sigma_{2}^{2}+(1-\lambda(t))\sigma_{\theta}^{2}},\quad s^{2}(t|\sigma_{\theta}^{2})=\frac{\sigma_{\theta}^{2}\sigma_{2}^{2}}{\lambda(t)\sigma_{2}^{2}+(1-\lambda(t))\sigma_{\theta}^{2}}.
\]

\section{Toy examples}

In this section, we consider two additional examples to investigate
the quality of the Gibbs flow approximation.

\subsection{Banana-shaped posterior}

First we consider a banana-shaped posterior distribution on $x=(x_{1},x_{2})\in\mathbb{R}^{2}$,
induced by the prior distribution 
\begin{equation}
\pi_{0}(x)=\mathcal{N}(x_{1};0,1)\mathcal{N}(x_{2};0,1)\label{eq:toy_prior}
\end{equation}
and the likelihood function 
\[
L(x)=\exp(-(\alpha-x_{1})^{2}-\beta(x_{2}-x_{1}^{2})^{2})
\]
that is defined by the Rosenbrock function. The log-likelihood function
is not concave and has a global maximum at $(x_{1},x_{2})=(\alpha,\alpha^{2})$.
Therefore the parameter $\alpha\geq0$ controls the overlap between
the prior distribution and the likelihood function. The parameter
$\beta\geq0$ specifies the strength of the dependency between the
variables $x_{1}$ and $x_{2}$: having $\beta=0$ would give an independent
posterior distribution $\pi(x)\propto\pi_{0}(x)L(x)$, while larger
values of $\beta$ would induce more dependent posterior distributions.
In the following, we will consider $\alpha=5$ and $\beta=10$.

\begin{figure}[htbp]
\includegraphics[scale=0.5]{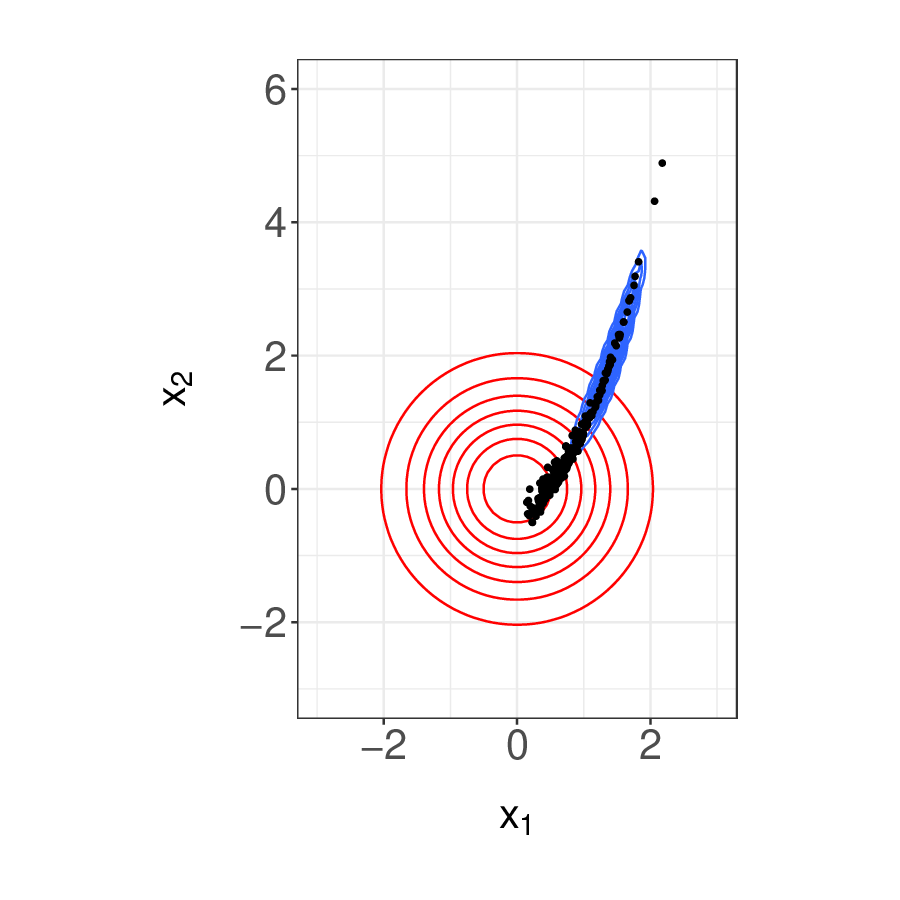}\includegraphics[scale=0.5]{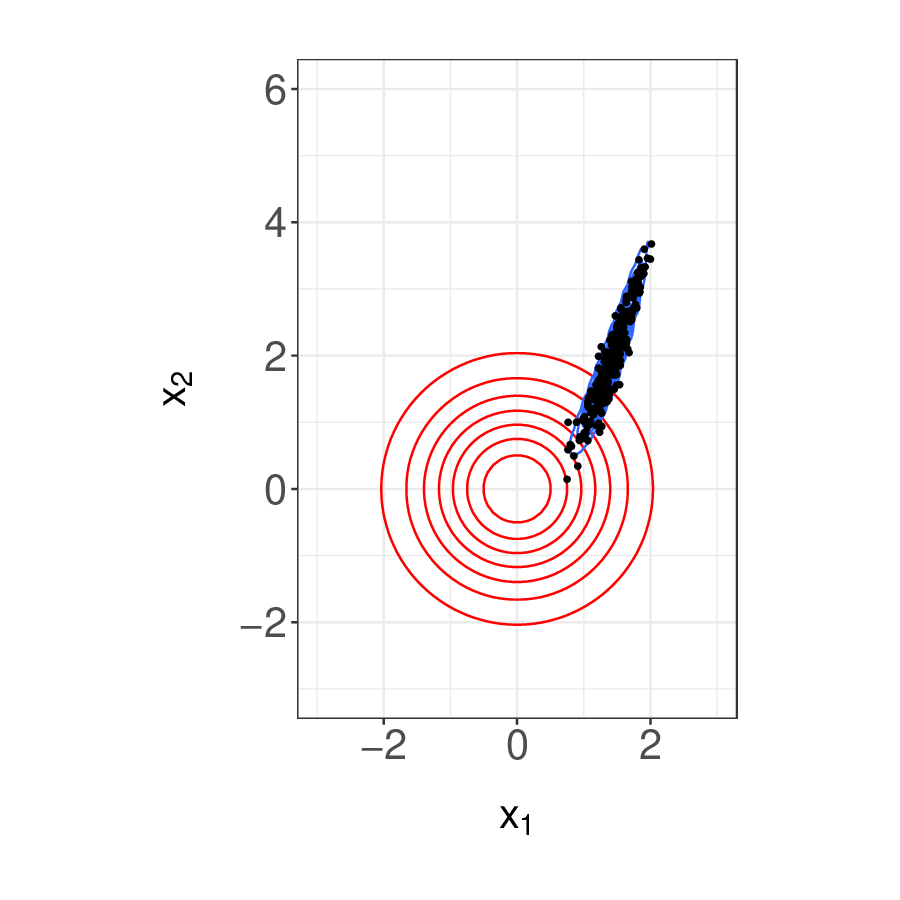}
\includegraphics[scale=0.5]{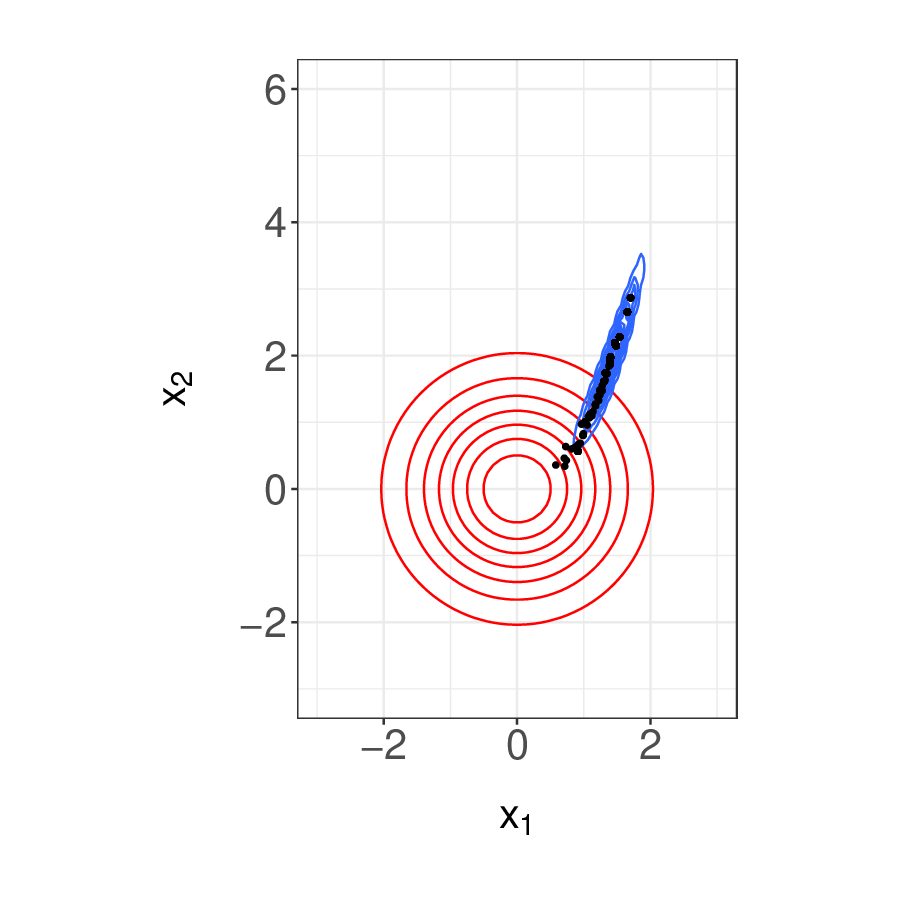}\includegraphics[scale=0.5]{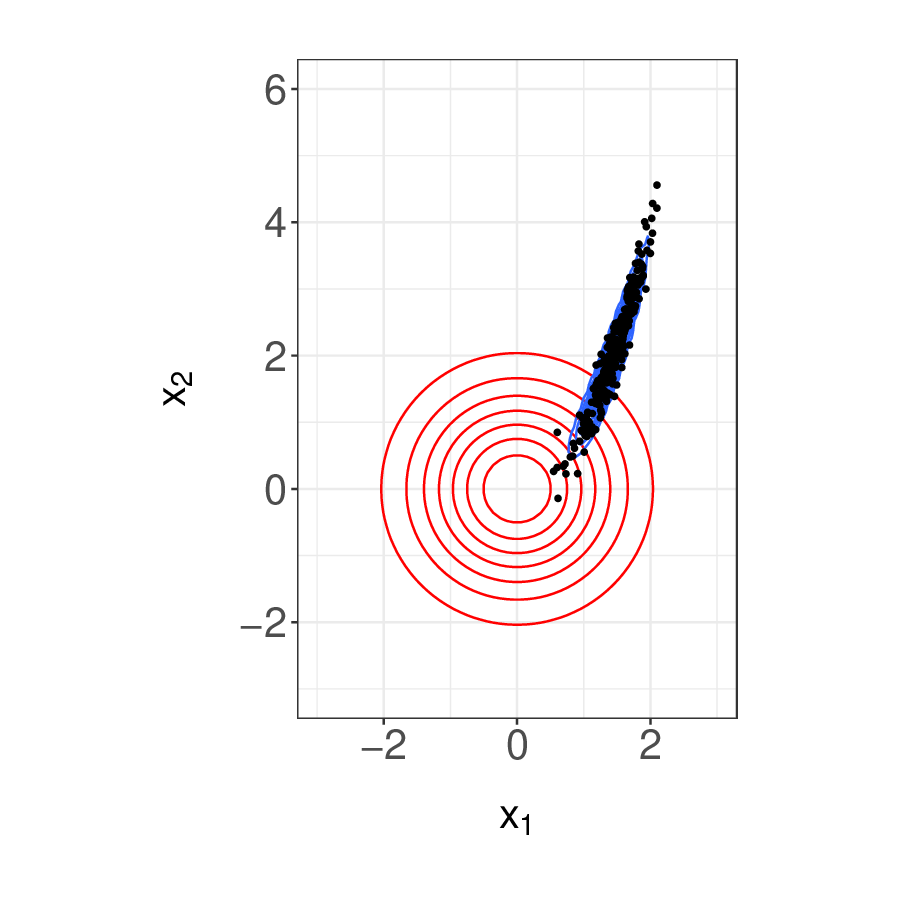}\caption{Terminal position of samples (\emph{black dots}) under GF-SIS (\emph{upper
left}), GF-AIS (\emph{upper right}), GF-SISR (\emph{lower left}) and
GF-SMC (\emph{lower right}) for a banana-shaped posterior distribution.
The superimposed red and blue contours correspond to the prior and
posterior densities respectively. The plots on the right column illustrate
the impact of adding MCMC moves, while the plots on the lower row
show the effect of weighting and resampling steps.}
\label{fig:banana}
\end{figure}

Using a composite trapezoidal rule with $R=200$ quadrature points
and the Euler discretization (\ref{eq:euler_discretization}) with
$M=200$ time steps, the upper left panel of Figure \ref{fig:banana}
shows the terminal position of $N=256$ samples against contours of
the prior and posterior densities. To improve performance, we either
combine the Gibbs flow with HMC kernels within GF-AIS (Algorithm \ref{alg:GF-AIS})
or include weighting and resampling steps within GF-SISR (Algorithm
\ref{alg:GF-SISR}). The upper right panel of Figure \ref{fig:banana}
illustrates the benefits of adding MCMC moves to prevent accumulation
of errors; the lower left panel of Figure \ref{fig:banana} shows
that weighting and resampling offers improvement at the expense of
sample diversity, as this results in a duplicate set of samples. Lastly,
the lower right panel of Figure \ref{fig:banana} demonstrates that
sample diversity can be rejuvenated by combining resampling with MCMC
moves within GF-SMC (Algorithm \ref{alg:GF-SMC}). 

\subsection{Gaussian mixture posterior}

Next we examine a multimodal posterior distribution on $x=(x_{1},x_{2})\in\mathbb{R}^{2}$,
given by the prior distribution (\ref{eq:toy_prior}) and a likelihood
function of the form 
\[
L(x)=\sum_{j=1}^{4}w_{j}\mathcal{N}(y_{j};x,R_{j}).
\]
The weights satisfy $w_{j}\geq0$ for $j=1,\ldots,4$ and $\sum_{j=1}^{4}w_{j}=1$,
the observations are 
\[
y_{1}=(-\xi,\xi),\quad y_{2}=(\xi,\xi),\quad y_{3}=(-\xi,-\xi),\quad y_{4}=(\xi,-\xi),
\]
for some location parameter $\xi\geq0$ and $R_{1}=R_{4}=R_{-}$,
$R_{2}=R_{3}=R_{+}$ with 
\[
R_{-}=\left(\begin{array}{cc}
1 & -\rho\\
-\rho & 1
\end{array}\right),\quad R_{+}=\left(\begin{array}{cc}
1 & \rho\\
\rho & 1
\end{array}\right),
\]
for some correlation parameter $\rho\in[0,1]$. It can be shown that
the posterior distribution is a Gaussian mixture 
\[
\pi(x)\propto\pi_{0}(x)L(x)\propto\sum_{j=1}^{4}w_{j}\mathcal{N}(x;\mu_{j},\Sigma_{j})
\]
with mean vectors $\mu_{j}=(I_{2}+R_{j})^{-1}y_{j}$ for $j=1,\ldots,4$
and covariance matrices 
\[
\Sigma_{1}=\Sigma_{4}=(I_{2}+R_{-}^{-1})^{-1},\quad\Sigma_{2}=\Sigma_{3}=(I_{2}+R_{+}^{-1})^{-1}.
\]
We will set $w_{1}=w_{3}=0.4$, $w_{2}=w_{4}=0.1$, $\xi=6$ and vary
$\rho\in\{0,0.15,0.30,0.45,0.60,0.75,0.90\}$.

\begin{figure}[htbp]
\includegraphics[scale=0.5]{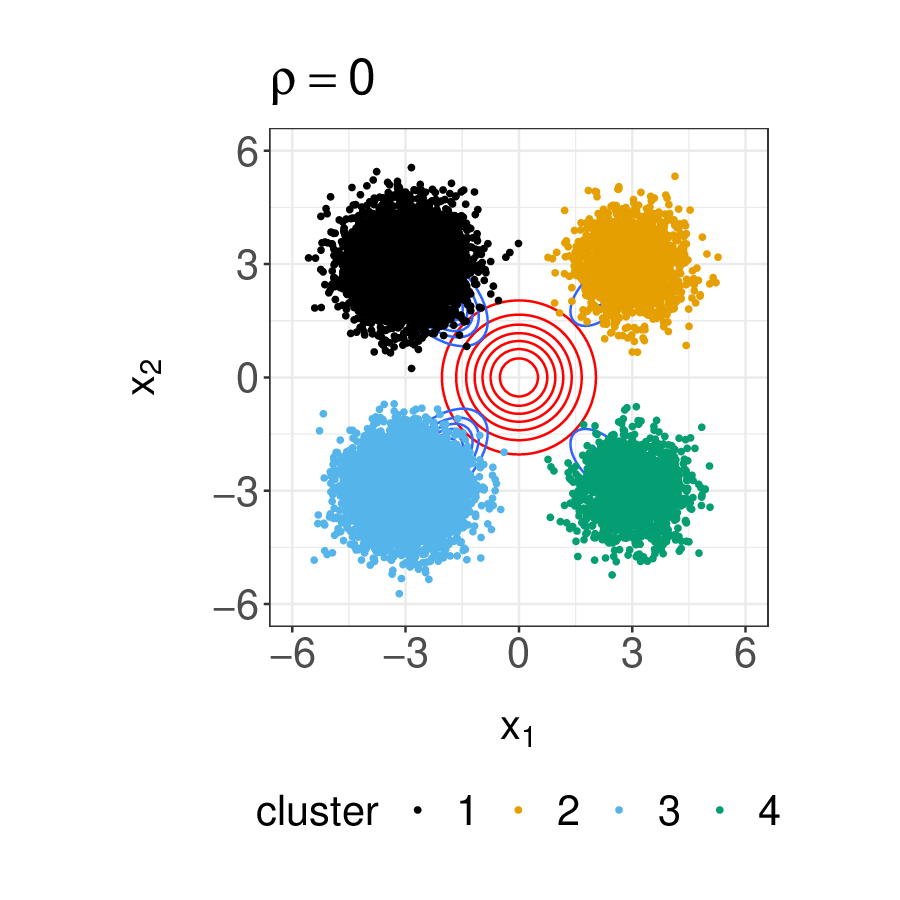}\includegraphics[scale=0.5]{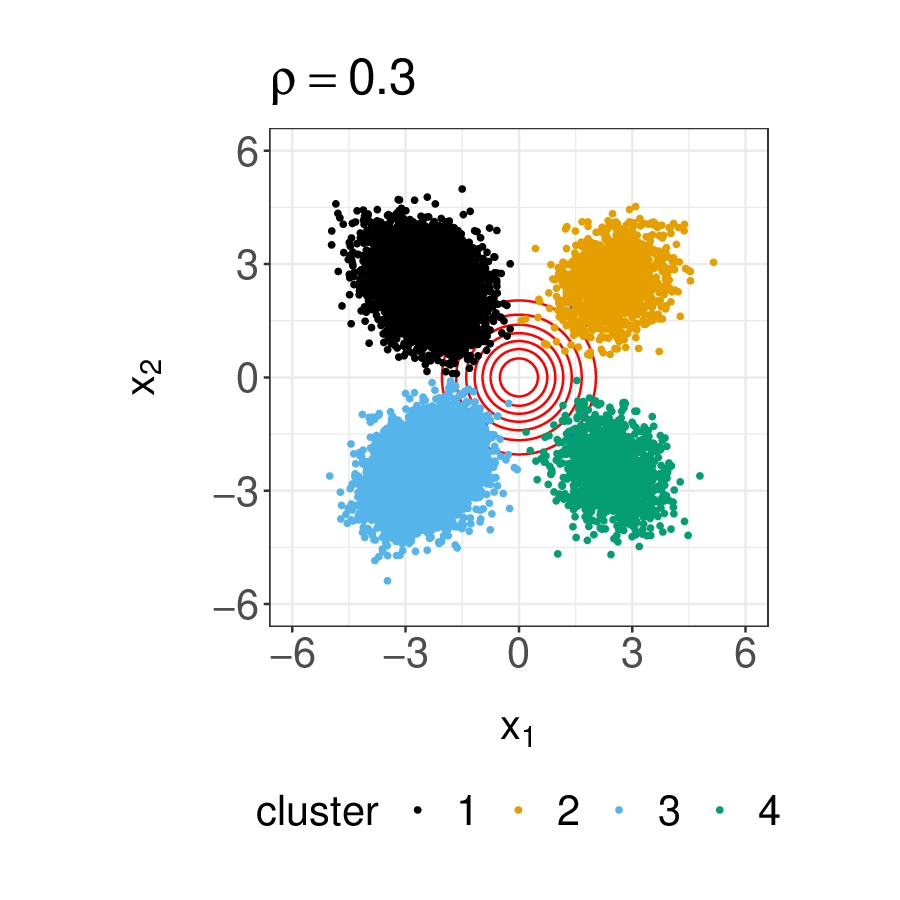}
\includegraphics[scale=0.5]{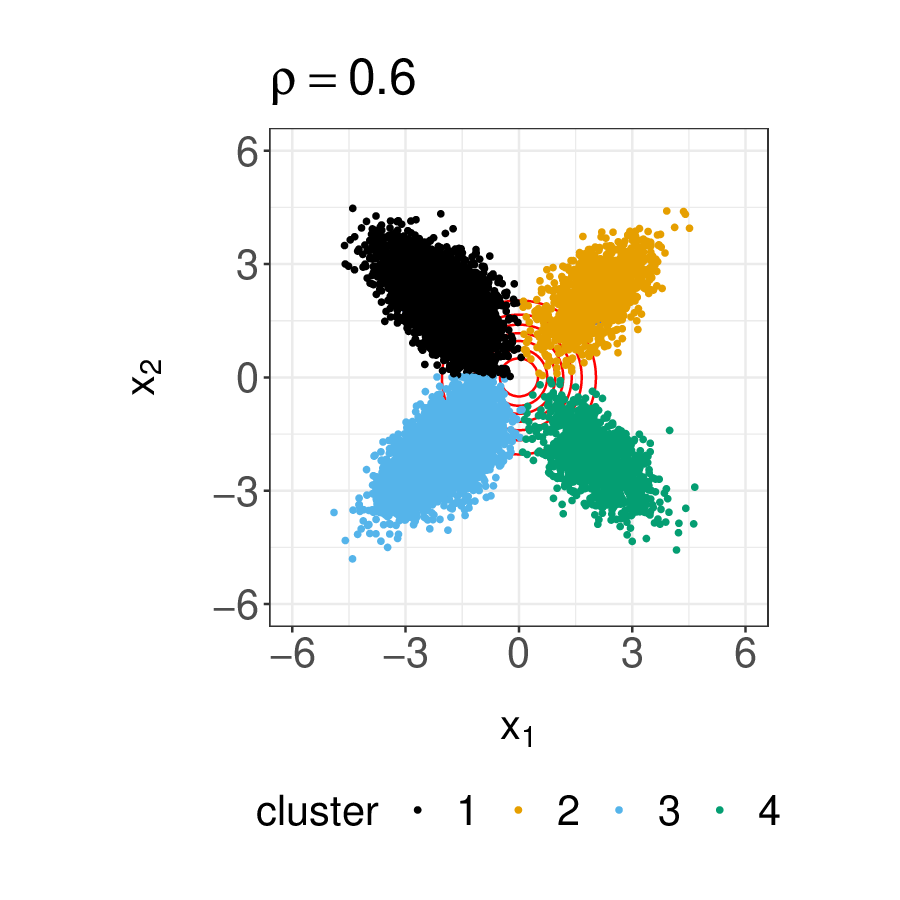}\includegraphics[scale=0.5]{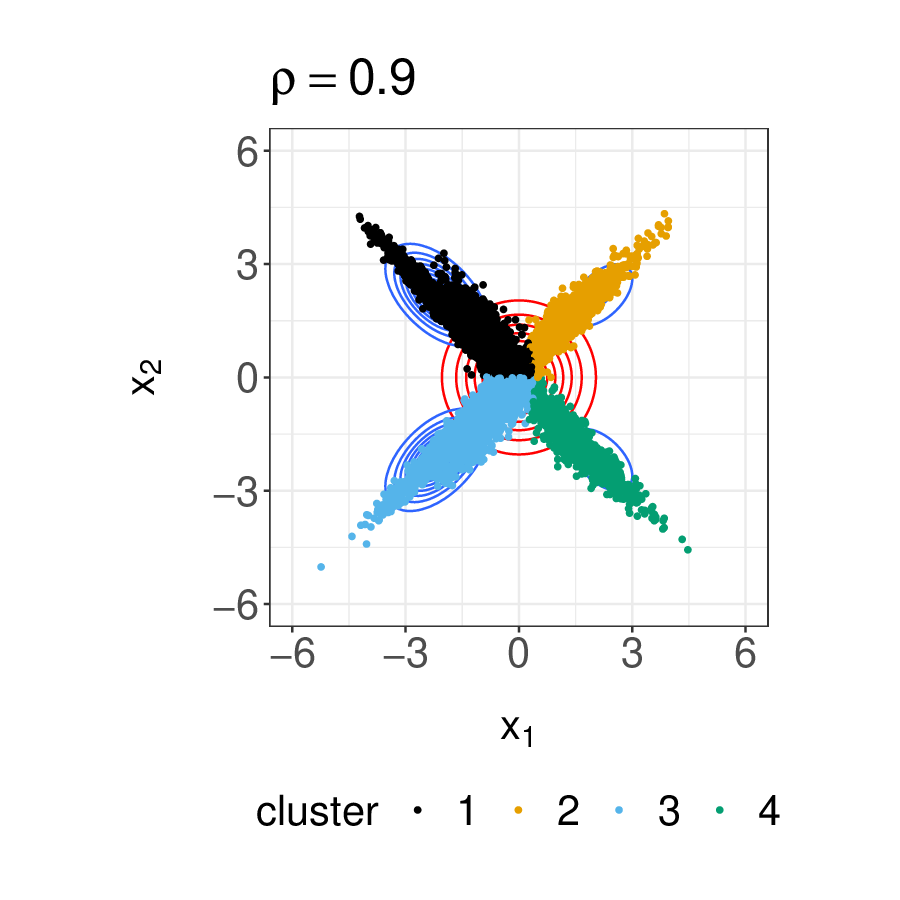}\caption{\label{fig:gmixture}Terminal position of Gibbs flow samples (\emph{dots})
for Gaussian mixture posterior distributions as the correlation parameter
$\rho$ varies. The colored dots represent a clustering obtained using
$K$-means.  The superimposed red and blue contours correspond to
the prior and posterior densities respectively. }
\end{figure}

To approximate the Gibbs flow, we employ a composite trapezoidal rule
with $R=200$ quadrature points and the default ODE solver from the
\texttt{deSolve} R package. To estimate the mixture weights $(w_{j})$
using the terminal positions of $N=16,384$ samples under the Gibbs
flow, we run a $K$-means clustering algorithm with $K=4$ clusters
and initialization at the posterior means $(\mu_{j})$. The proportion
of samples in each of the four clusters are taken as estimates of
$(w_{j})$ and reported in Table \ref{tab:Empirical-proportion} for
the values of $\rho$ that are considered. We also compute the p-values
of Pearson\textquoteright s Chi-squared goodness-of-fit tests under
the null hypothesis that the population weights are equal to $(w_{1},w_{2},w_{3},w_{4})=(0.4,0.1,0.4,0.1)$.
Figure \ref{fig:gmixture} displays the terminal position of samples
with color-coding to show the clustering for various values of $\rho$,
against contours of the prior and posterior densities.

It is apparent from Table \ref{tab:Empirical-proportion} that the
estimation error increases with the correlation parameter $\rho$,
which is to be expected. The relative error in estimating $w_{1}=w_{3}=0.4$
ranges from $0.05\%$ to $8.2\%$, while that of $w_{2}=w_{4}=0.1$
ranges from $0.3\%$ to $33.5\%$. The p-values that were computed
with $N=16,384$ samples, on the other hand, indicate poor approximation
of the true values $(w_{1},w_{2},w_{3},w_{4})=(0.4,0.1,0.4,0.1)$
for all non-zero values of $\rho$. Lastly, we note that one can improve
the quality of the Gibbs flow approximation by adding MCMC moves or
weighting and resampling steps, as discussed above.

\begin{table}
\begin{centering}
\begin{tabular}{|c|c|c|c|c|c|}
\hline 
$\rho$ & $w_{1}$ & $w_{2}$ & $w_{3}$ & $w_{4}$ & p-value\tabularnewline
\hline 
$0$ & $0.4002$ & $0.1007$ & $0.3994$ & $0.0997$ & $9.9\times10^{-1}$\tabularnewline
\hline 
$0.15$ & $0.4090$ & $0.0957$ & $0.4031$ & $0.0922$ & $8.4\times10^{-4}$\tabularnewline
\hline 
$0.30$ & $0.4113$ & $0.0851$ & $0.4151$ & $0.0885$ & $1.3\times10^{-15}$\tabularnewline
\hline 
$0.45$ & $0.4048$ & $0.0878$ & $0.4222$ & $0.0852$ & $2.2\times10^{-16}$\tabularnewline
\hline 
$0.60$ & $0.4185$ & $0.0823$ & $0.4180$ & $0.0813$ & $2.2\times10^{-16}$\tabularnewline
\hline 
$0.75$ & $0.4207$ & $0.0779$ & $0.4247$ & $0.0767$ & $2.2\times10^{-16}$\tabularnewline
\hline 
$0.90$ & $0.4317$ & $0.0665$ & $0.4327$ & $0.0691$ & $2.2\times10^{-16}$\tabularnewline
\hline 
\end{tabular}
\par\end{centering}
\caption{\label{tab:Empirical-proportion}Empirical estimates of Gaussian mixture
weights $(w_{j})$ based on $K$-means clustering as the correlation
parameter $\rho$ varies. The reported p-values correspond to Pearson\textquoteright s
Chi-squared goodness-of-fit tests under the null hypothesis that the
population weights are equal to $(w_{1},w_{2},w_{3},w_{4})=(0.4,0.1,0.4,0.1)$.}
\end{table}

\end{document}